\documentclass{emulateapj}

\usepackage{natbib}
\usepackage{amsmath}
\usepackage{url}
\usepackage{longtable}

\newcommand{\bvri}{\protect\hbox{$BV\!RI$} }
\newcommand{\sbvri}{\protect\hbox{$BV\!ri$} }

\newcommand{\ubvgri}{\protect\hbox{$uBV\!gri$} }
\newcommand{\ubvgrrii}{\protect\hbox{$uBV\!grRiI$} }

\newcommand{\vri}{\protect\hbox{$V\!RI$} }

\newcommand{\about}{$\sim\!\!$~}
\newcommand{\kms}{\,km\,s$^{-1}$}

\newcommand{\err}[2]{\ensuremath{^{+#1}_{-#2}}}

\def\lsim{\hbox{\rlap{\raise 0.425ex\hbox{$<$}}\lower 0.65ex\hbox{$\sim$}}}
\def\gsim{\hbox{\rlap{\raise 0.425ex\hbox{$>$}}\lower 0.65ex\hbox{$\sim$}}}

\def\arcdeg{\mbox{$^\circ$}}

\shorttitle{Type I\lowercase{ax} Supernovae}
\shortauthors{Foley et~al.}

\begin{document}

\title{Type I\lowercase{ax} Supernovae: A New Class of Stellar
  Explosion\footnotemark[1]}\footnotetext[1]{This paper is dedicated
  to the memory of our friend and colleague, Dr. Weidong Li, a pioneer
  in the identification and detailed study of this class of objects.}

\def\cfa{2}
\def\clay{3}
\def\berk{4}
\def\li{5}
\def\lco{6}
\def\ab{7}
\def\aar{8}
\def\ut{9}
\def\thca{10}
\def\chile{11}
\def\car{12}
\def\rut{13}
\def\unc{14}
\def\am{15}
\def\mitch{16}

\author{
{Ryan~J.~Foley}\altaffilmark{\cfa,\clay}, 
{P.~J.~Challis}\altaffilmark{\cfa},
{R.~Chornock}\altaffilmark{\cfa},
{M.~Ganeshalingam}\altaffilmark{\berk},
{W.~Li}\altaffilmark{\berk,\li},
{G.~H.~Marion}\altaffilmark{\cfa},
{N.~I.~Morrell}\altaffilmark{\lco},\\
{G.~Pignata}\altaffilmark{\ab},
{M.~D.~Stritzinger}\altaffilmark{\aar},
{J.~M.~Silverman}\altaffilmark{\berk, \ut},
{X.~Wang}\altaffilmark{\thca},
{J.~P.~Anderson}\altaffilmark{\chile},
{A.~V.~Filippenko}\altaffilmark{\berk},\\
{W.~L.~Freedman}\altaffilmark{\car},
{M.~Hamuy}\altaffilmark{\chile},
{S.~W.~Jha}\altaffilmark{\rut},
{R.~P.~Kirshner}\altaffilmark{\cfa},
{C.~McCully}\altaffilmark{\rut},
{S.~E.~Persson}\altaffilmark{\car},\\
{M.~M.~Phillips}\altaffilmark{\lco},
{D.~E.~Reichart}\altaffilmark{\unc}, \and
{A.~M.~Soderberg}\altaffilmark{\cfa}
}

\altaffiltext{\cfa}{
Harvard-Smithsonian Center for Astrophysics,
60 Garden Street, 
Cambridge, MA 02138, USA
}
\altaffiltext{\clay}{
Clay Fellow. Electronic address rfoley@cfa.harvard.edu .
}
\altaffiltext{\berk}{
Department of Astronomy,
University of California,
Berkeley, CA 94720-3411, USA
}
\altaffiltext{\li}{
Deceased 12 December 2011
}
\altaffiltext{\lco}{
Carnegie Observatories,
Las Campanas Observatory,
La Serena, Chile 
}
\altaffiltext{\ab}{
Departamento de Ciencias Fisicas,
Universidad Andres Bello,
Avda.\ Republica 252,
Santiago, Chile
}
\altaffiltext{\aar}{
Department of Physics and Astronomy,
Aarhus University,
Ny Munkegade,
DK-8000 Aarhus C, Denmark
}
\altaffiltext{\ut}{
Department of Astronomy,
University of Texas,
Austin, TX 78712-0259, USA
}
\altaffiltext{\thca}{
Physics Department and Tsinghua Center for Astrophysics (THCA),
Tsinghua University,
Beijing 100084, China
}
\altaffiltext{\chile}{
Departamento de Astronom\'{i}a,
Universidad de Chile,
Casilla 36-D, Santiago, Chile
}
\altaffiltext{\car}{
Observatories of the Carnegie Institution of Washington,
813 Santa Barbara St.,
Pasadena, CA 91101, USA
}
\altaffiltext{\rut}{
Department of Physics and Astronomy,
Rutgers, the State University of New Jersey,
136 Frelinghuysen Road,
Piscataway, NJ 08854, USA
}
\altaffiltext{\unc}{
Department of Physics and Astronomy,
University of North Carolina at Chapel Hill,
Chapel Hill, NC, USA
}

\begin{abstract}
  We describe observed properties of the Type Iax class of supernovae
  (SNe~Iax), consisting of SNe observationally similar to its
  prototypical member, SN~2002cx.  The class currently has 25 members,
  and we present optical photometry and/or optical spectroscopy for
  most of them.  SNe~Iax are spectroscopically similar to SNe~Ia, but
  have lower maximum-light velocities ($2000 \lesssim |v| \lesssim
  8000$~\kms), typically lower peak magnitudes ($-14.2 \ge M_{V, {\rm
  peak}} \gtrsim -18.9$~mag), and most have hot photospheres.
  Relative to SNe~Ia, SNe~Iax have low luminosities for their
  light-curve shape.  There is a correlation between luminosity and
  light-curve shape, similar to that of SNe~Ia, but offset from that
  of SNe~Ia and with larger scatter.  Despite a host-galaxy morphology
  distribution that is highly skewed to late-type galaxies without any
  SNe~Iax discovered in elliptical galaxies, there are several
  indications that the progenitor stars are white dwarfs (WDs):
  evidence of C/O burning in their maximum-light spectra, low
  (typically \about 0.5~$M_{\sun}$) ejecta masses, strong Fe lines in
  their late-time spectra, a lack of X-ray detections, and deep limits
  on massive stars and star formation at the SN sites.  However, two
  SNe~Iax show strong He lines in their spectra.  The progenitor
  system and explosion model that best fits all of the data is a
  binary system of a C/O WD that accretes matter from a He star and
  has a deflagration.  At least some of the time, this explosion will
  not disrupt the WD.  The small number of SNe in this class prohibit
  a detailed analysis of the homogeneity and heterogeneity of the
  entire class.  We estimate that in a given volume there are
  $31\err{17}{13}$ SNe~Iax for every 100 SNe~Ia, and for every
  1~$M_{\sun}$ of iron generated by SNe~Ia at $z = 0$, SNe~Iax
  generate \about 0.036~$M_{\sun}$.  Being the largest class of
  peculiar SNe, thousands of SNe~Iax will be discovered by LSST.
  Future detailed observations of SNe~Iax should further our
  understanding of both their progenitor systems and explosions as
  well as those of SNe~Ia.
\end{abstract}



\section{Introduction}\label{s:intro}

Most thermonuclear supernovae are spectroscopically defined as Type
Ia.  These supernovae (SNe) lack hydrogen and helium in their spectra
(except perhaps from circumstellar interaction), and most have strong
lines from intermediate mass elements (IMEs) in their
near-maximum-light spectra (see \citealt{Filippenko97} for a review of
SN classification).  The bulk of observational diversity within this
group can be described by a single parameter that relates peak
luminosity with light-curve shape \citep[a width-luminosity relation
or WLR;][]{Phillips93}, intrinsic color \citep{Riess96}, and $^{56}$Ni
mass \citep{Mazzali07}.  However, there is additional diversity
related to ejecta velocity \citep[e.g.,][]{Benetti05, Foley11:vel,
Ganeshalingam11}.  The ability to collapse the observational
diversity of this class to one or two parameters suggests that most
SNe~Ia have similar progenitor stars (although not necessarily
progenitor systems, as some SN~Ia observables correlate with their
progenitor environment; \citealt{Foley12:csm}) and explosion
mechanisms.  There are also examples of particular thermonuclear SNe
that do not follow this parameterization \citep[e.g.,][]{Li01:00cx,
Howell06, Foley10:06bt, Ganeshalingam12}, which may be the result of
these SNe having different progenitors and/or explosion mechanisms
than most SNe~Ia, rather than being extreme examples of the normal
SN~Ia progenitor system and explosion mechanism.

Almost all thermonuclear SNe that are outliers to the trends defined
by SNe~Ia are part of a single, relatively large class.  Members of
this class, previously labeled ``SN~2002cx-like'' after the
prototypical object \citep{Li03:02cx}, have peak magnitudes
$\gtrsim$1~mag below that of normal SNe~Ia, spectra that show
low-velocity ejecta, and maximum-light spectra that typically resemble
those of the high-luminosity SN~Ia 1991T (blue continua and absorption
from higher-ionization species consistent with a hot photosphere).
Studying thermonuclear outliers can both help determine what
progenitors and explosions mechanisms do {\it not} produce normal
SNe~Ia and constrain various models by examining the extremes of the
population.

In addition to the properties mentioned above, the SN~2002cx-like
class has several observational properties that distinguish it from
that of normal SNe~Ia: low luminosity for its light-curve shape
\citep[e.g.,][]{Li03:02cx}, no observed second maximum in the
near-infrared (NIR) bands \citep[e.g.,][]{Li03:02cx}, late-time
spectra dominated by narrow permitted \ion{Fe}{2} lines
\citep{Jha06:02cx, Sahu08}, but can occasionally have strong
[\ion{Fe}{2}] emission \citep{Foley10:08ge}, strong mixing of the
ejecta \citep{Jha06:02cx, Phillips07}, and a host-galaxy morphology
distribution highly skewed to late-type galaxies, and no member of
this class has been discovered in an elliptical galaxy
\citep{Foley09:08ha, Valenti09}.  Additionally, some members of the
class, such as SN~2007J, display strong \ion{He}{1} lines in their
spectra \citep{Foley09:08ha}.

Because of these physical distinctions as well as others discussed in
this paper, we designate this class of objects ``Type Iax supernovae''
or SNe~Iax.  This designation indicates the observational and physical
similarities to SNe~Ia, but also emphasizes the physical differences
between SNe~Iax and normal SNe~Ia (e.g., SNe~Iax are not simply a
subclass of SNe~Ia) and will hopefully reduce confusion within the
literature.

Several of the extreme characteristics of some members of this class,
including low kinetic energy and significant mixing in the ejecta, may
be consistent with a full deflagration of a white dwarf (WD)
\citep{Branch04, Phillips07}, rather than a deflagration that
transitions into a detonation as expected for normal SNe~Ia
\citep{Khokhlov91}.  Because of their low velocities, which eases line
identification and helps probe the deflagration process, which is
essential to all SN~Ia explosions, this class is particularly useful
for understanding typical SN~Ia explosions.

An extreme member of this class, SN~2008ha \citep{Foley09:08ha,
Valenti09, Foley10:08ha}, was much fainter (peaking at $M_{V} =
-14.2$~mag) and had a significantly lower velocity ($|v| \approx
2000$~\kms) than the typical member.  Although its maximum-light
spectrum indicates that the object underwent C/O burning
\citep{Foley10:08ha}, certain observations are consistent with a
massive-star progenitor \citep{Foley09:08ha, Valenti09, Moriya10}.
Nonetheless, a massive-star progenitor is inconsistent with other
observables \citep{Foley10:08ha}.  SN~2008ha generated \about
$10^{-3}$~$M_{\sun}$ of $^{56}$Ni and ejected \about 0.3~$M_{\sun}$ of
material \citep{Foley10:08ha}, suggesting that the most plausible
explanation was a failed deflagration of a WD \citep{Foley09:08ha,
  Foley10:08ha}.

Furthermore, another member of this class, SN~2008ge, was hosted in an
S0 galaxy with no signs of star formation or massive stars, including
at the SN position in pre-explosion {\it Hubble Space Telescope} ({\it
HST}) images \citep{Foley10:08ge}.  SN~2008ge most likely had a WD
progenitor.  SN~2008ha had an inferred ejecta mass of \about $0.3
M_{\sun}$, which is less than the total mass of any WD expected to
explode as a SN.  If SN~2008ha had a WD progenitor, then its
progenitor star was not completely disrupted during the explosion
\citep{Foley08:08ha, Foley09:08ha, Foley10:08ha}.

Although there have been a number of papers on individual members of
this class \citep{Li03:02cx, Branch04, Chornock06, Jha06:02cx,
Phillips07, Sahu08, Foley09:08ha, Foley10:08ha, Foley10:08ge,
Valenti09, Maund10:05hk, McClelland10, Narayan11, Kromer12}, a
holistic view of the entire class has not yet been published.  Here we
present new data for several SNe and examine the properties of all
known members of the class, totaling 25 SNe, with the intention of
further understanding the relations between observational properties
of the class, their progenitor systems, and their explosions.

The manuscript is structured in the following way.
Section~\ref{s:class} outlines the criteria for membership in the
class and details the members of the class.  We present previously
published and new observations of the SNe in Section~\ref{s:obs}.  We
describe the photometric and spectroscopic properties of the class in
Sections~\ref{s:phot} and \ref{s:spec}, respectively.  In
Section~\ref{s:rates}, we provide estimates of the relative rate of
SNe~Iax to normal SNe~Ia and the Fe production from SNe~Iax.  We
summarize the observations and constrain possible progenitor systems
in Section~\ref{s:prog}, and we conclude in Section~\ref{s:conc}.  UT
dates are used throughout the paper.


\section{Members of the Class}\label{s:class}

SNe have historically been classified spectroscopically
\citep{Minkowski41}, with the presence or absence of particular
spectral features being the fundamental distinction between various
classes.  Occasionally, photometric properties are used to subclassify
various classes (e.g., SNe~IIL and IIP).  Recently, the underlying
diversity of SNe combined with the recent surge in discoveries has
blurred clear lines in SN classification.

Observational classification has several advantages over
``theoretical'' classification.  Observational classification is
purely empirical, which has previously linked physically unrelated
objects.  For instance, SNe~I used to be a single class.  Even with
these potential mistakes, this kind of classification has been
exceedingly useful for determining the physical underpinnings of
stellar explosions.  On the other hand, ascribing particular
theoretical models to perform classification indicates a ``correct''
model and can hinder further advancements.  Here we attempt to provide
an observational classification scheme for the SN~Iax class.

The primary motivation of this classification scheme is to include all
SNe physically similar to SN~2002cx without including those with
significantly different progenitors or explosion mechanisms; however,
further refinements may be necessary in the future.  We exclusively
use observational properties of the SNe to provide the classification.

\begin{figure}
\begin{center}
\epsscale{1.3}
\rotatebox{90}{
\plotone{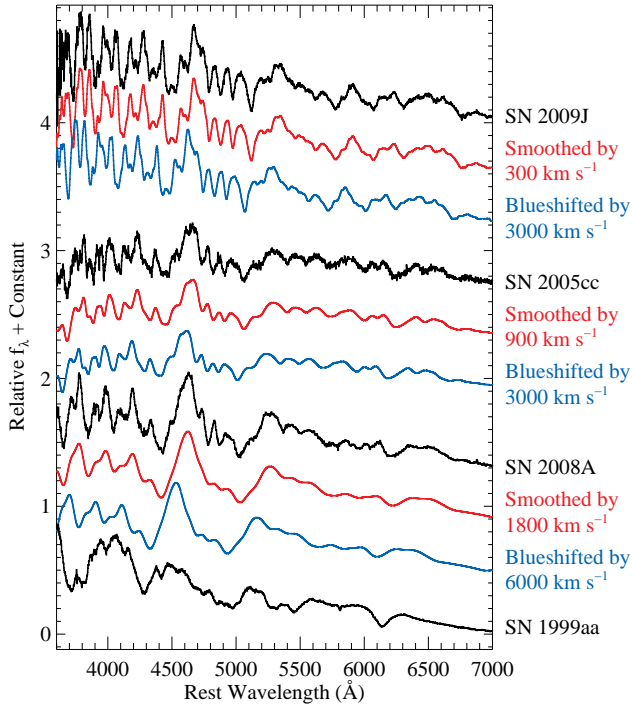}}
\caption{Spectra of SNe~Iax, from lowest to highest ejecta velocity
  (SNe~2009J, 2005cc, and 2008A, respectively), in comparison with the
  high-luminosity SN~Ia 1999aa \citep{Li01:pec, Garavini04}.  The
  black spectra are unaltered.  The red spectra are smoothed using a
  Gaussian filter of 300, 900, and 1800~\kms\ for SNe~2009J, 2005cc,
  and 2008A, respectively.  The blue spectra are the smoothed spectra
  after being blueshifted by 3000, 3000, and 6000~\kms, respectively.
  The smoothed, blueshifted SN~2009J spectrum is visually similar to
  the unaltered SN~2005cc spectrum; the smoothed, blueshifted
  SN~2005cc spectrum is visually similar to the unaltered SN~2008A
  spectrum; and the smoothed, blueshifted SN~2008A spectrum is
  visually similar to the unaltered SN~1999aa
  spectrum.}\label{f:sniax}
\end{center}
\end{figure}

As noted above, SNe~Iax are somewhat spectroscopically similar to the
high-luminosity SN~Ia 1991T.  Figure~\ref{f:sniax} shows the
near-maximum-brightness spectra of three SNe~Iax.  These SNe have
ejecta velocities which range from $|v| = 2200$ to $6900$~\kms.  By
artificially smoothing the spectra \citep{Blondin07} to broaden the
spectral features and also blueshifting the spectra, one can produce
spectra that are similar to what one would expect if the SN simply had
a higher ejecta velocity.  Doing this exercise, one can see the
continuum from SN~2009J (with $|v| = 2200$~\kms) through SN~2005cc
(with $|v| = 5200$~\kms) to SN~2008A (with $|v| = 6900$~\kms).  All
SNe~Iax seem to have similar composition with varying ejecta velocity.
SNe~Iax also appear to be similar to high-luminosity SNe~Ia, such as
SN~1991T and SN~1999aa \citep{Li01:pec, Garavini04}, near maximum
brightness after accounting for ejecta velocity.  Since the individual
narrow lines in SNe~Iax are the same lines seen as blends in SNe~Ia,
identifying features in SNe~Iax helps with interpreting SN~Ia spectra.

To be a member of the SN~Iax class, we require (1) no evidence of
hydrogen in any spectrum, (2) a maximum-brightness photospheric
velocity lower than that of any normal SN~Ia at maximum brightness
($|v| \lesssim 8000$~\kms), (3) if near-maximum light curves are
available, an absolute magnitude that is low for a normal SN~Ia given
its light-curve shape (i.e., falling below the WLR for SNe~Ia).  By
the first criterion, we exclude all SNe~II and any SN that obviously
has a hydrogen envelope.  The second criterion will exclude all
``normal'' SNe~Ia, Ib, and Ic, including high-luminosity and
low-luminosity SNe~Ia similar to SN~1991T \citep{Filippenko92:91T,
Phillips92} and SN~1991bg \citep{Filippenko92:91bg, Leibundgut93},
respectively, as well as the unique SN~2000cx \citep{Li01:00cx}, SNe
similar to SN~2006bt \citep{Foley10:06bt}, SN~2010X
\citep{Kasliwal10}, and ultraluminous SNe~I
\citep[e.g.,][]{Pastorello10, Quimby11, Chomiuk11}.  No known
core-collapse SN passes these first two criteria.  The third criterion
excludes all ``super-Chandrasekhar'' SNe~Ia \citep[e.g.,][]{Howell06},
which can have low ejecta velocities, but have high peak luminosities.

A final criterion is that the spectra need to be similar to those of
SN~2002cx at comparable epochs.  This last criterion is somewhat
subjective, yet necessary.  It is also a primary criterion, as the
first two criteria listed above naturally result from spectral
similarity.  We note that this criterion is no more subjective than
the one used to distinguish between SNe~Ia and Ic.  However,
exclusively using the previously listed criteria for classification
would include specific SNe that do not appear to be physically related
to SN~2002cx.  For instance, SN~2005E \citep{Perets10:05e} lacks
hydrogen, has a low ejecta velocity, and a low luminosity for its
light-curve shape, but is clearly spectroscopically different from
SN~2002cx at all epochs.  SNe~2005E and 2008ha have somewhat similar
spectra at \about 2 months after maximum brightness.  Both have strong
[\ion{Ca}{2}] and \ion{Ca}{2} emission, but there are several
significant differences including SN~2008ha lacking [\ion{O}{1}]
\citep{Foley09:08ha}.  Significant differences in the host-galaxy
morphologies for SNe similar to SNe~2002cx and 2005E further suggest
that these SNe have significantly different progenitor systems
\citep{Foley09:08ha, Perets10:05e}.  SNe~2002es
\citep{Ganeshalingam12} and PTF~09dav\citep[which fails our first
criterion]{Sullivan11:09dav} have low luminosity and low velocities,
but also have significantly cooler spectra (lacking \ion{Fe}{3} and
having strong \ion{Ti}{2} features) than SN~2002cx near maximum
brightness.Although SN~2002es and PTF~09dav may be physically related
to the SN~2002cx-like class in the same way that SN~1991bg is similar
to the hotter, more common ``Branch-normal SNe~Ia''
\citep{Branch93:normal}, we do not currently link these SNe to
SNe~Iax.  The progenitor environments of the SN~2002es and PTF~09dav
are also suggestive of older progenitor systems: SN~2002es was hosted
in an S0 galaxy, a fairly unusual host for a SN~Iax
\citep{Foley09:08ha}, and PTF~09dav was found \about 40~kpc from its
host, which may indicate a particularly old progenitor system, unlike
what is inferred for the majority of SNe~Iax.  We also exclude
SN~2002bj \citep{Poznanski10}, which is somewhat spectroscopically
similar to SN~2002cx, but also different in several ways.  Its light
curve was extremely fast ($\Delta m_{15} > 4$~mag), yet it was still
fairly luminous at peak ($M \approx -18$~mag).  We consider there to
be too many significant differences to include SN~2002bj in the class.
A list of our criteria and how various SN classes and particular
objects pass or fail the criteria is presented in Table~\ref{t:class}.

\begin{deluxetable*}{lcccc}
\tabletypesize{\scriptsize}
\tablewidth{0pt}
\tablecaption{Classification Criteria for SNe~Iax\label{t:class}}
\tablehead{
\colhead{SN Class} &
\colhead{Has Hydrogen?} &
\colhead{$|v| \lesssim 8$000~\kms?} &
\colhead{Low $L$ for LC Shape} &
\colhead{Spec.\ like SN~2002cx}}

\startdata

{\bf SN~Iax}  & {\bf N} & {\bf Y} & {\bf Y} & {\bf Y} \\
SN~II         & Y       & Some    & N/A     & N \\
SN~Ib/c       & N       & N       & Y       & N \\
SLSN~I        & N       & Y       & N       & N \\
Normal SN~Ia  & N       & N       & N       & N \\
Super-Chandra & N       & Y       & N       & N \\
SN~1991T      & N       & N       & N       & Somewhat \\
SN~1991bg     & N       & N       & N       & N \\
SN~2000cx     & N       & N       & Y       & N \\
SN~2002bj     & N       & Y       & N       & Somewhat \\
SN~2002es     & N       & Y       & Y       & Somewhat \\
SN~2002ic     & Y       & N       & N       & N \\
SN 2005E      & N       & Y       & Y       & N \\
SN~2006bt     & N       & N       & Y       & N \\
SN~2010X      & N       & N       & Y       & N \\
PTF~09dav     & Y       & Y       & Y       & Somewhat

\enddata

\end{deluxetable*}

When applying the criteria to our sample, we note that three of the
four criteria can be determined from a single spectrum near maximum
light.  We also note that there are no SNe that are spectroscopically
similar to SN~2002cx, and also have luminosities equal to or larger
than SNe~Ia with the same light-curve shape.  Therefore, a single
near-maximum-light spectrum appears to be sufficient for
classification.  However, because of the spectral similarities with
other SNe~Ia (except for the lower ejecta velocities; see
Figure~\ref{f:sniax}), SNe~Iax are easily mistaken as normal SNe~Ia in
initial classifications.  In particular, spectral classification
software such as SNID \citep{Blondin07} can easily misclassify a
SN~Iax as a SN~1991T-like SN~Ia if one does not know the redshift of
the SN or does not restrict the redshift to be the host-galaxy
redshift.  Misclassification was even more common before the
recognition of SN~2002cx as being distinct from SNe~Ia and before a
substantial set of examples was assembled.  We therefore do not
believe that our sample is complete, and it could be significantly
incomplete (especially for SNe discovered before 2002).  However, the
CfA and Berkeley Supernova Ia Program (BSNIP) spectral samples have
been searched for misclassified SNe~Iax \citep{Blondin12,
Silverman12:bsnip}, with BSNIP identifying one new SN~Iax. When
determining membership in the class, we first examine our own and
published data to match the criteria described above.  For some
objects, we use data from IAU Circulars and The Astronomer's
Telegrams.  There is no definitively identified SN~Iax for which we do
not have access to previously published data or our own data published
here.

There are currently 25 known SNe~Iax\footnote{In the final stages of
the preparation of this manuscript, LSQ12fhs, SN~2006ct, and
PS1-12bwh were identified as potential SNe~Iax.  At the time of
publication, we had not verified the classifications \citep{Copin12,
Quimby12, Wright12}.}.  Fifteen of these SNe were considered
members of the class by \citet{Foley09:08ha}.  Since that publication,
6 additional members have been discovered and 4 previously known SNe
have been identified as members.  The sample size is considerable; the
original Type I/II SN classification and the definition of the
SN~IIP/IIL, SN~IIn, SN~IIb, and broad-lined SN~Ic classes were all
performed with significantly fewer members of each class.  A list of
the members and some of their basic properties are in
Table~\ref{t:prop}.

\begin{deluxetable*}{lrrcccccc}
\tabletypesize{\scriptsize}
\tablewidth{0pt}
\tablecaption{Properties of SNe~Iax\label{t:prop}}
\tablehead{
\colhead{SN} &
\colhead{R.A.} &
\colhead{Dec.} &
\colhead{Refs.} &
\colhead{$t_{\rm max} (V)$} &
\colhead{$M_{V \rm{, peak}}$} &
\colhead{$\Delta m_{15} (V)$} &
\colhead{$v_{\rm peak}$} &
\colhead{He?} \\
\colhead{Name} &
\colhead{(J2000)} &
\colhead{(J2000)} &
\colhead{} &
\colhead{(JD $-$ 2,450,000)} &
\colhead{(mag)} &
\colhead{(mag)} &
\colhead{(\kms)} &
\colhead{}}

\startdata

1991bj & 03:41:30.47 & $-04$:39:49.5 & 1,2,3       & \nodata        & $\lesssim -15.4$   & \nodata    & \nodata & N \\
1999ax & 14:03:57.92 & $+15$:51:09.2 & 4,5         & \nodata        & $\lesssim -16.4$   & \nodata    & \nodata & N \\
2002bp & 11:19:18.20 & $+20$:48:23.1 & 6           & \nodata        & $\lesssim -16.1$   & \nodata    & \nodata & N \\
2002cx & 13:13:49.72 & $+06$:57:31.9 & 7,8,9       & 2418.31        & $-17.63$           & 0.84       & $-5600$ & N \\ 
2003gq & 22:53:20.68 & $+32$:07:57.6 & 9,10        & 2852.56        & $-17.29$           & 0.98       & $-5200$ & N \\
2004cs & 17:50:14.38 & $+14$:16:59.5 & 4,11        & \about3185     & $\sim -16.2$       & \about1.4  & \nodata & Y \\
2004gw & 05:08:48.41 & $+62$:26:20.7 & 1,12,13     & \nodata        & $\lesssim -16.4$   & \nodata    & \nodata & N \\
2005P  & 14:06:34.01 & $-05$:27:42.6 & 4,9,14      & \nodata        & $\lesssim -15.3$   & \nodata    & \nodata & N \\
2005cc & 13:57:04.85 & $+41$:50:41.8 & 15,16       & 3522.10        & $-16.48$           & 0.97       & $-5000$ & N \\
2005hk & 00:27:50.89 & $-01$:11:53.3 & 17,18,19    & 3689.81        & $-18.37$           & 0.92       & $-4500$ & N \\
2006hn & 11:07:18.67 & $+76$:41:49.8 & 1,20,21     & $>$3895.0      & $< -17.7$          & \nodata    & \nodata & N \\
2007J  & 02:18:51.70 & $+33$:43:43.3 & 1,4,22,23   & 4075.7--4114.3 & $\lesssim -15.4$   & \nodata    & \nodata & Y \\
2007ie & 22:17:36.69 & $+00$:36:48.0 & 25,26       & $<$4348.5      & $\sim -18.2$       & \nodata    & \nodata & N \\
2007qd & 02:09:33.56 & $-01$:00:02.2 & 24          & 4353.9--4404.4 & \nodata            & \nodata    & \nodata & N \\
2008A  & 01:38:17.38 & $+35$:22:13.7 & 16,27,28,29 & 4483.61        & $-18.46$           & 0.82       & $-6400$ & N \\
2008ae & 09:56:03.20 & $+10$:29:58.8 & 4,27,30,31  & 4513.52        & $-17.67$           & 0.94       & $-6100$ & N \\
2008ge & 04:08:24.68 & $-47$:53:47.4 & 14          & 4725.77        & $-17.60$           & 0.34       & \nodata & N \\
2008ha & 23:34:52.69 & $+18$:13:35.4 & 1,32,33     & 4785.24        & $-14.19$           & 1.22       & $-3200$ & N \\
2009J  & 05:55:21.13 & $-76$:55:20.8 & 4,34        & $>$4836.6      & $\lesssim -16.6$   & \nodata    & $-2200$ & N \\
2009ku & 03:29:53.23 & $-28$:05:12.2 & 35,36       & \nodata        & $-18.94$           & 0.38       & \nodata & N \\
2010ae & 07:15:54.65 & $-57$:20:36.9 & 4,37        & $>$5244.6      & $\lesssim -14.9$   & \nodata    & \nodata & N \\
2010el & 04:19:58.83 & $-54$:56:38.5 & 38          & $>$5350.6      & $\lesssim -14.8$   & \nodata    & \nodata & N \\
2011ay & 07:02:34.06 & $+50$:35:25.0 & 4,39        & 5651.81        & $-18.40$           & 0.75       & $-5600$ & N \\
2011ce & 18:55:35.84 & $-53$:43:29.1 & 4,40        & 5658--5668     & $-17.8$ -- $-18.9$ & 0.4 -- 1.3 & \nodata & N \\
2012Z  & 03:22:05.35 & $-15$:23:15.6 & 4,41        & $>$5955.7      & $\lesssim -16.8$   & \nodata    & \nodata & N

\enddata

\tablerefs{1 = \citet{Foley09:08ha}, 2 = \citet{Gomez96}, 3 =
  \citet{Stanishev07:05hk}, 4 = This Paper, 5 = \citet{Gal-Yam08}, 6 =
  \citet{Silverman12:bsnip}, 7 = \citet{Li03:02cx}, 8 =
  \citet{Branch04}, 9 = \citet{Jha06:02cx}, 10 =
  \citet{Filippenko03:03gq2}, 11 = \citet{Rajala05}, 12 =
  \citet{Foley05:05e}, 13 = \citet{Filippenko05:04gw}, 14 =
  \citet{Foley10:08ge}, 15 = \citet{Antilogus05}, 16 =
  \citet{Ganeshalingam10}, 17 = \citet{Chornock06}, 18 =
  \citet{Phillips07}, 19 = \citet{Sahu08}, 20 = \citet{Foley06:06hn},
  21 = \citet{Hicken09:lc}, 22 = \citet{Filippenko07:07J1}, 23 =
  \citet{Filippenko07:07J2}, 24 = \citet{McClelland10}, 25 =
  \citet{Bassett07:07ie}, 26 = \citet{Ostman11}, 27 =
  \citet{Blondin08:08A}, 28 = \citet{Hicken12}, 29 = McCully et~al.,
  in preparation, 30 = \citet{Blondin08:08ae}, 31 = \citet{Milne10},
  32 = \citet{Foley10:08ha}, 33 = \citet{Valenti09}, 34 =
  \citet{Stritzinger09}, 35 = \citet{Rest09}, 36 = \citet{Narayan11},
  37 = \citet{Stritzinger10:10ae2}, 38 = \citet{Bessell10}, 39 =
  \citet{Silverman11:11ay}, 40 = \citet{Anderson11}, 41 =
  \citet{Cenko12}.}

\end{deluxetable*}

Below we present details for the SNe.  For additional information, see
\citet{Foley09:08ha}, references therein, and the references listed
for each SN below.

\subsection{SN~1991bj}

SN~1991bj is the oldest known member of the class, although it has
only recently been identified as a member \citep{Foley09:08ha}.  It
was originally classified as a SN~Ia by two separate teams
\citep{Pollas92}.  \citet{Gomez96} presented a spectrum of SN~1991bj
and noted its low ejecta velocity.  \citet{Stanishev07:05hk} first
identified SN~1991bj as a possible SN~Iax.  \citet{Foley09:08ha} used
a Lick spectrum obtained by A. V. Filippenko to classify SN~1991bj as
a SN~Iax.

\subsection{SN~1999ax}

SN~1999ax was discovered by the Wise Observatory Optical Transient
Search \citep[WOOTS;][]{Gal-Yam08} in the field of Abell 1852, which
has a redshift $z = 0.181$ \citep{Gal-Yam99}.  Spectra of the SN
suggested that SN~1999ax was a SN~Ia at $z \approx 0.05$
\citep{Gal-Yam00}.  A spectrum was also presented by
\citet{Gal-Yam08}, where they used template matching to find $z
\approx 0.05.$\footnote{\citet{Gal-Yam08} also note that their
  spectrum of SN~1999ax is ``somewhat peculiar,'' but do not expand
  further since it was not the focus of their study.}

\begin{figure}
\begin{center}
\epsscale{0.9}
\rotatebox{90}{
\plotone{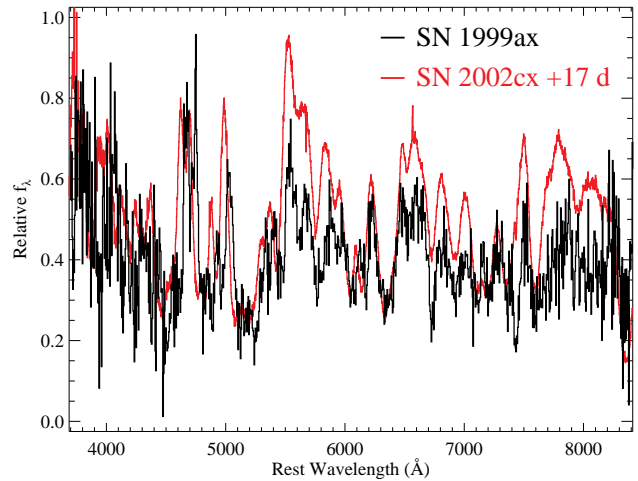}}
\caption{Spectra of SN~1999ax (black) and SN~2002cx (red).  Both
  spectra have been divided by a fifth-order polynomial to remove the
  poor flux calibration of the SN~1999ax spectrum and to make an
  appropriate comparison.  The rest-frame phase relative to $V$
  maximum is marked for SN~2002cx.}\label{f:99ax_spec}
\end{center}
\end{figure}

However, an SDSS spectrum revealed that its host galaxy, SDSS
J140358.27+155101.2, is at $z = 0.023$ \citep{Abazajian09}.  We
obtained the spectrum of SN~1999ax from 6 April 1999,\footnote{Spectra
are available at\\
  \tt{http://www.physto.se/{\raise.17ex\hbox{$\scriptstyle\sim$}}snova/private/near-z/spectroscopy/reduced\_data/}.}
but its flux calibration does not appear to be correct.  To make a
comparison to other SNe, we divided the flux by a fifth-order
polynomial and compared the resulting spectrum to other SN spectra
that were similarly modified.  Using the SDSS redshift and correcting
the flux, it is clear that SN~1999ax is similar to SN~2002cx (see
Figure~\ref{f:99ax_spec}).

This is the first time that SN~1999ax has been considered a member of
the SN~Iax class.

\subsection{SN~2002bp}

SN~2002bp was discovered by the Puckett Observatory Supernova Search
\citep[POSS;][]{Puckett02}, but remained unclassified for several
years.  Finally, while analyzing the large BSNIP sample of SN~Ia
spectra, \citet{Silverman12:bsnip} determined that SN~2002bp was
similar to SN~2008ha, another member of this class.

\subsection{SN~2002cx}

SN~2002cx was the first SN in this class recognized as being peculiar
and is its namesake.  \citet{Wood-Vasey02} discovered SN~2002cx on 12
May 2002.  Basic observational information derived from near-maximum
data were originally presented by \citet{Li03:02cx}.  The photometric
data were re-evaluated by \citet{Phillips07}.  Late-time spectra were
presented by \citet{Jha06:02cx}.  \citet{Branch04} performed a
detailed spectral analysis of its maximum-light spectra.

\subsection{SN~2003gq}

SN~2003gq was independently discovered \citep{Graham03, Puckett03} by
the Lick Observatory Supernova Search \citep[LOSS;][]{Li00,
Filippenko01} and POSS.  It was originally classified as a SN~Ia by
\citet{Filippenko03:03gq1} and was later revised as a SN~Iax
\citep{Filippenko03:03gq2}.  As part of the LOSS photometric follow-up
effort, filtered photometry of SN~2003gq was obtained and presented by
\citet{Ganeshalingam10}.  \citet{Blondin12} presented a spectrum of
SN~2003gq.

\subsection{SN~2004cs}

\begin{figure}
\begin{center}
\epsscale{0.9}
\rotatebox{90}{
\plotone{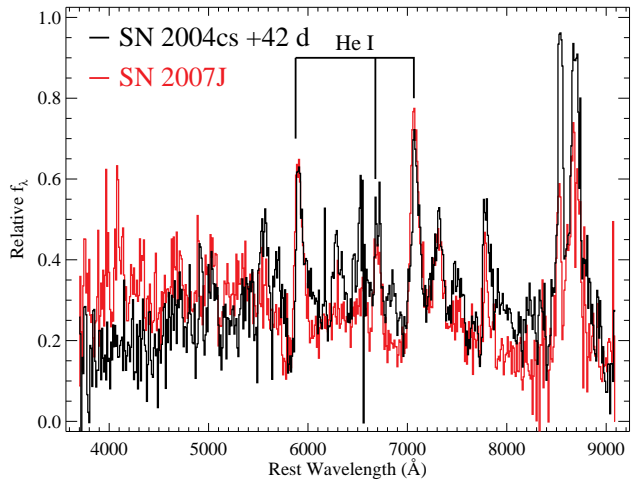}}
\caption{Spectra of SN~2004cs (black) and SN~2007J (red).  The
  approximate rest-frame phase relative to $V$ maximum is marked for
  SN~2004cs.}\label{f:04cs_spec}
\end{center}
\end{figure}

\citet{Li04} discovered SN~2004cs as part of LOSS.  \citet{Rajala05}
presented its spectrum and classified it as a SN~IIb, identifying both
H$\alpha$ and strong \ion{He}{1} features.  We obtained this spectrum
(via D.\ Leonard), and show a comparison of SNe~2004cs and 2007J in
Figure~\ref{f:04cs_spec}.  Both SNe are very similar, and SN~2004cs
clearly has \ion{He}{1} features.  However, we do not identify
H$\alpha$ in the spectrum; there are clear residuals from galaxy
subtraction at the position of H$\alpha$, but the peak of the feature
is also blueward of H$\alpha$.  SN~2007J was identified as a member of
the SN~Iax class, but showed \ion{He}{1} lines \citep{Foley09:08ha}.
SN~2004cs demonstrates that there is more than one member of this
class that exhibits He in its spectrum.

This is the first time that SN~2004cs has been considered a SN~Iax.
In Section~\ref{ss:phot}, we present a previously unpublished
unfiltered light curve obtained with the robotic 0.76~m Katzman
Automatic Imaging Telescope \citep[KAIT;][]{Filippenko01} at Lick
Observatory.

\subsection{SN~2004gw}

SN~2004gw was discovered by POSS on 29 December 2004
\citep{Puckett04}.  \citet{Gal-Yam05} originally classified it as a
SN~I with some indications that it was of Type Ic.
\citet{Foley05:05e} suggested that it was a SN~Ia which ``exhibits a
number of spectral peculiarities.''  \citet{Filippenko05:04gw} later
confirmed that it was of Type Ia.  Finally, with the aid of a larger
comparison sample, \citet{Foley09:08ha} showed that SN~2004gw was a
SN~Iax.

\subsection{SN~2005P}

SN~2005P was discovered by LOSS on 21 January 2005
\citep{Burket05:05P}.  Visual inspection of unpublished data taken on
22 January 2005 by Schmidt \& Salvo (private communication) indicates
that it is a SN~Iax.  The SN remained unclassified in the literature
for over a year.  Based on a late-time spectrum, \citet{Jha06:02cx}
classified it as a SN~Iax.  \citet{Foley10:08ge} later suggested that
SN~2005P was spectroscopically most similar to SN~2008ge having a
late-time spectrum with relatively broad lines and relatively strong
forbidden Fe lines.  We present a previously unpublished KAIT
unfiltered light curve in Section~\ref{ss:phot}.

\subsection{SN~2005cc}

SN~2005cc was discovered by POSS on 19 May 2005 \citep{Puckett05}.
Several spectra indicated that it was a young SN similar to SN~2002cx
\citep{Antilogus05}.  As a part of the LOSS photometric follow-up
effort, filtered photometry of SN~2005cc was obtained and presented by
\citet{Ganeshalingam10}.  Spectra of SN~2005cc were presented by
\citet{Blondin12}.

\subsection{SN~2005hk}

SN~2005hk, which was independently discovered by both LOSS
\citep{Burket05:05hk} and SDSS-II \citep{Barentine05:05hk}, is the
best-observed SN~Iax.  \citet{Phillips07} and \citet{Sahu08} presented
extensive data near maximum brightness.  \citet{Kromer12} presented
NIR spectra and late-time photometry of SN~2005hk, while
\citet{Sahu08} and \citet{Valenti09} published late-time spectra.
\citet{Chornock06} and \citet{Maund10:05hk} showed spectropolarimetric
observations indicating that SN~2005hk had low polarization near
maximum brightness.  Late-time spectroscopy and {\it HST} photometry
will be presented by McCully et~al.\ (in preparation).

\subsection{SN~2006hn}

POSS discovered SN~2006hn on 28 September 2006 \citep{Sehgal06}.
\citet{Foley06:06hn} classified SN~2006hn as a SN~Ia, and
\citet{Foley09:08ha} noted that it was a SN~Iax.  Photometry of
SN~2006hn was published as part of the CfA3 data release
\citep{Hicken09:lc}.

\subsection{SN~2007J}

SN~2007J, which was independently discovered by both LOSS and POSS
\citep{Lee07}, was the first known member of the SN~Iax class to
display \ion{He}{1} lines.  Initially, these lines were weak, and
SN~2007J appeared to be very similar to SN~2002cx
\citep{Filippenko07:07J1}; however, the \ion{He}{1} lines became
stronger with time, causing \citet{Filippenko07:07J2} to reclassify
SN~2007J as a peculiar SN~Ib.  \citet{Foley09:08ha} re-examined the
spectra, showing that besides the \ion{He}{1} lines, SN~2007J is
indeed very similar to SN~2002cx.  We therefore consider SN~2007J to
be a peculiar SN~Iax.  We present previously unpublished KAIT filtered
and unfiltered light curves in Section~\ref{ss:phot}.

\subsection{SN~2007ie}

\begin{figure}
\begin{center}
\epsscale{0.9}
\rotatebox{90}{
\plotone{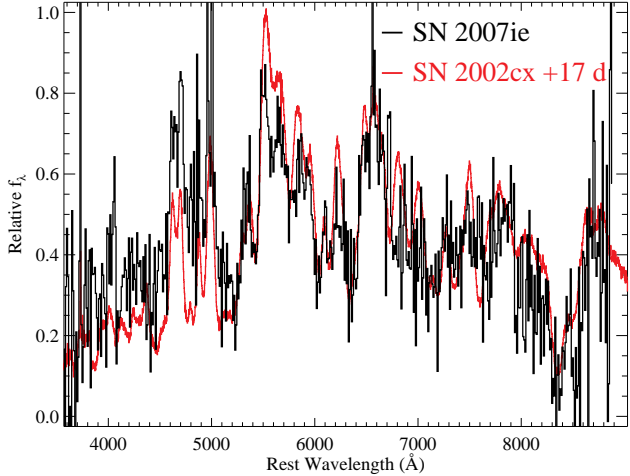}}
\caption{Spectra of SN~2007ie (black; galaxy subtracted) and SN~2002cx
  (red).  The rest-frame phase relative to $V$ maximum is marked for
  SN~2002cx.}\label{f:07ie_comp}
\end{center}
\end{figure}

SN~2007ie was discovered by SDSS-II \citep{Bassett07:07ie}, who also
classified it as a probable SN~Ia.  \citet{Ostman11} presented a
spectrum of SN~2007ie and indicated that it was a probable SN~Iax.
They noted the low velocity and spectral similarities to SNe~2002cx
and 2005hk, but allowed the possibility that it was a normal SN~Ia.
However, restricting the redshift of the comparison spectra to that of
SN~2007ie, only SNe~Iax provide reasonable matches.  Additionally, the
peak magnitude is $M \approx -18.2$~mag, similar to that of SN~2002cx,
although the peak of the light curve was not covered in their
photometry.  Figure~\ref{f:07ie_comp} shows the galaxy-subtracted
spectrum of SN~2007ie compared to SN~2002cx.  We consider SN~2007ie to
be a clear member of the SN~Iax class.

\subsection{SN~2007qd}

SN~2007qd was discovered by SDSS-II \citep{Bassett07:07qd}.  It is a
relatively faint SN~Iax with ejecta velocity between those of
SNe~2002cx and 2008ha \citep{McClelland10}.  \citet{McClelland10} used
observations of SNe~2002cx, 2005hk, 2007qd, and 2008ha to argue that
there was a relationship between ejecta velocity and peak absolute
magnitude for SNe~Iax.  \citet{Narayan11} showed that SN~2009ku was a
prominent outlier to this trend.

\subsection{SN~2008A}

SN~2008A was discovered on 2 January 2008 by \citet{Nakano08}. It was
classified as a SN~Iax by \citet{Blondin08:08A}.  As a part of the
LOSS photometric follow-up effort, filtered photometry of SN~2008A was
obtained and presented by \citet{Ganeshalingam10}.  Photometry of
SN~2008A was also published as part of the CfA4 data release
\citep{Hicken12}.  The SN was observed photometrically by {\it Swift};
\citep{Milne10} found that SNe~Iax have very blue UV colors relative
to normal SNe~Ia.  Spectroscopy and late-time {\it HST} photometry
will be presented by McCully et~al.\ (in preparation).
\citet{Blondin12} presented several spectra of SN~2008A.

\subsection{SN~2008ae}

POSS discovered SN~2008ae on 9 February 2008 \citep{Sostero08}.
\citet{Blondin08:08ae} classified it as a SN~Iax, and they further
note that SN~2008ae is relatively luminous ($M < -17.7$~mag) a few
days before maximum brightness.  Optical and UV photometry of
SN~2008ae was also published as part of the CfA4 data release
\citep{Hicken12} and the {\it Swift} photometry compilation
\citep{Milne10}, respectively.  \citet{Blondin12} presented several
spectra of SN~2008ae.  We present previously unpublished KAIT and
Carnegie Supernova Project (CSP) light curves in
Section~\ref{ss:phot}.

\subsection{SN~2008ge}

SN~2008ge was very nearby and bright.  It was discovered by CHASE
\citep{Pignata08:08ge} well past maximum brightness, but CHASE had
several pre-discovery images from which a light curve could be
generated.  Its host galaxy, NGC~1527, is an S0 galaxy with no signs
of star formation to deep limits \citep[$< 7.2 \times
10^{-3}$~$M_{\sun}$~yr$^{-1}$;][]{Foley10:08ge}.  It was also imaged
by {\it HST} before SN~2008ge occurred, and analysis showed that there
were no massive stars near the SN site or any indication of star
formation in the host galaxy \citep{Foley10:08ge}.

SN~2008ge had a relatively broad light curve and (unlike SNe~2002cx
and 2005hk) strong [\ion{Fe}{2}] emission lines in its late-time
spectra \citep{Foley10:08ge}.  The lack of massive stars near the SN
site, the strict limit on the star-formation rate, and the presumably
large generated $^{56}$Ni mass all suggest a WD progenitor
\citep{Foley10:08ge}.

\subsection{SN~2008ha}

SN~2008ha is an extreme SN~Iax, being less luminous than any other
member and having lower ejecta velocity than most members of the class
\citep{Foley09:08ha, Foley10:08ha, Valenti09}.  It was discovered by
POSS \citep{Puckett08}.

Several studies have been devoted to SN~2008ha \citep{Foley09:08ha,
Foley10:08ha, Valenti09}, and details of the SN are presented in
those works.  Notably, the total inferred ejecta mass is significantly
below the Chandrasekhar mass.  Although SN~2008ha may have had a
massive-star progenitor \citep{Foley09:08ha, Valenti09}, carbon/oxygen
burning products in its maximum-light spectrum, and the energy/ejecta
mass balancing necessary to create a low-velocity, low-luminosity SN
like SN~2008ha make that scenario unlikely \citep{Foley09:08ha,
  Foley10:08ha}.

\subsection{SN~2009J}

\begin{figure}
\begin{center}
\epsscale{0.9}
\rotatebox{90}{
\plotone{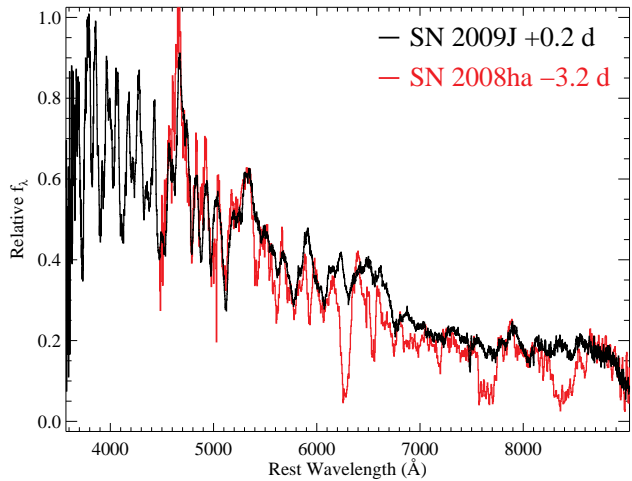}}
\caption{Spectra of SN~2009J (black) and SN~2008ha (red).  The
  rest-frame phase relative to $V$ maximum is marked for both
  SNe.}\label{f:09j_comp}
\end{center}
\end{figure}

CHASE discovered SN~2009J on 13 January 2009 \citep{Pignata09:09J}.
Follow-up spectroscopy revealed that it was a SN~Iax
\citep{Stritzinger09}.  SN~2009J is a particularly low-velocity SN,
similar to, but even lower velocity than SN~2008ha
(Figure~\ref{f:09j_comp}).  We present previously unpublished CSP and
CHASE light curves in Section~\ref{ss:phot} and previously unpublished
spectra in Section~\ref{ss:spec}.

\subsection{SN~2009ku}

SN~2009ku was discovered by Pan-STARRS1 \citep[PS1;][]{Rest09}, and
\citet{Narayan11} presented detailed observations.  SN~2009ku is a
relatively luminous SN~Iax ($M_{V \rm{, peak}} \approx -18.4$~mag),
but had ejecta velocity comparable to that of the extremely
low-luminosity SN~2008ha.  This showed that, contrary to what was
presented by \citet{McClelland10}, velocity and luminosity are not
strongly correlated for all SNe~Iax \citep{Narayan11}.

\subsection{SN~2010ae}

\begin{figure}
\begin{center}
\epsscale{0.9}
\rotatebox{90}{
\plotone{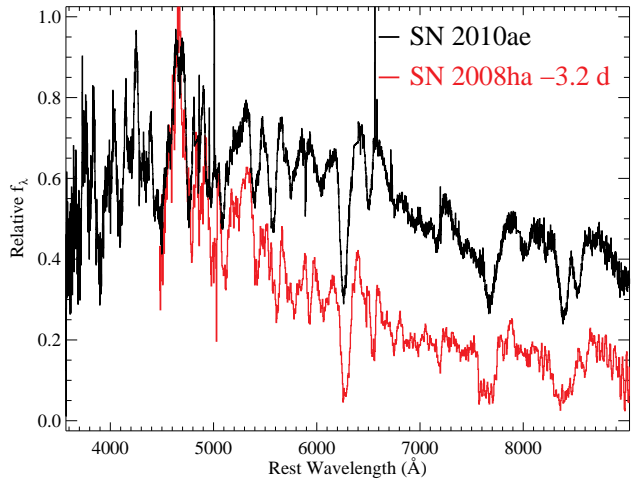}}
\caption{Spectra of SN~2010ae (black) and SN~2008ha (red).  Although
  the continua have different shapes, the spectral features are
  similar.  The rest-frame phase relative to $V$ maximum is marked for
  SN~2008ha.}\label{f:10ae_comp}
\end{center}
\end{figure}

SN~2010ae was discovered on 23 February 2010 in ESO~162-G017
\citep{Pignata10}.  \citet{Stritzinger10:10ae1} originally classified
it as a peculiar SN~Ia similar to the possible ``super-Chandrasekhar''
SN~2006gz \citep{Hicken07}.  Using additional data,
\citet{Stritzinger10:10ae2} determined that SN~2010ae was most similar
to SN~2008ha.  In Figure~\ref{f:10ae_comp}, we present a spectrum at
an epoch similar to those from \citet{Stritzinger10:10ae2} that shows
this similarity to SN~2008ha (although with a slightly different
continuum shape).  A full analysis of SN~2010ae will be presented by
Stritzinger et~al.\ (in prep.).

\subsection{SN~2010el}

SN~2010el was discovered on 19 June 2010 in NGC~1566 \citep{Monard10}.
\citet{Bessell10} determined that SN~2010el was spectroscopically
similar to SN~2008ha.  A full analysis of SN~2010el will be presented
by Valenti et~al. (in prep.).

\subsection{SN~2011ay}

\begin{figure}
\begin{center}
\epsscale{0.9}
\rotatebox{90}{
\plotone{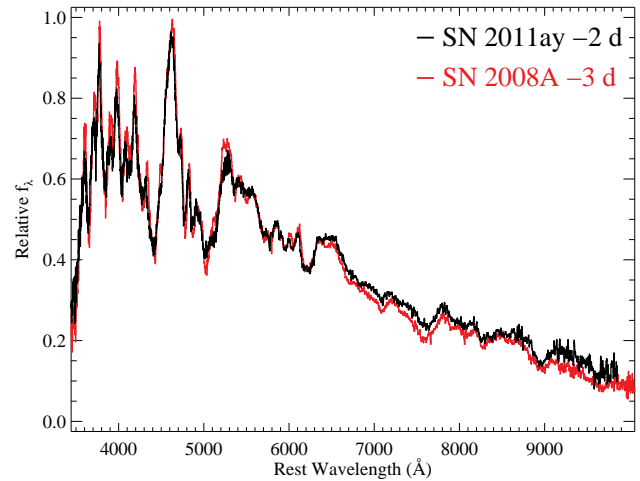}}
\caption{Spectra of SN~2011ay (black) and SN~2008A (red).  The
  rest-frame phase relative to $V$ maximum is marked for both
  SNe.}\label{f:11ay_comp}
\end{center}
\end{figure}

SN~2011ay was discovered on 18 March 2011 in NGC~2314 as part of LOSS
\citep{Blanchard11}.  \citet{Pogge11} classified SN~2011ay as a SN~Ia
similar to the high-luminosity SN~1999aa \citep{Li01:pec, Garavini04}.
However, \citet{Silverman11:11ay} later classified SN~2011ay as a
SN~Iax.  Spectra of SNe~2011ay and 2008A are shown in
Figure~\ref{f:11ay_comp}.  SN~2011ay is clearly a SN~Iax.  We present
previously unpublished CfA light curves in Section~\ref{ss:phot} and
previously unpublished spectra in Section~\ref{ss:spec}.

\subsection{SN~2011ce}\label{ss:11ce}

SN~2011ce was discovered on 26 March 2011 in NGC~6708 \citep{Maza11},
although it was not announced until about a month later.  There was a
nondetection on 4 March 2011.  Soon after the announcement, on 24
April 2011, \citet{Anderson11} classified SN~2011ce as a SN~Iax with a
phase of about 20~days after maximum brightness.

We present a previously unpublished, unfiltered CHASE light curve in
Section~\ref{ss:phot}.  Using the light curve of SN~2005hk, and
allowing for a range of light-curve stretches, we find that SN~2011ce
likely peaked between 7 and 17 April 2011 with a peak unfiltered
magnitude of $-17.6$ to $-18.7$ (corresponding roughly to $-17.8 \ge
M_{V} \ge -18.3$~mag).  We obtained optical spectra on 24 April 2011
and on 20 April 2012, more than a year after discovery.  The spectra
are 47/409 and 37/360 rest-frame days after the last nondetection and
the first detection, respectively, and \about 42 and 371 rest-frame
days after maximum brightness, respectively.

\begin{figure}
\begin{center}
\epsscale{0.9}
\rotatebox{90}{
\plotone{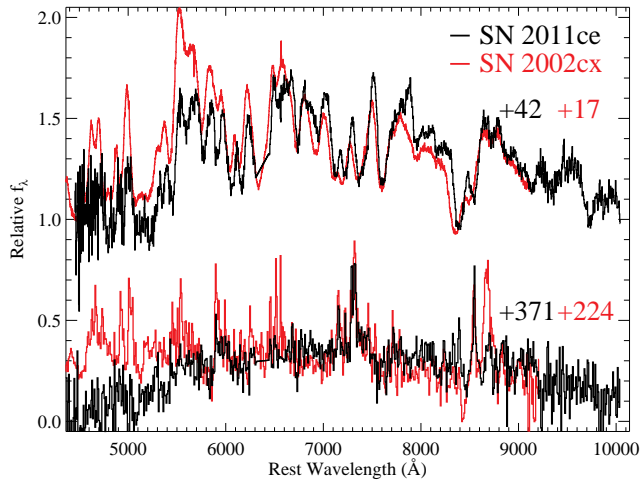}}
\caption{Spectra of SN~2011ce (black) and SN~2002cx (red).  The
  approximate rest-frame phase relative to $V$ maximum is marked for
  all spectra.}\label{f:11ce_comp}
\end{center}
\end{figure}

A comparison of our SN~2011ce spectra and those of SN~2002cx is shown
in Figure~\ref{f:11ce_comp}.  SN~2011ce is clearly a SN~Iax, at early
times being similar to SN~2002cx, and at late times displaying
low-velocity P-Cygni profiles of Fe as well as strong [\ion{Ca}{2}]
and \ion{Ca}{2} emission lines.

\subsection{SN~2012Z}

\begin{figure}
\begin{center}
\epsscale{0.9}
\rotatebox{90}{
\plotone{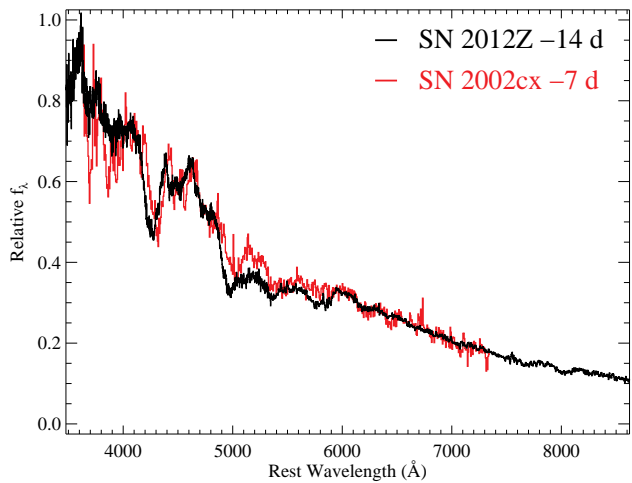}}
\caption{Spectra of SN~2012Z (black) and SN~2002cx (red).  The
  rest-frame phase relative to $V$ maximum is marked for both
  SNe.}\label{f:12z_comp}
\end{center}
\end{figure}

SN~2012Z was discovered on 29 January 2012 in NGC~1309 by LOSS
\citep{Cenko12}.  \citet{Cenko12} also report spectroscopic
observations indicating that SN~2012Z is a SN~Iax.  A detailed study
of this SN and its progenitor will be presented by Fong et~al.\ (in
preparation).  A comparison of the spectra of SNe~2002cx and 2012Z is
shown in Figure~\ref{f:12z_comp}.  We present previously unpublished
CfA light curves in Section~\ref{ss:phot}.

\subsection{Possible Subclasses}

Based on the criteria outlined above, we have selected the members of
the SN~Iax class in the previous subsections.  As mentioned above, the
inclusion and exclusion of particular SNe is somewhat subjective.
Here we focus on possible subclasses of SNe~Iax that could be
physically different from the other members.  The two obvious groups
(which may be considered subclasses of SNe~Iax) are the SNe similar to
SNe~2007J and 2008ha, respectively.

Unlike other members of the class, SNe~2004cs and 2007J both have
strong \ion{He}{1} lines in their spectra.  It is still unclear if
this difference is the result of different ejecta (SNe~2004cs and
2007J have significantly more helium in their ejecta than other
members), different physical conditions (such as ionization at a given
time), or an observational effect (only particular viewing angles show
\ion{He}{1}, or the lines are only present at particular phases).
Additional observations with significantly more data should help
disentangle these possibilities.  Until there is a clear physical
difference, we include these objects in the class, with the
\ion{He}{1} lines being an interesting constraint on the progenitor
systems and explosion.

Several members of the class (SNe~2002bp, 2009ku, 2010ae, and 2010el)
have extremely low velocity ($|v| \approx 3$,000~\kms) similar to
SN~2008ha.  Although this is a possible indication of a distinct
subclass, other observations conflict with this possibility.
Specifically, there is a range of ejecta velocities for the class (see
Section~\ref{s:spec}), and there appears to be a continuous sequence
of velocities \citep{McClelland10}.  Additionally, SN~2009ku had very
low velocities but a luminosity similar to that of SN~2002cx
\citep{Narayan11}, indicating that the SNe with very low velocity are
not distinct in all properties.  We consider these objects to be
extreme members of the class rather than a separate group.


\section{Observations and Data Reduction}\label{s:obs}

\subsection{Photometry}\label{ss:phot}

Below, we present new and previously published photometry of several
SNe~Iax.  SNe~2003gq, 2005cc, and 2008A have previously published
filtered photometry from KAIT \citep{Ganeshalingam10}.  The KAIT
photometry is supplemented by previously published photometry of
SNe~2006hn, 2008A, and 2008ae from the CfA3 \citep{Hicken09:lc} and
CfA4 \citep{Hicken12} samples.  We present previously unpublished
filtered KAIT photometry of SN~2007J and unfiltered (similar to $R$
band) KAIT photometry of SNe~2004cs, 2005P, and 2007J.  We also
present previously unpublished filtered CSP photometry of SNe~2008ae
and 2009J, unpublished filtered CfA photometry of SNe~2011ay and
2012Z, and unpublished filtered CHASE photometry of SNe 2009J and
2011ce.

Unfiltered (as part of LOSS) and broadband \bvri\ photometry of
several SNe was obtained using KAIT at Lick Observatory.  The data
were reduced using a mostly automated pipeline developed for KAIT
images \citep{Ganeshalingam10}.  Images are bias-corrected and
flatfielded at the telescope.  Using galaxy templates obtained a year
and a half after discovery, the data images are galaxy-subtracted to
remove galaxy flux at the position of the SN.  The flux of the SN and
the local field star are measured using the point-spread function
(PSF) fitting photometry package \texttt{DAOPHOT} in
IRAF.\footnote{IRAF: The Image Reduction and Analysis Facility is
distributed by the National Optical Astronomy Observatory, which is
operated by the Association of Universities for Research in
Astronomy (AURA) under cooperative agreement with the National
Science Foundation (NSF).}  Instrumental magnitudes are
color-corrected to the \citet{Landolt92} system using the average
color terms measured on multiple photometric nights.

The majority of the CSP imaging was obtained at the Las Campanas
Observatory (LCO) with the Henrietta Swope 1.0~m telescope equipped
with the ``SITe3'' direct optical camera.  These data are accompanied
with additional images taken with the Ir\'{e}n\'{e}e du~Pont 2.5~m
telescope.  Optical du~Pont images were obtained with the direct CCD
camera known as ``Tek 5''; see \citet{Hamuy06} for details regarding
these instruments.  An in-depth description of CSP observational
procedures, data-reduction techniques, and the computation of
definitive photometry in the natural photometric system is given by
\citet{Contreras10}, with additional descriptions in
\citet{Stritzinger12}.  SN photometry is computed differentially with
respect to a local sequence of stars, and reported in the natural
system.  Conversions to standard-system magnitudes are presented by
\citet{Stritzinger11}.

Photometry of several SNe was also obtained by the 0.41~m Panchromatic
Robotic Optical Monitoring and Polarimetry Telescope
\citep[PROMPT][]{Reichart05}.  Instrumental magnitudes were measured
using the template-subtraction technique with a code based on the ISIS
package \citep{Alard98, Alard00}, and we report magnitudes in the
natural system.

\begin{figure}
\begin{center}
\epsscale{1.3}
\rotatebox{90}{
\plotone{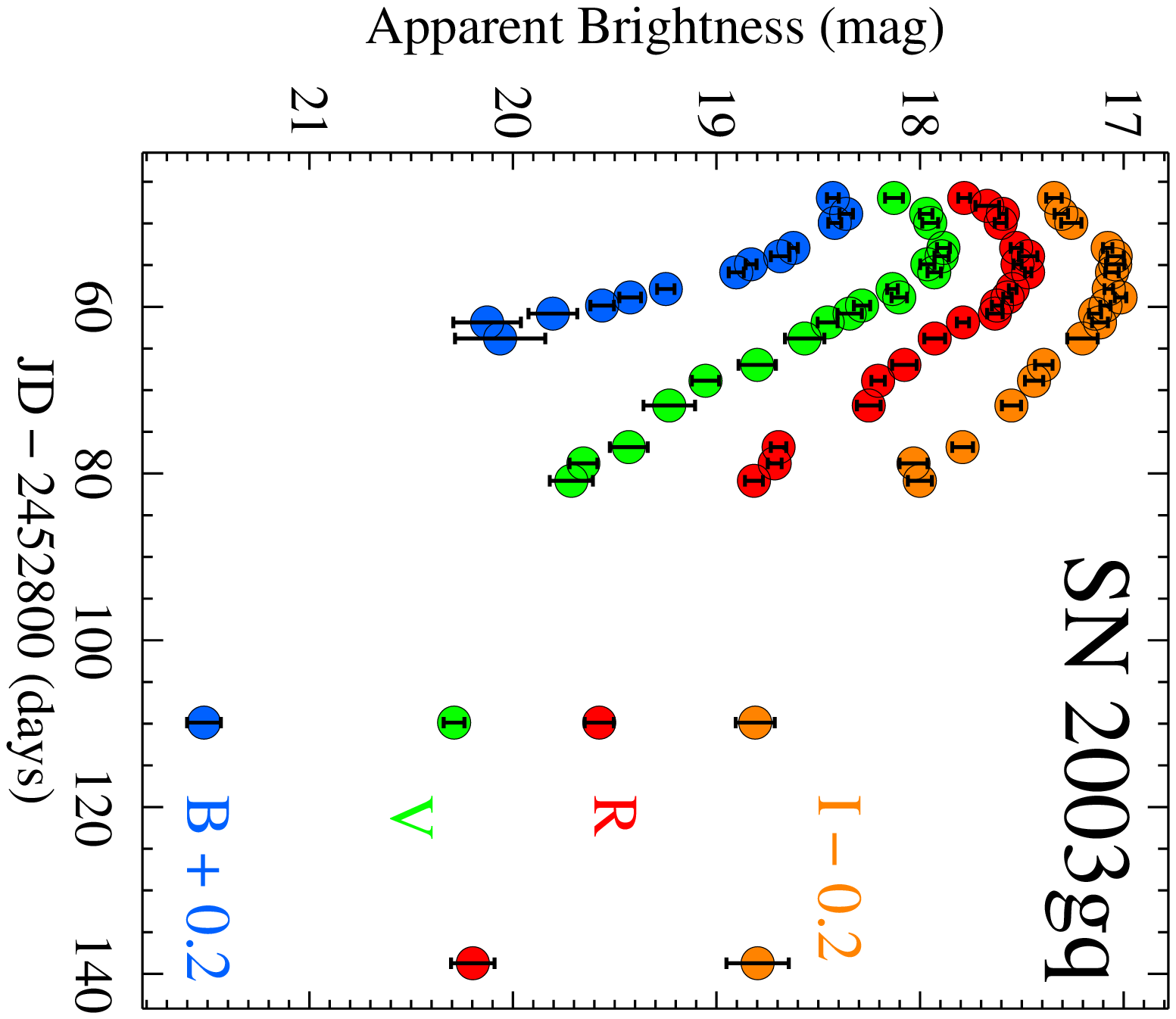}}
\caption{KAIT \bvri\ (blue, green, red, and orange, respectively)
  light curves of SN~2003gq \citep{Ganeshalingam10}.  The
  uncertainties for most data points are smaller than the plotted
  symbols.}\label{f:03gq_lc}
\end{center}
\end{figure}

\begin{figure}
\begin{center}
\epsscale{1.3}
\rotatebox{90}{
\plotone{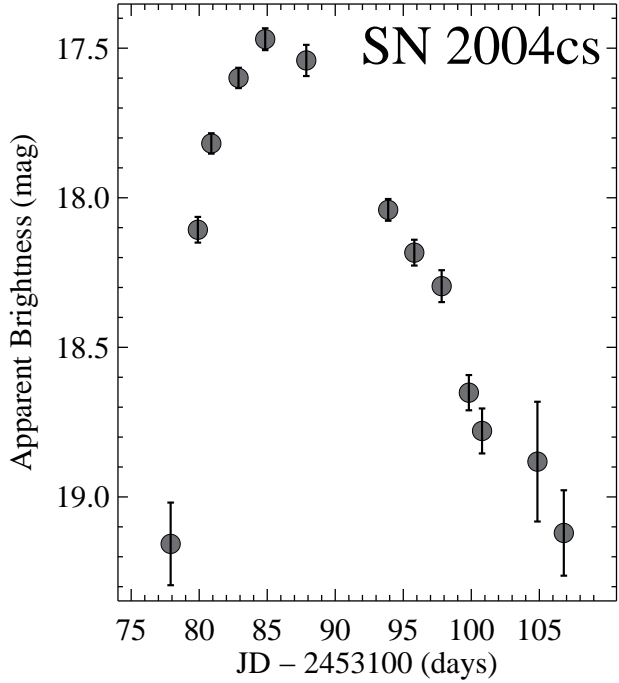}}
\caption{KAIT unfiltered light curve of SN~2004cs.}\label{f:04cs_lc}
\end{center}
\end{figure}

\begin{figure}
\begin{center}
\epsscale{1.3}
\rotatebox{90}{
\plotone{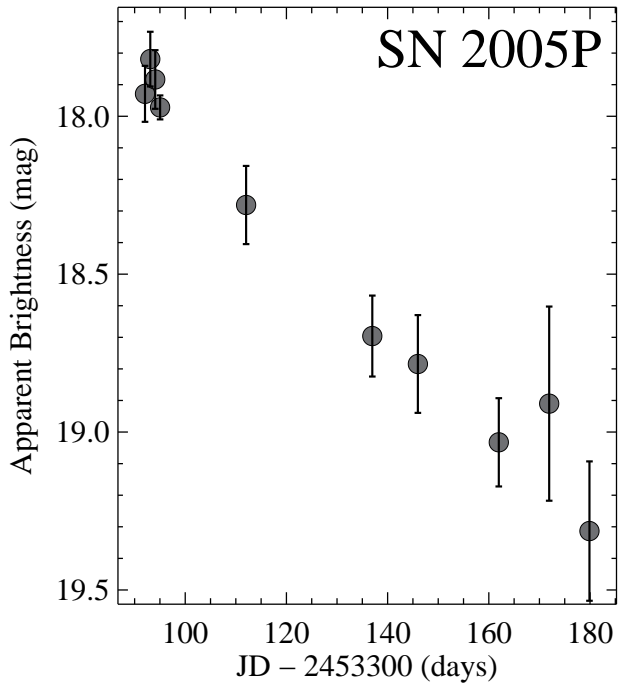}}
\caption{KAIT unfiltered light curve of SN~2005P.}\label{f:05p_lc}
\end{center}
\end{figure}

\begin{figure}
\begin{center}
\epsscale{1.3}
\rotatebox{90}{
\plotone{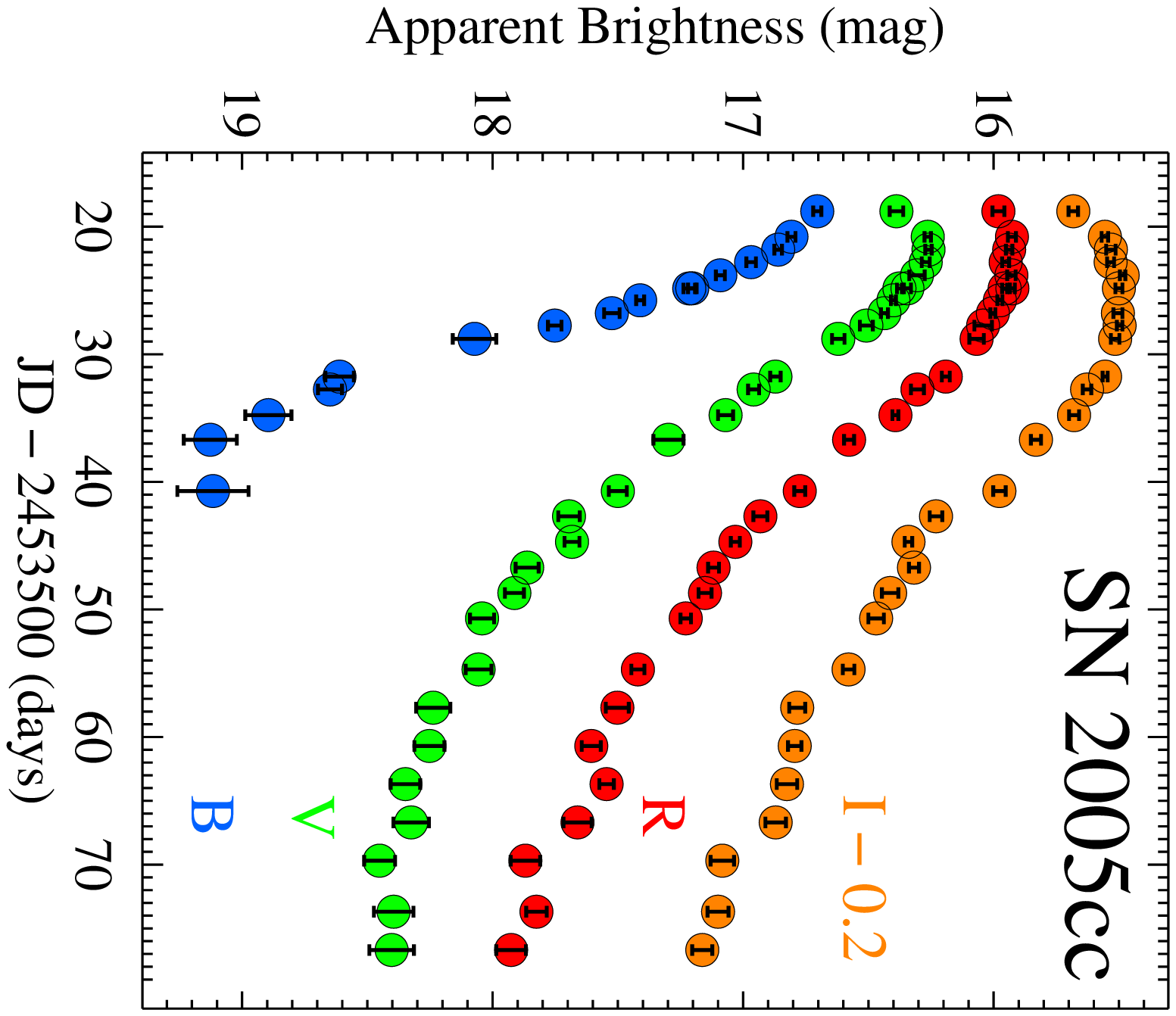}}
\caption{KAIT \bvri\ (blue, green, red, and orange, respectively)
  light curves of SN~2005cc \citep{Ganeshalingam10}.  The
  uncertainties for most data points are smaller than the plotted
  symbols.}\label{f:05cc_lc}
\end{center}
\end{figure}

\begin{figure}
\begin{center}
\epsscale{1.3}
\rotatebox{90}{
\plotone{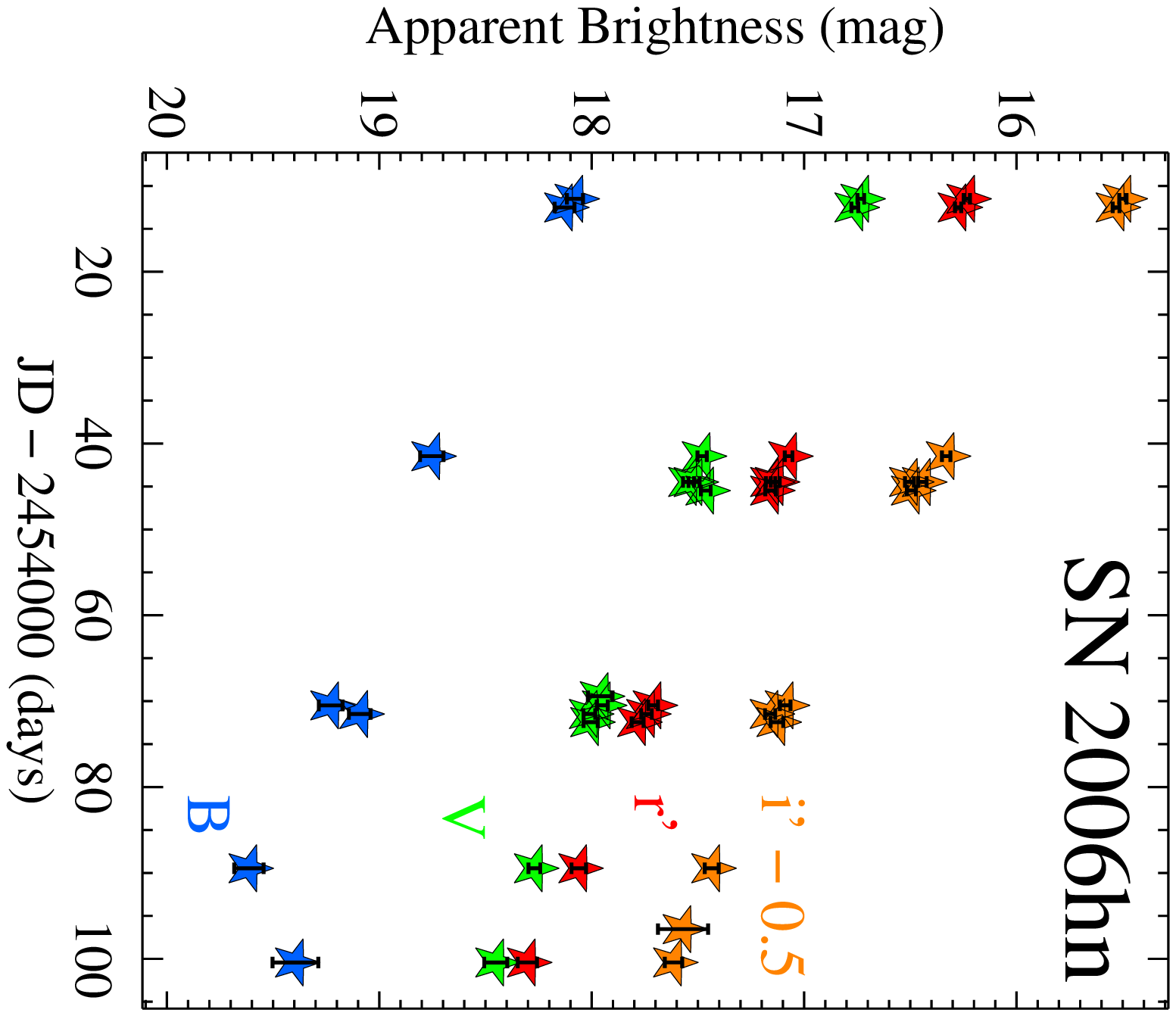}}
\caption{CfA3 \sbvri\ (blue, green, red, and orange, respectively)
  light curves of SN~2006hn \citep{Hicken09:lc}.  The uncertainties
  for most data points are smaller than the plotted
  symbols.}\label{f:06hn_lc}
\end{center}
\end{figure}

\begin{figure}
\begin{center}
\epsscale{1.3}
\rotatebox{90}{
\plotone{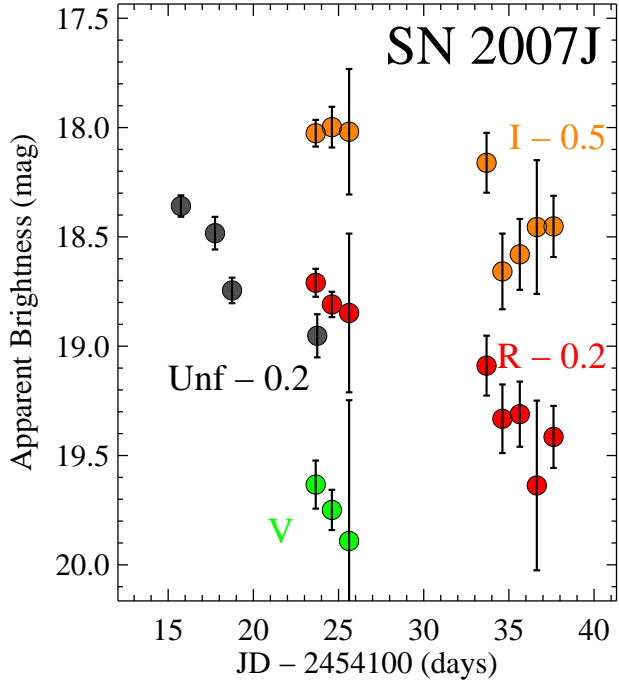}}
\caption{KAIT \vri (green, red, and orange, respectively) and
  unfiltered (grey) light curves of SN~2007J.  There is an upper limit
  unfiltered nondetection point 40~days before the first detection
  that is not plotted.}\label{f:07j_lc}
\end{center}
\end{figure}

\begin{figure}
\begin{center}
\epsscale{1.3}
\rotatebox{90}{
\plotone{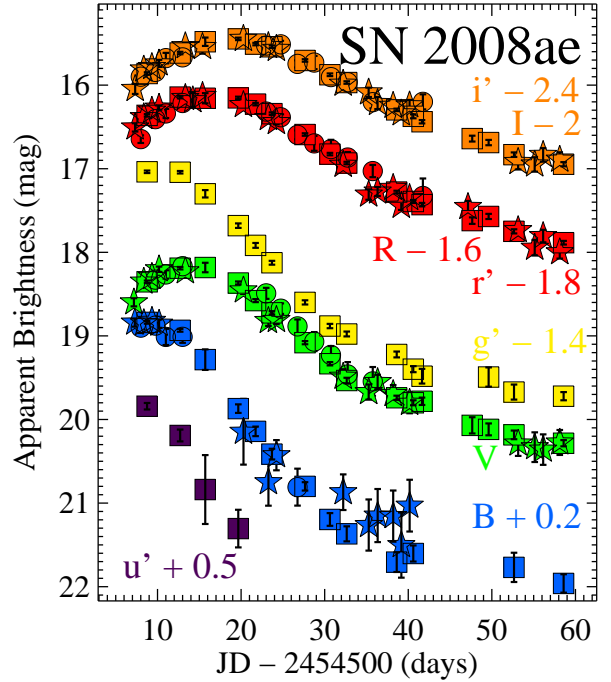}}
\caption{\ubvgrrii\ (purple, blue, green, yellow, red, red, orange,
  and orange, respectively) light curves of SN~2008ae.  KAIT, CfA4,
  and CSP photometry are represented by circles, stars, and squares,
  respectively.  The KAIT and CfA4 data were previously published by
  \citet{Ganeshalingam10} and \citet{Hicken12}, respectively.  The
  uncertainties for most data points are smaller than the plotted
  symbols.}\label{f:08ae_lc}
\end{center}
\end{figure}

\begin{figure}
\begin{center}
\epsscale{1.3}
\rotatebox{90}{
\plotone{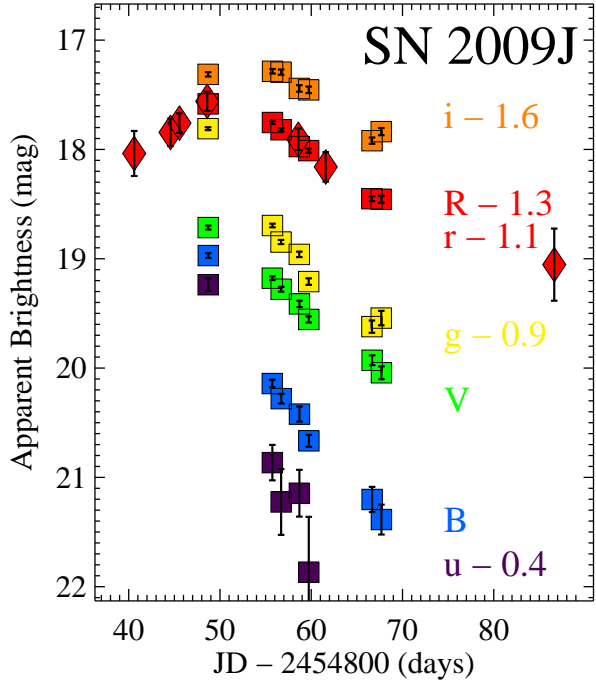}}
\caption{\ubvgri\ (purple, blue, green, yellow, red, and orange,
  respectively) light curves of SN~2009J.  CSP and CHASE photometry
  are represented by squares and diamonds, respectively.  The
  uncertainties for most data points are smaller than the plotted
  symbols.}\label{f:09j_lc}
\end{center}
\end{figure}

\begin{figure}
\begin{center}
\epsscale{1.3}
\rotatebox{90}{
\plotone{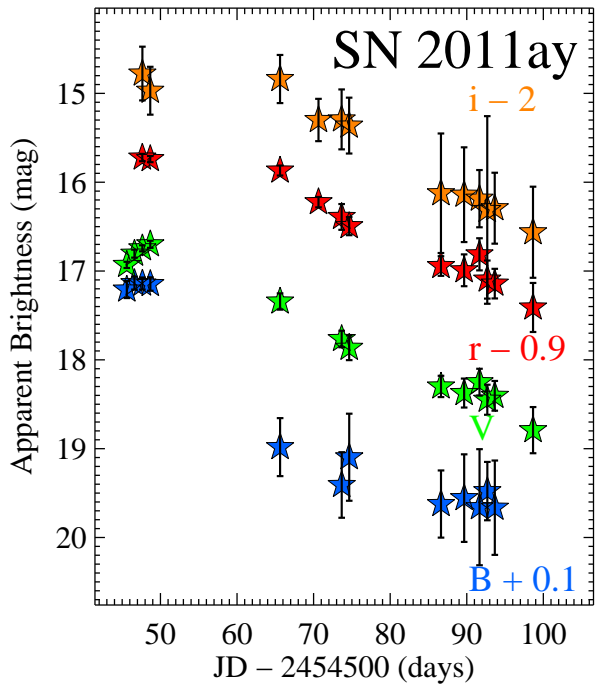}}
\caption{CfA \sbvri\ (blue, green, red, and orange, respectively)
  light curves of SN~2011ay.}\label{f:11ay_lc}
\end{center}
\end{figure}

\begin{figure}
\begin{center}
\epsscale{1.3}
\rotatebox{90}{
\plotone{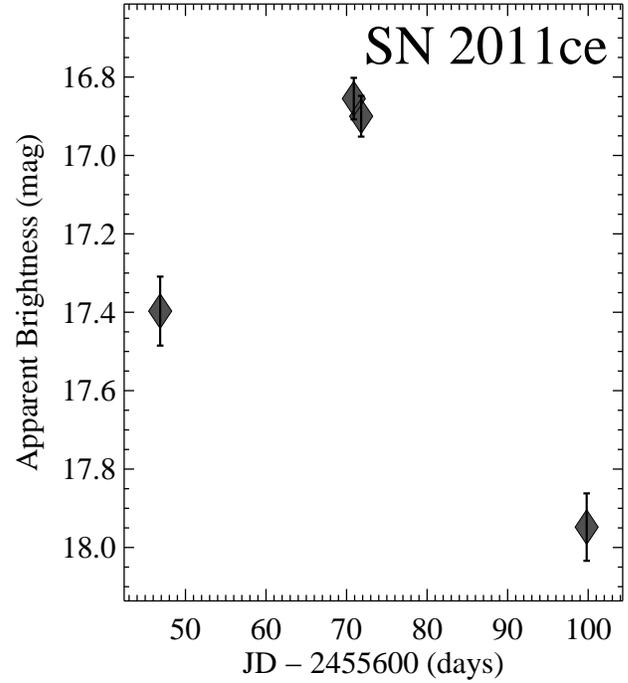}}
\caption{CHASE unfiltered light curve of SN~2011ce.}\label{f:11ce_lc}
\end{center}
\end{figure}

\begin{figure}
\begin{center}
\epsscale{1.3}
\rotatebox{90}{
\plotone{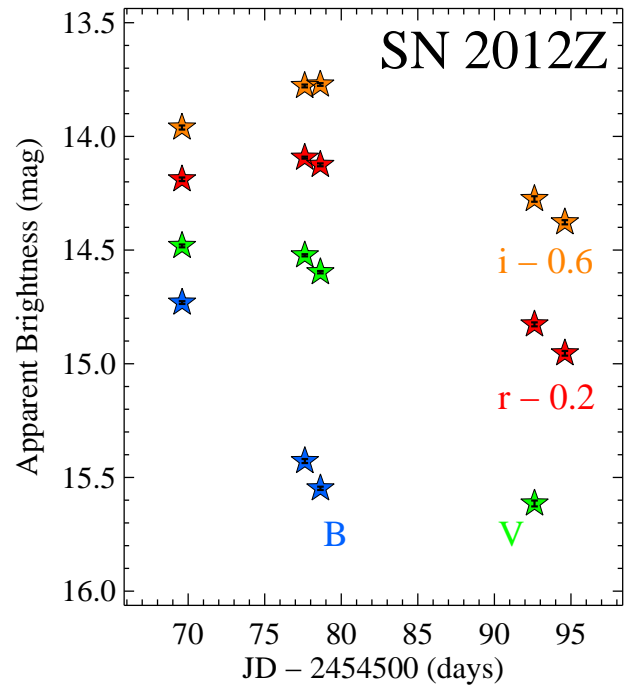}}
\caption{CfA \sbvri\ (blue, green, red, and orange, respectively)
  light curves of SN~2012Z.  The uncertainties for most data points
  are smaller than the plotted symbols.}\label{f:12z_lc}
\end{center}
\end{figure}

All previously unpublished photometry is listed in Table~\ref{t:phot}.
Light curves for SNe~2003gq, 2004cs, 2005P, 2005cc, 2006hn, 2007J,
2008ae, 2009J, 2011ay, and 2012Z are shown in
Figures~\ref{f:03gq_lc}--\ref{f:12z_lc}, respectively.

\begin{deluxetable}{ccc}
\tabletypesize{\scriptsize}
\tablewidth{0pt}
\tablecaption{Photometry of SNe~Iax\label{t:phot}}
\tablehead{
\colhead{JD $-$ 2450000} &
\colhead{Magnitude} &
\colhead{Uncertainty}} \\
 
\startdata
 
\multicolumn{3}{c}{SN~2004cs} \\
\tableline
\tableline
\multicolumn{3}{c}{KAIT Unfiltered} \\
\tableline
3177.90 & 19.16 & 0.14 \\
3179.90 & 18.11 & 0.04 \\
3180.89 & 17.82 & 0.03 \\
3182.89 & 17.60 & 0.03 \\
3184.85 & 17.47 & 0.04 \\
3187.87 & 17.54 & 0.05 \\
3193.89 & 18.04 & 0.04 \\
3195.81 & 18.18 & 0.04 \\
3197.82 & 18.30 & 0.05 \\
3199.82 & 18.65 & 0.06 \\
3200.79 & 18.78 & 0.08 \\
3204.86 & 18.88 & 0.20 \\
3206.80 & 19.12 & 0.14 \\
\tableline
\tableline
\multicolumn{3}{c}{SN~2005P} \\
\tableline
\tableline
\multicolumn{3}{c}{KAIT Unfiltered} \\
\tableline
3392.05 & 17.93 & 0.09 \\
3393.09 & 17.82 & 0.09 \\
3394.07 & 17.88 & 0.09 \\
3395.03 & 17.97 & 0.04 \\
3412.03 & 18.28 & 0.12 \\
3436.97 & 18.70 & 0.13 \\
3445.98 & 18.78 & 0.15 \\
3461.97 & 19.03 & 0.14 \\
3471.93 & 18.91 & 0.31 \\
3479.89 & 19.31 & 0.22 \\
\tableline
\tableline
\multicolumn{3}{c}{SN~2007J} \\
\tableline
\tableline
\multicolumn{3}{c}{KAIT Unfiltered} \\
\tableline
4115.75 & 18.56 & 0.05 \\
4117.75 & 18.68 & 0.07 \\
4118.75 & 18.95 & 0.06 \\
4123.75 & 19.15 & 0.10 \\
\tableline
\multicolumn{3}{c}{KAIT $V$} \\
\tableline
4123.66 & 19.63 & 0.11 \\
4124.61 & 19.75 & 0.09 \\
4125.62 & 19.89 & 0.64 \\
\tableline
\multicolumn{3}{c}{KAIT $R$} \\
\tableline
4123.66 & 18.91 & 0.06 \\
4124.61 & 19.01 & 0.06 \\
4125.62 & 19.05 & 0.36 \\
4133.69 & 19.29 & 0.14 \\
4134.62 & 19.53 & 0.16 \\
4135.64 & 19.51 & 0.15 \\
4136.64 & 19.84 & 0.39 \\
4137.62 & 19.61 & 0.14 \\
\tableline
\multicolumn{3}{c}{KAIT $I$} \\
\tableline
4123.66 & 18.53 & 0.06 \\
4124.61 & 18.50 & 0.09 \\
4125.62 & 18.52 & 0.29 \\
4133.69 & 18.66 & 0.14 \\
4134.62 & 19.16 & 0.17 \\
4135.64 & 19.08 & 0.16 \\
4136.64 & 18.95 & 0.31 \\
4137.62 & 18.95 & 0.14 \\
\tableline
\tableline
\multicolumn{3}{c}{SN~2008ae} \\
\tableline
\tableline
\multicolumn{3}{c}{CSP $u$} \\
\tableline
4508.75 & 19.34 & 0.04 \\
4512.71 & 19.69 & 0.07 \\
4515.71 & 20.34 & 0.41 \\
4519.67 & 20.80 & 0.23 \\
\tableline
\multicolumn{3}{c}{CSP $B$} \\
\tableline
4508.75 & 18.64 & 0.02 \\
4512.71 & 18.73 & 0.02 \\
4515.71 & 19.08 & 0.12 \\
4519.67 & 19.67 & 0.05 \\
4521.73 & 19.94 & 0.06 \\
4523.69 & 20.21 & 0.06 \\
4527.65 & 20.60 & 0.05 \\
4532.64 & 21.17 & 0.09 \\
4552.65 & 21.57 & 0.17 \\
4558.58 & 21.76 & 0.11 \\
4564.58 & 21.72 & 0.11 \\
4591.54 & 22.21 & 0.17 \\
4530.63 & 21.00 & 0.08 \\
4538.62 & 21.51 & 0.12 \\
4540.60 & 21.40 & 0.10 \\
\tableline
\multicolumn{3}{c}{CSP $V$} \\
\tableline
4508.75 & 18.35 & 0.01 \\
4512.71 & 18.19 & 0.01 \\
4515.71 & 18.18 & 0.06 \\
4519.67 & 18.37 & 0.02 \\
4521.73 & 18.58 & 0.02 \\
4523.69 & 18.72 & 0.02 \\
4527.65 & 19.08 & 0.02 \\
4532.64 & 19.53 & 0.03 \\
4541.65 & 19.78 & 0.04 \\
4547.65 & 20.07 & 0.10 \\
4549.61 & 20.12 & 0.07 \\
4552.65 & 20.18 & 0.05 \\
4558.58 & 20.28 & 0.04 \\
4564.58 & 20.36 & 0.04 \\
4569.64 & 20.57 & 0.10 \\
4574.56 & 20.46 & 0.13 \\
4576.53 & 20.43 & 0.10 \\
4591.54 & 20.92 & 0.07 \\
4597.58 & 21.06 & 0.11 \\
4530.63 & 19.33 & 0.02 \\
4538.62 & 19.74 & 0.03 \\
4540.60 & 19.80 & 0.03 \\
\tableline
\multicolumn{3}{c}{CSP $g$} \\
\tableline
4508.75 & 18.44 & 0.01 \\
4512.71 & 18.44 & 0.01 \\
4515.71 & 18.70 & 0.04 \\
4519.67 & 19.08 & 0.03 \\
4521.73 & 19.32 & 0.03 \\
4523.69 & 19.52 & 0.02 \\
4527.65 & 20.00 & 0.03 \\
4532.64 & 20.38 & 0.03 \\
4541.65 & 20.88 & 0.09 \\
4549.61 & 20.89 & 0.12 \\
4552.65 & 21.07 & 0.09 \\
4558.58 & 21.12 & 0.05 \\
4564.58 & 21.17 & 0.05 \\
4569.64 & 21.22 & 0.13 \\
4591.54 & 21.62 & 0.07 \\
4597.58 & 21.73 & 0.16 \\
4530.63 & 20.28 & 0.03 \\
4538.62 & 20.62 & 0.04 \\
4540.60 & 20.80 & 0.04 \\
\tableline
\multicolumn{3}{c}{CSP $r$} \\
\tableline
4508.75 & 18.16 & 0.01 \\
4512.71 & 17.94 & 0.01 \\
4515.71 & 17.96 & 0.06 \\
4519.67 & 17.95 & 0.01 \\
4521.73 & 18.03 & 0.01 \\
4523.69 & 18.14 & 0.01 \\
4527.65 & 18.39 & 0.01 \\
4532.64 & 18.73 & 0.01 \\
4541.65 & 19.23 & 0.02 \\
4547.65 & 19.42 & 0.04 \\
4549.61 & 19.37 & 0.03 \\
4552.65 & 19.55 & 0.03 \\
4558.58 & 19.68 & 0.02 \\
4564.58 & 19.84 & 0.03 \\
4569.64 & 19.90 & 0.04 \\
4574.56 & 20.02 & 0.06 \\
4576.53 & 19.98 & 0.05 \\
4591.54 & 20.43 & 0.04 \\
4597.58 & 20.65 & 0.08 \\
4530.63 & 18.62 & 0.01 \\
4538.62 & 19.08 & 0.02 \\
4540.60 & 19.19 & 0.02 \\
\tableline
\multicolumn{3}{c}{CSP $i$} \\
\tableline
4508.75 & 18.26 & 0.01 \\
4512.71 & 18.02 & 0.01 \\
4515.71 & 17.88 & 0.05 \\
4519.67 & 17.85 & 0.01 \\
4521.73 & 17.91 & 0.01 \\
4523.69 & 17.94 & 0.01 \\
4527.65 & 18.11 & 0.01 \\
4532.64 & 18.37 & 0.01 \\
4541.65 & 18.84 & 0.02 \\
4547.65 & 19.04 & 0.04 \\
4549.61 & 19.09 & 0.03 \\
4552.65 & 19.23 & 0.03 \\
4558.58 & 19.35 & 0.03 \\
4564.58 & 19.54 & 0.03 \\
4569.64 & 19.56 & 0.04 \\
4574.56 & 19.70 & 0.06 \\
4576.53 & 19.70 & 0.05 \\
4591.54 & 20.08 & 0.04 \\
4597.58 & 20.25 & 0.08 \\
4530.63 & 18.28 & 0.02 \\
4538.62 & 18.70 & 0.01 \\
4540.60 & 18.77 & 0.02 \\
\tableline
\tableline
\multicolumn{3}{c}{SN~2009J} \\
\tableline
\tableline
\multicolumn{3}{c}{CSP $u$} \\
\tableline
4848.71 & 19.64 & 0.06 \\
4855.74 & 21.27 & 0.16 \\
4856.70 & 21.62 & 0.30 \\
4858.70 & 21.55 & 0.21 \\
4859.73 & 22.27 & 0.51 \\
\tableline
\multicolumn{3}{c}{CSP $B$} \\
\tableline
4848.72 & 18.97 & 0.02 \\
4855.73 & 20.14 & 0.04 \\
4856.71 & 20.28 & 0.04 \\
4858.71 & 20.42 & 0.07 \\
4859.74 & 20.67 & 0.06 \\
4866.67 & 21.20 & 0.12 \\
4867.69 & 21.39 & 0.14 \\
\tableline
\multicolumn{3}{c}{CSP $V$} \\
\tableline
4848.71 & 18.71 & 0.02 \\
4855.73 & 19.18 & 0.02 \\
4856.70 & 19.28 & 0.02 \\
4858.71 & 19.41 & 0.03 \\
4859.73 & 19.55 & 0.03 \\
4866.67 & 19.93 & 0.04 \\
4867.69 & 20.04 & 0.06 \\
\tableline
\multicolumn{3}{c}{CSP $g$} \\
\tableline
4848.69 & 18.71 & 0.01 \\
4855.75 & 19.60 & 0.02 \\
4856.69 & 19.75 & 0.02 \\
4858.69 & 19.86 & 0.03 \\
4859.72 & 20.11 & 0.03 \\
4866.65 & 20.52 & 0.06 \\
4867.67 & 20.44 & 0.07 \\
\tableline
\multicolumn{3}{c}{CSP $r$} \\
\tableline
4848.70 & 18.68 & 0.01 \\
4855.75 & 18.85 & 0.01 \\
4856.69 & 18.92 & 0.01 \\
4858.69 & 19.07 & 0.02 \\
4859.72 & 19.11 & 0.02 \\
4866.65 & 19.55 & 0.02 \\
4867.67 & 19.56 & 0.03 \\
\tableline
\multicolumn{3}{c}{CSP $i$} \\
\tableline
4848.70 & 18.91 & 0.02 \\
4855.76 & 18.89 & 0.02 \\
4856.69 & 18.89 & 0.02 \\
4858.70 & 19.04 & 0.03 \\
4859.72 & 19.05 & 0.03 \\
4866.66 & 19.52 & 0.03 \\
4867.67 & 19.44 & 0.03 \\
\tableline
\multicolumn{3}{c}{CHASE $R$} \\
\tableline
4840.60 & 19.34 & 0.21 \\
4844.58 & 19.14 & 0.12 \\
4845.56 & 19.06 & 0.09 \\
4848.60 & 18.86 & 0.09 \\
4858.59 & 19.21 & 0.10 \\
4861.59 & 19.46 & 0.14 \\
4886.63 & 20.35 & 0.33 \\
\tableline
\tableline
\multicolumn{3}{c}{SN~2011ay} \\
\tableline
\tableline
\multicolumn{3}{c}{CfA $B$} \\
\tableline
5645.61 & 17.11 & 0.10 \\
5646.61 & 17.04 & 0.07 \\
5647.64 & 17.04 & 0.07 \\
5648.68 & 17.05 & 0.08 \\
5665.63 & 18.88 & 0.33 \\
5673.67 & 19.31 & 0.37 \\
5674.65 & 19.00 & 0.49 \\
5686.64 & 19.52 & 0.38 \\
5689.66 & 19.46 & 0.49 \\
5691.66 & 19.56 & 0.65 \\
5692.69 & 19.38 & 0.33 \\
5693.66 & 19.56 & 0.53 \\
\tableline
\multicolumn{3}{c}{CfA $V$} \\
\tableline
5645.60 & 16.93 & 0.03 \\
5646.60 & 16.80 & 0.04 \\
5647.64 & 16.74 & 0.03 \\
5648.67 & 16.70 & 0.04 \\
5665.62 & 17.34 & 0.09 \\
5673.66 & 17.77 & 0.09 \\
5674.65 & 17.86 & 0.14 \\
5686.63 & 18.30 & 0.12 \\
5689.66 & 18.37 & 0.16 \\
5691.66 & 18.25 & 0.15 \\
5692.68 & 18.45 & 0.17 \\
5693.66 & 18.41 & 0.17 \\
5698.66 & 18.79 & 0.26 \\
\tableline
\multicolumn{3}{c}{CfA $r$} \\
\tableline
5647.63 & 16.62 & 0.04 \\
5648.67 & 16.64 & 0.03 \\
5665.62 & 16.77 & 0.06 \\
5670.64 & 17.12 & 0.06 \\
5673.66 & 17.29 & 0.15 \\
5674.64 & 17.39 & 0.10 \\
5686.63 & 17.84 & 0.11 \\
5689.66 & 17.89 & 0.18 \\
5691.66 & 17.71 & 0.18 \\
5692.68 & 18.00 & 0.21 \\
5693.66 & 18.04 & 0.17 \\
5698.65 & 18.31 & 0.28 \\
\tableline
\multicolumn{3}{c}{CfA $i$} \\
\tableline
5647.64 & 16.78 & 0.30 \\
5648.67 & 16.97 & 0.27 \\
5665.62 & 16.84 & 0.27 \\
5670.64 & 17.30 & 0.24 \\
5673.66 & 17.29 & 0.34 \\
5674.64 & 17.36 & 0.31 \\
5686.63 & 18.12 & 0.67 \\
5689.65 & 18.14 & 0.53 \\
5691.66 & 18.19 & 0.32 \\
5692.68 & 18.31 & 1.06 \\
5693.65 & 18.29 & 0.40 \\
5698.65 & 18.56 & 0.51 \\
5852.98 & 20.20 & 0.36 \\
5855.95 & 20.24 & 0.33 \\
\tableline
\tableline
\multicolumn{3}{c}{SN~2012Z} \\
\tableline
\tableline
\multicolumn{3}{c}{CfA $B$} \\
\tableline
5969.61 & 14.73 & 0.01 \\
5977.62 & 15.43 & 0.01 \\
5978.64 & 15.55 & 0.01 \\
\tableline
\multicolumn{3}{c}{CfA $V$} \\
\tableline
5969.60 & 14.48 & 0.01 \\
5977.62 & 14.52 & 0.01 \\
5978.64 & 14.60 & 0.01 \\
5992.62 & 15.61 & 0.01 \\
\tableline
\multicolumn{3}{c}{CfA $r$} \\
\tableline
5969.60 & 14.39 & 0.01 \\
5977.61 & 14.29 & 0.01 \\
5978.63 & 14.33 & 0.01 \\
5992.62 & 15.03 & 0.01 \\
5994.60 & 15.15 & 0.01 \\
\tableline
\multicolumn{3}{c}{CfA $i$} \\
\tableline
5969.60 & 14.56 & 0.01 \\
5977.61 & 14.38 & 0.01 \\
5978.63 & 14.37 & 0.01 \\
5992.61 & 14.88 & 0.01 \\
5994.60 & 14.98 & 0.01 \\
\tableline
\tableline
\multicolumn{3}{c}{SN~2011ce} \\
\tableline
\tableline
\multicolumn{3}{c}{CHASE Unfiltered} \\
\tableline
5646.86 & 17.40 & 0.09 \\
5670.90 & 16.86 & 0.05 \\
5671.82 & 16.90 & 0.05 \\
5699.81 & 17.95 & 0.09 \\
\tableline
\tableline
 
\enddata
 
\end{deluxetable}

\subsection{Spectroscopy}\label{ss:spec}

We have obtained several low-resolution optical spectra with the FAST
spectrograph \citep{Fabricant98} on the FLWO 1.5~m telescope, the Kast
double spectrograph \citep{Miller93} on the Shane 3~m telescope at
Lick Observatory, the EMMI spectrograph \citep{Dekker86} on the New
Technology Telescope at La Silla Observatory, the EFOSC spectrograph
\citep{Buzzoni84} on the ESO 3.6~m telescope, the LDSS3
spectrograph\footnote{http://www.lco.cl/telescopes-information/magellan/instruments-1/ldss-3-1/
.} on the Magellan Clay 6.5~m telescope, the IMACS spectrograph
\citep{Dressler11} on the Magellan Baade 6.5~m telescope, and the Low
Resolution Imaging Spectrometer \citep[LRIS;][]{Oke95} on the 10~m
Keck~I telescope.

Standard CCD processing and spectrum extraction were accomplished with
IRAF.  The data were extracted using the optimal algorithm of
\citet{Horne86}.  Low-order polynomial fits to calibration-lamp
spectra were used to establish the wavelength scale, and small
adjustments derived from night-sky lines in the object frames were
applied.  We employed our own IDL routines to flux calibrate the data
and remove telluric lines using the well-exposed continua of the
spectrophotometric standard stars \citep{Wade88, Foley03}.  Details of
our spectroscopic reduction techniques are described by
\citet{Silverman12:bsnip}.

\begin{figure}
\begin{center}
\epsscale{1.25}
\rotatebox{90}{
\plotone{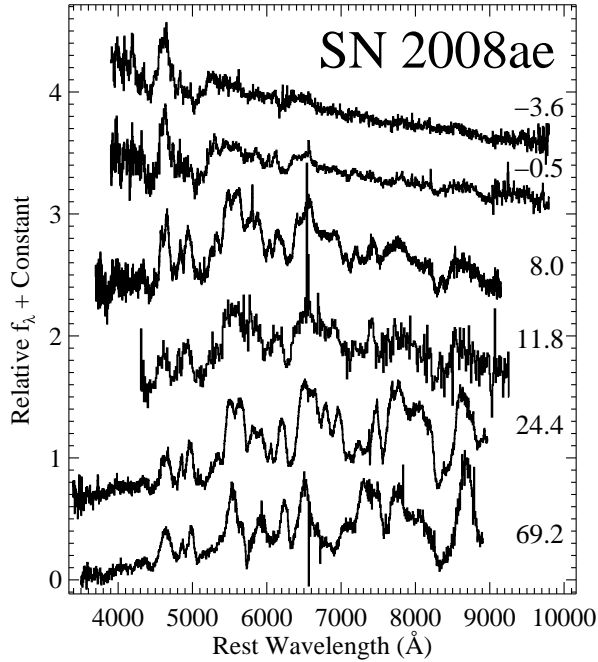}}
\caption{Optical spectra of SN~2008ae.  Rest-frame phases relative to
  $V$ maximum are listed to the right of each
  spectrum.}\label{f:08ae_spec}
\end{center}
\end{figure}

\begin{figure}
\begin{center}
\epsscale{1.25}
\rotatebox{90}{
\plotone{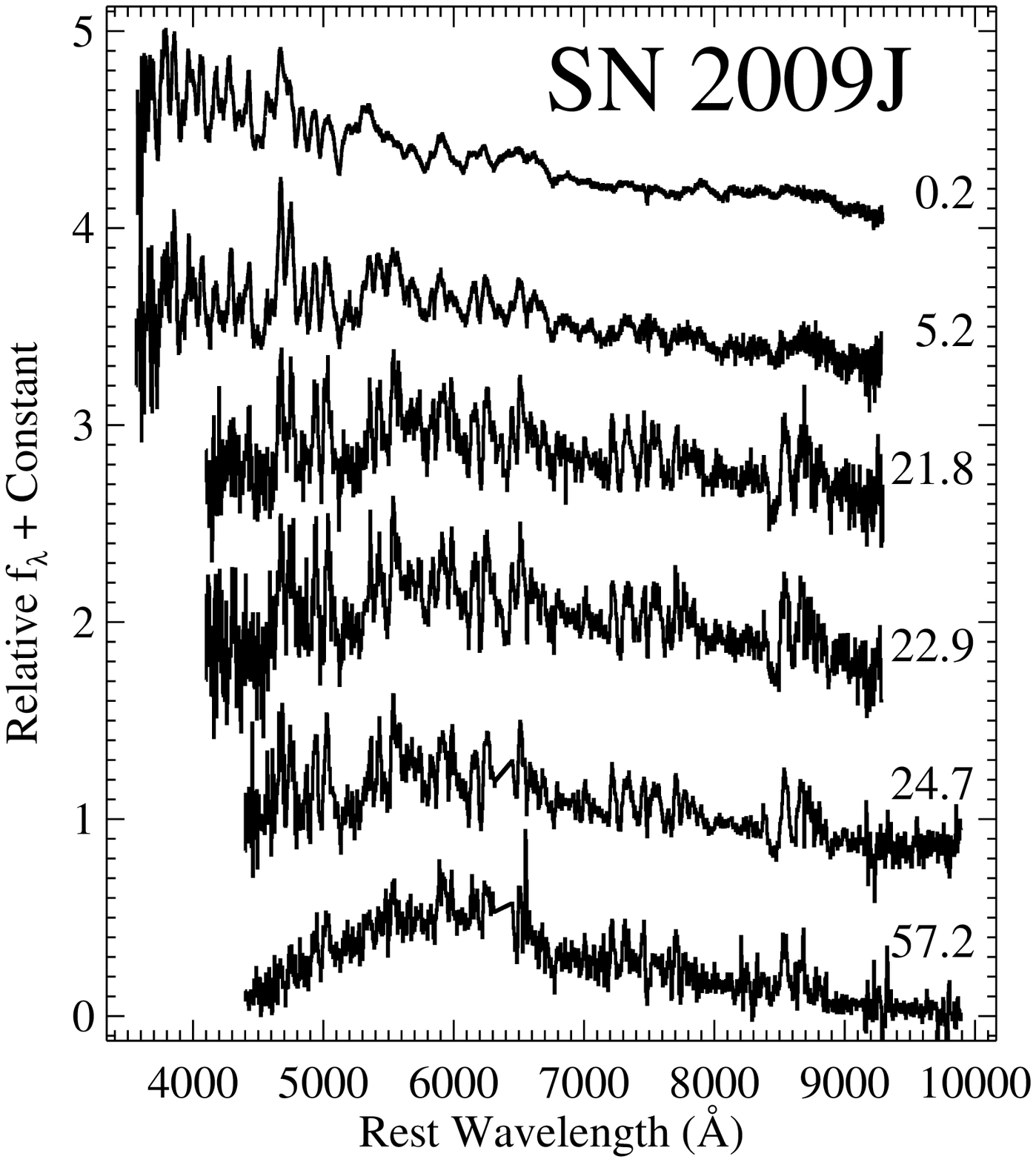}}
\caption{Optical spectra of SN~2009J.  Rest-frame phases relative to
  $V$ maximum are listed to the right of each
  spectrum.}\label{f:09j_spec}
\end{center}
\end{figure}

\begin{figure}
\begin{center}
\epsscale{1.75}
\rotatebox{90}{
\plotone{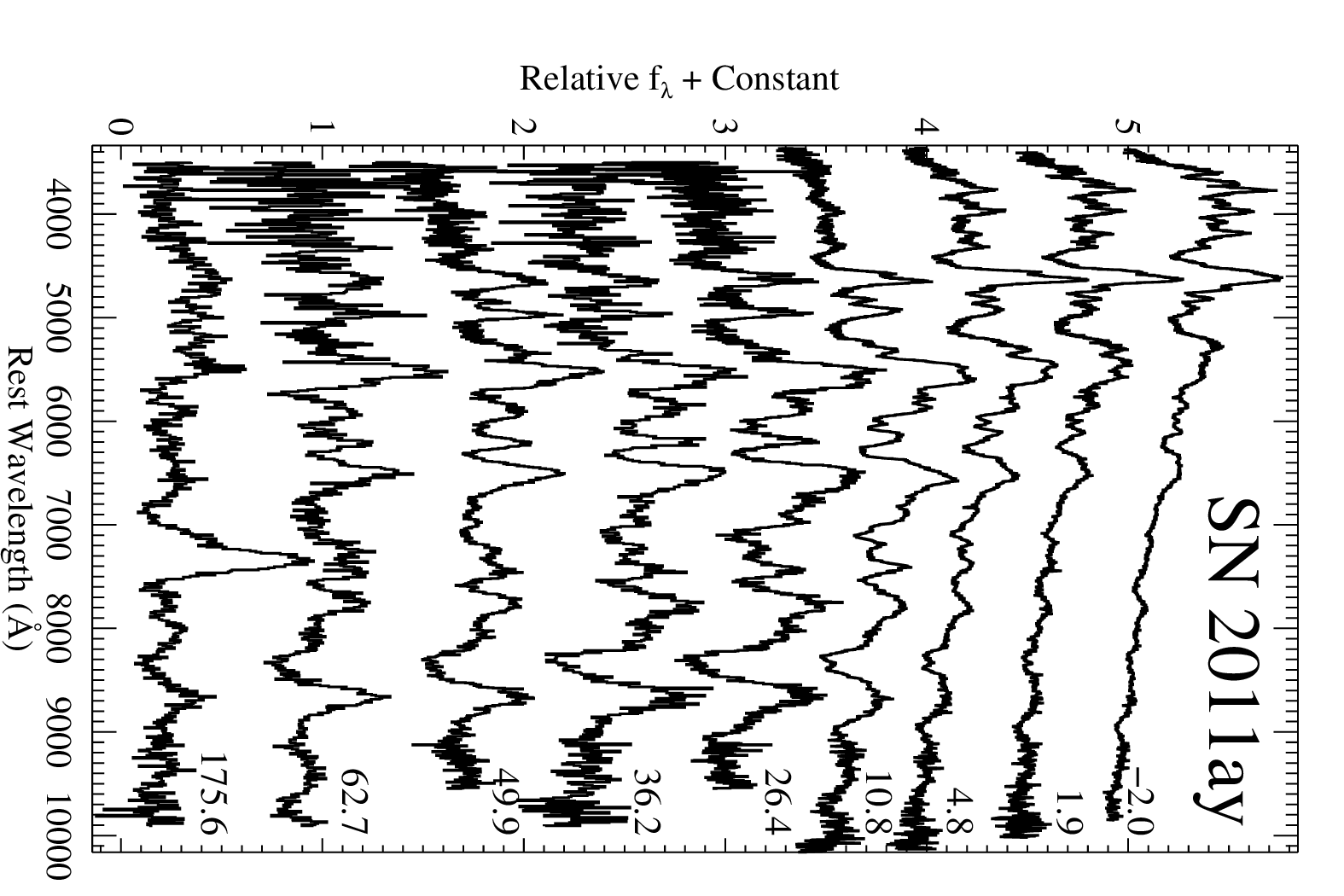}}
\caption{Optical spectra of SN~2011ay.  Rest-frame phases relative to
  $V$ maximum are listed to the right of each
  spectrum.}\label{f:11ay_spec}
\end{center}
\end{figure}

\begin{figure}
\begin{center}
\epsscale{1.4}
\rotatebox{90}{
\plotone{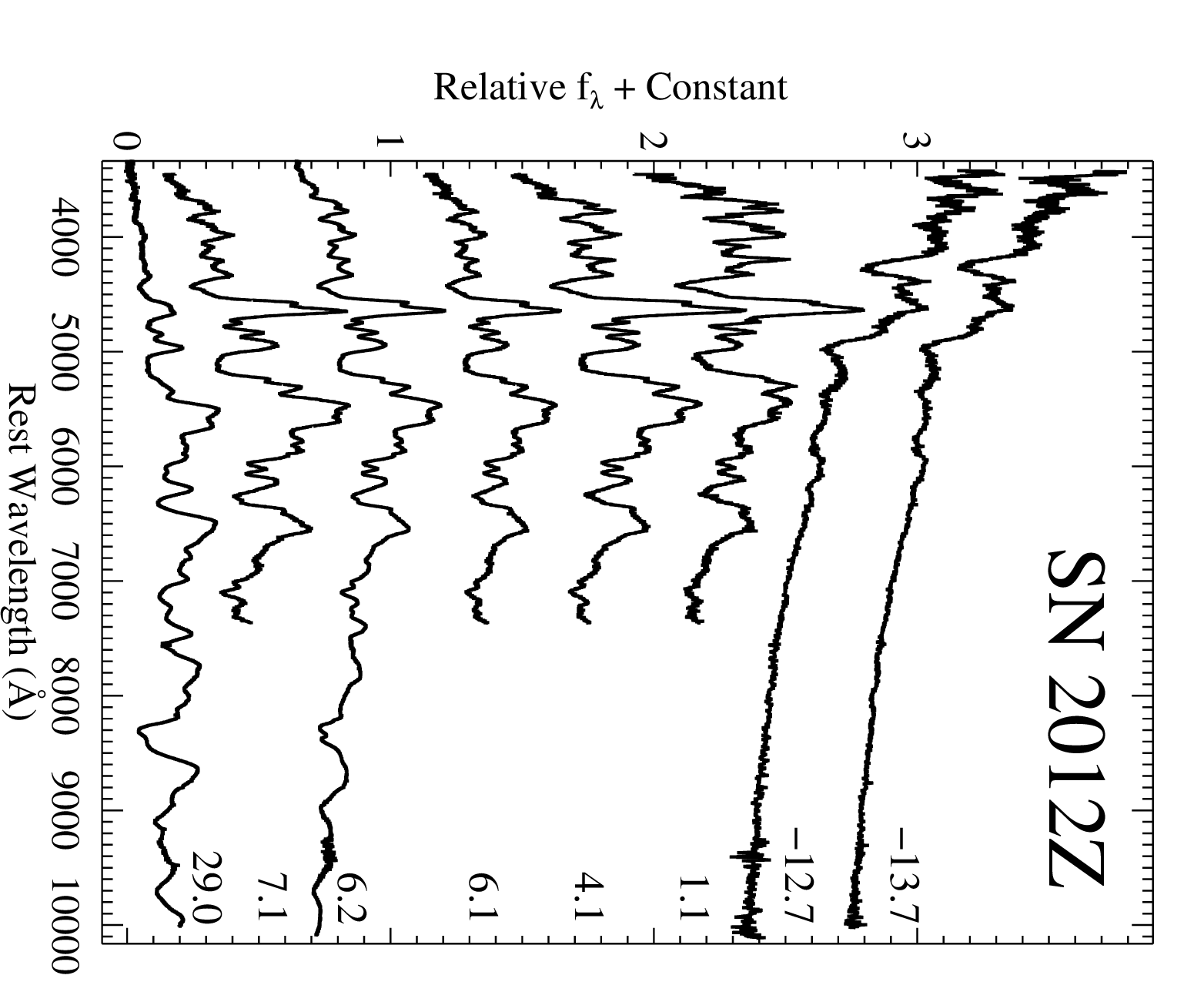}}
\caption{Optical spectra of SN~2012Z.  Rest-frame phases relative to
  $V$ maximum are listed to the right of each
  spectrum.}\label{f:12z_spec}
\end{center}
\end{figure}

A log of all previously unpublished spectral observations is presented
in Table~\ref{t:spec}.  Spectral sequences for SNe~2008ae, 2009J,
2011ay, and 2012Z are shown in Figures~\ref{f:08ae_spec} --
\ref{f:12z_spec}, respectively.  We only display previously
unpublished data.  We do not display our single previously unpublished
spectrum of SN~2008ge since it was obtained the same night as the
first spectrum presented by \citet{Foley10:08ge}.

\begin{deluxetable}{r@{ }l@{ }l@{ }r@{ }l}
\tabletypesize{\scriptsize}
\tablewidth{0pt}
\tablecaption{Log of Spectral Observations\label{t:spec}}
\tablehead{
\colhead{} &
\colhead{} &
\colhead{Telescope /} &
\colhead{Exp.} &
\colhead{} \\
\colhead{Phase\tablenotemark{a}} &
\colhead{UT Date} &
\colhead{Instrument} &
\colhead{(s)} &
\colhead{Observer\tablenotemark{b}}}

\startdata

 \multicolumn{5}{c}{SN~2007J} \\
\tableline
 2--42  & 2007 Jan.\ 17.2 & FLWO/FAST        & 1800                 & PB1 \\
 6--46  & 2007 Jan.\ 21.3 & Lick/Kast        & 2100                 & MG, TS \\
 6--46  & 2007 Jan.\ 21.4 & Keck/LRIS        &  300                 & AF, JS, RF \\
30--70  & 2007 Feb.\ 14.3 & Keck/LRIS        & 1200                 & AF, JS, RC, RF \\
62--102 & 2007 Mar.\ 18.3 & Keck/LRIS        &  900                 & JS, RC \\
\tableline
 \multicolumn{5}{c}{SN~2008ae} \\
\tableline
 $-3.6$ & 2008 Feb.\ 13.3 & NTT/EMMI         &  900                 & GF \\
 $-0.5$ & 2008 Feb.\ 16.4 & Lick/Kast        & 1800                 & JS, MG, NL, TS \\
   8.0  & 2008 Feb.\ 25.3 & Clay/LDSS        & 1800                 & GP, NM \\
  11.8  & 2008 Feb.\ 29.2 & Lick/Kast        & 1800                 & JS, MG, NL, TS \\
  24.4  & 2008 Mar.\ 13.1 & ESO/EFOSC        & 1200                 & GF \\
  69.2  & 2008 Apr.\ 28.3 & Keck/LRIS        &  900                 & AF, DP, JS, TS \\
\tableline
 \multicolumn{5}{c}{SN~2008ge} \\
\tableline
  41.1  & 2008 Oct.\ 27.5 & Keck/LRIS        &  105                 & AF, BT, JS, TS \\
\tableline
 \multicolumn{5}{c}{SN~2009J} \\
\tableline
   0.2  & 2009 Jan.\ 17.2 & Clay/LDSS        &  700                 & MP, MS \\
   5.2  & 2009 Jan.\ 22.2 & Clay/LDSS        &  700                 & NM \\
  21.8  & 2009 Feb.\  8.1 & Clay/LDSS        &  700                 & NM \\
  22.9  & 2009 Feb.\  9.2 & Clay/LDSS        &  700                 & NM \\
  24.7  & 2009 Feb.\ 11.1 & Baade/IMACS      &  600                 & NM \\
  57.2  & 2009 Mar.\ 16.1 & Baade/IMACS      & 1200                 & NM \\
\tableline
 \multicolumn{5}{c}{SN~2010ae} \\
\tableline
\nodata & 2010 Feb.\ 26.1 & Baade/IMACS      & 1800                 & AS1 \\
\tableline
 \multicolumn{5}{c}{SN~2011ay} \\
\tableline
 $-2.0$ & 2008 Mar.\ 29.2 & Lick/Kast        & 1500                 & AB, JW, RA \\
   1.9  & 2008 Apr.\  2.2 & Lick/Kast        & 1921                 & SH \\
   4.8  & 2008 Apr.\  5.2 & Lick/Kast        & 2100                 & JR, KC \\
  10.8  & 2008 Apr.\ 11.2 & Lick/Kast        & 1800                 & AS2, SH \\
  26.4  & 2008 Apr.\ 27.2 & Lick/Kast        & 1800                 & VB \\
  36.2  & 2008 May    7.2 & Lick/Kast        & 1800                 & RA \\
  49.9  & 2008 May   21.2 & Lick/Kast        & 2100                 & AD, GC, ML \\
  62.7  & 2008 Jun.\  3.2 & Keck/LRIS        &  450                 & AF, BC, JS \\
 175.6  & 2008 Sep.\ 26.6 & Keck/LRIS        & 1850                 & AF, BC, JS \\
\tableline
 \multicolumn{5}{c}{SN~2011ce} \\
\tableline
  42\tablenotemark{c} & 2011 Apr.\ 24.2 & Baade/IMACS      &  450        & JA \\
 371\tablenotemark{c} & 2012 Apr.\ 20.4 & Baade/IMACS      & 3600        & RF \\
\tableline
 \multicolumn{5}{c}{SN~2012Z} \\
\tableline
$-13.7$ & 2012 Feb.\  1.2 & Lick/Kast        &  842                 & BC, DC \\
$-12.7$ & 2012 Feb.\  2.2 & Lick/Kast        & 2400                 & KC, PB2 \\
   1.1  & 2012 Feb.\ 16.1 & FLWO/FAST        & 1800                 & JI \\
   4.1  & 2012 Feb.\ 19.1 & FLWO/FAST        & 1800                 & JI \\
   6.1  & 2012 Feb.\ 21.1 & FLWO/FAST        & 1800                 & JI \\
   6.2  & 2012 Feb.\ 21.3 & Keck/LRIS        &  300                 & AM1, AM2, JS, PN \\
   7.1  & 2012 Feb.\ 22.1 & FLWO/FAST        & 1800                 & JI \\
  29.0  & 2012 Mar.\ 15.2 & Keck/LRIS        &  510                 & AM1, BC, JS, PN

\enddata

\tablenotetext{a}{Rest-frame days since $V$ maximum.}

\tablenotetext{b}{AB = A.\ Barth, AD = A.\ Diamond-Stanic, AF = A.\
  Filippenko, AM1 = A.\ Miller, AM1 = A.\ Morgan, AS1 = A.\ Soderberg,
  AS2 = A.\ Sonnenfeld, BC = B.\ Cenko, BT = B.\ Tucker, DC = D.\
  Cohen, DP = D.\ Poznanski, GC = G.\ Canalizo, GF = G.\ Folatelli, GP
  = G.\ Pignata, JA = J.\ Anderson, JI = J.\ Irwin, JR = J.\ Rex, JS =
  J.\ Silverman, JW = J.\ Walsh, KC = K.\ Clubb, MG = M.\
  Ganeshalingam, ML = M.\ Lazarova, MP = M.\ Phillips, MS = M.\
  Stritzinger, NL = N.\ Lee, NM = N.\ Morrell, PB1 = P.\ Berlind, PB2
  = P.\ Blanchard, PN = P.\ Nugent, RA = R.\ Assef, RC = R.\ Chornock,
  RF = R.\ Foley, SH = S.\ Hoenig, TS = T.\ Steele, VB = V.\ Bennert}

\tablenotetext{c}{Approximate phase; see Section~\ref{ss:11ce} for details.}

\end{deluxetable}

\section{Photometric Properties}\label{s:phot}

In this section, we examine the photometric properties of SNe~Iax.  A
minority of the full sample has filtered photometry that covers
maximum brightness.  Since observations near maximum brightness are
critical for direct comparisons between SNe, we focus on the SNe
having those data.  For the SNe in our sample with light curves around
maximum brightness, we can derive several light curve parameters: time
of maximum brightness, peak brightness, peak absolute brightness, and
decline rate.  We fit low-order polynomials to each light curve near
maximum brightness to derive these values for each available band.
For some SNe and some bands, we can only place limits or broad ranges
on the derived values.  We present our measurements in
Table~\ref{t:lc}.

\begin{deluxetable*}{lccccccc}
\tabletypesize{\scriptsize}
\tablewidth{0pt}
\tablecaption{Light-Curve Properties of SNe~Iax\label{t:lc}}
\tablehead{
\colhead{SN Name} &
\colhead{$B$ (mag)} &
\colhead{$V$ (mag)} &
\colhead{$R$ (mag)} &
\colhead{$I$ (mag)} &
\colhead{$g$ (mag)} &
\colhead{$r$ (mag)} &
\colhead{$i$ (mag)}}

\startdata

 & \multicolumn{7}{c}{$t_{\max}$ (JD $-$ 2,450,000)} \\
\tableline

2002cx & 2413.34 (0.25) & 2418.31 (0.39)    & 2420.95 (0.94) & 2421.79 (2.83) & \nodata          & \nodata          & \nodata          \\
2003gq & 2848.78 (0.26) & 2852.56 (0.28)    & 2854.49 (0.35) & 2857.01 (0.40) & \nodata          & \nodata          & \nodata          \\
2005cc & $<$3519.29     & 3522.10 (0.18)    & 3523.01 (0.11) & 3526.88 (0.30) & \nodata          & \nodata          & \nodata          \\
2005hk & 3685.09 (0.44) & 3689.81 (0.63)    & 3690.04 (1.93) & 3694.86 (0.27) & 3685.48 (0.74)   & 3691.61 (0.27)   & 3694.52 (0.47)   \\
2007qd & \nodata        & \nodata           & \nodata        & \nodata        & 4355.42--4405.89 & 4355.42--4405.89 & 4355.42--4405.89 \\
2008A  & 4478.23 (0.23) & 4483.61 (0.31)    & 4485.03 (0.36) & 4488.28 (0.54) & \nodata          & 4483.15 (1.69)   & 4485.67 (1.67)   \\
2008ae & 4508.40 (1.71) & 4513.52 (0.11)    & \nodata        & \nodata        & \nodata          & 4515.73 (0.53)   & 4518.30 (0.71)   \\
2008ge & \nodata        & 4725.77 (1.69)    & \nodata        & \nodata        & \nodata          & \nodata          & \nodata          \\
2008ha & 4783.23 (0.16) & 4785.24 (0.30)    & 4787.30 (0.18) & 4787.95 (0.28) & \nodata          & \nodata          & \nodata          \\
20009J & $<$4848.72     & $<$4848.71        & 4850.03 (1.45) & \nodata        & $<$4848.69       & $<$4848.70       & 4852.39 (2.43)   \\
2009ku & \nodata        & \nodata           & \nodata        & \nodata        & 5097.87 (0.59)   & 5098.33 (0.88)   & 5099.54 (1.70)   \\
2011ay & 5647.64 (1.54) & 5651.81 (1.70   ) & \nodata        & \nodata        & \nodata          & 5653.77 (1.53)   & 5653.12 (4.96)   \\
2012Z  & $<$5969.61     & $\lesssim$5969.60 & \nodata        & \nodata        & \nodata          & 5974.91 (0.78)   & 5977.27 (1.89)   \\

\tableline
 & \multicolumn{7}{c}{Peak Magnitude} \\
\tableline

2002cx & 17.78 (0.23) & 17.72 (0.09)    & 17.58 (0.02) & 17.42 (0.19) & \nodata      & \nodata      & \nodata      \\
2003gq & 18.19 (0.01) & 17.88 (0.01)    & 17.48 (0.02) & 17.25 (0.01) & \nodata      & \nodata      & \nodata      \\
2005cc & $<$16.70     & 16.27 (0.01)    & 15.92 (0.01) & 15.68 (0.01) & \nodata      & \nodata      & \nodata      \\
2005hk & 15.92 (0.01) & 15.73 (0.01)    & 15.51 (0.01) & 15.37 (0.01) & 15.80 (0.03) & 15.67 (0.01) & 15.79 (0.01) \\
2007qd & \nodata      & \nodata         & \nodata      & \nodata      & $<$21.90     & $<$21.72     & $<$21.35     \\
2008A  & 16.39 (0.01) & 16.11 (0.01)    & 15.82 (0.01) & 15.67 (0.01) & \nodata      & 16.01 (0.02) & 16.06 (0.02) \\
2008ae & 18.62 (0.03) & 18.17 (0.02)    & \nodata      & \nodata      & \nodata      & 17.93 (0.05) & 17.85 (0.02) \\
2008ge & \nodata      & 13.77 (0.12)    & \nodata      & \nodata      & \nodata      & \nodata      & \nodata      \\
2008ha & 18.23 (0.01) & 17.68 (0.01)    & 17.54 (0.01) & 17.36 (0.01) & \nodata      & \nodata      & \nodata      \\
2009J  & $<$18.97     & $<$18.72        & 18.90 (0.16) & \nodata      & $<$18.71     & $<$18.68     & 18.79 (0.09) \\
2009ku & \nodata      & \nodata         & \nodata      & \nodata      & 19.60 (0.01) & 19.26 (0.01) & 19.24 (0.01) \\
2011ay & 17.04 (0.01) & 16.62 (0.04)    & \nodata      & \nodata      & \nodata      & 16.54 (0.11) & 16.83 (0.11) \\
2012Z  & $<$14.73     & $\lesssim$14.48 & \nodata      & \nodata      & \nodata      & 14.27 (0.03) & 14.27 (0.11) \\

\tableline
 & \multicolumn{7}{c}{Peak Absolute Magnitude} \\
\tableline

2002cx & $-17.48$ (0.28) & $-17.52$ (0.18)   & $-17.64$ (0.15) & $-17.77$ (0.15) & \nodata         & \nodata         & \nodata         \\
2003gq & $-16.76$ (0.15) & $-17.01$ (0.15)   & $-17.37$ (0.15) & $-17.56$ (0.15) & \nodata         & \nodata         & \nodata         \\
2005cc & $<$ $-16.05$    & $-16.79$ (0.15)   & $-17.13$ (0.15) & $-17.38$ (0.15) & \nodata         & \nodata         & \nodata         \\
2005hk & $-17.69$ (0.15) & $-17.86$ (0.15)   & $-18.07$ (0.15) & $-18.19$ (0.15) & $-18.12$ (0.36) & $-17.91$ (0.15) & $-17.78$ (0.15) \\
2007qd & \nodata         & \nodata           & \nodata         & \nodata         & $<$ $-14.50$    & $<$ $-14.61$    & $<$ $-14.99$    \\
2008A  & $-17.92$ (0.15) & $-18.16$ (0.15)   & $-18.41$ (0.15) & $-18.53$ (0.15) & \nodata         & $-18.23$ (0.15) & $-18.15$ (0.15) \\
2008ae & $-17.10$ (0.15) & $-17.53$ (0.15)   & \nodata         & \nodata         & \nodata         & $-17.76$ (0.16) & $-17.82$ (0.15) \\
2008ge & \nodata         & $-17.60$ (0.25)   & \nodata         & \nodata         & \nodata         & \nodata         & \nodata         \\
2008ha & $-13.70$ (0.15) & $-14.18$ (0.15)   & $-14.28$ (0.15) & $-14.41$ (0.15) & \nodata         & \nodata         & \nodata         \\
2009J  & $<$ $-15.53$    & $<$ $-15.70$      & $-15.47$ (0.22) & \nodata         & $<$ $-15.71$    & $<$ $-15.70$    & $-15.54$ (0.17) \\
2009ku & \nodata         & \nodata           & \nodata         & \nodata         & $-18.36$ (0.15) & $-18.70$ (0.15) & $-18.71$ (0.15) \\
2011ay & $-18.05$ (0.15) & $-18.40$ (0.16)   & \nodata         & \nodata         & \nodata         & $-18.43$ (0.19) & $-18.10$ (0.19) \\
2012Z  & $<$ $-17.96$    & $\lesssim -18.18$ & \nodata         & \nodata         & \nodata         & $-18.37$ (0.09) & $-18.25$ (0.14) \\

\tableline
 & \multicolumn{7}{c}{$\Delta m_{15}$ (mag)} \\
\tableline

2002cx & 1.23 (0.06) & 0.84 (0.09) & 0.54 (0.06) & 0.38 (0.06) & \nodata     & \nodata     & \nodata     \\
2003gq & 1.83 (0.02) & 0.98 (0.03) & 0.71 (0.10) & 0.52 (0.04) & \nodata     & \nodata     & \nodata     \\
2005cc & \nodata     & 0.97 (0.01) & 0.65 (0.01) & 0.61 (0.06) & \nodata     & \nodata     & \nodata     \\
2005hk & 1.54 (0.05) & 0.92 (0.01) & 0.52 (0.13) & 0.52 (0.01) & 1.22 (0.15) & 0.67 (0.01) & 0.56 (0.03) \\
2008A  & 1.26 (0.07) & 0.82 (0.06) & 0.51 (0.01) & 0.38 (0.03) & \nodata     & 0.56 (0.05) & 0.47 (0.03) \\
2008ae & 1.35 (0.23) & 0.94 (0.02) & \nodata     & \nodata     & \nodata     & 0.71 (0.13) & 0.50 (0.03) \\
2008ge & \nodata     & 0.34 (0.24) & \nodata     & \nodata     & \nodata     & \nodata     & \nodata     \\
2008ha & 2.17 (0.02) & 1.22 (0.03) & 0.97 (0.02) & 0.65 (0.02) & \nodata     & \nodata     & \nodata     \\
2009J  & \nodata     & \nodata     & 0.79 (0.05) & \nodata     & \nodata     & \nodata     & 0.69 (0.10) \\
2009ku & \nodata     & \nodata     & \nodata     & \nodata     & 0.43 (0.08) & 0.25 (0.03) & 0.18 (0.09) \\
2011ay & 1.34 (0.42) & 0.75 (0.12) & \nodata     & \nodata     & \nodata     & 0.44 (0.14) & 0.26 (0.16) \\
2012Z  & \nodata     & \nodata     & \nodata     & \nodata     & \nodata     & 0.59 (0.01) & 0.48 (0.11)

\enddata

\end{deluxetable*}

\subsection{Photometric Relations}\label{ss:phot_rel}

\begin{figure}
\begin{center}
\epsscale{1.3}
\rotatebox{90}{
\plotone{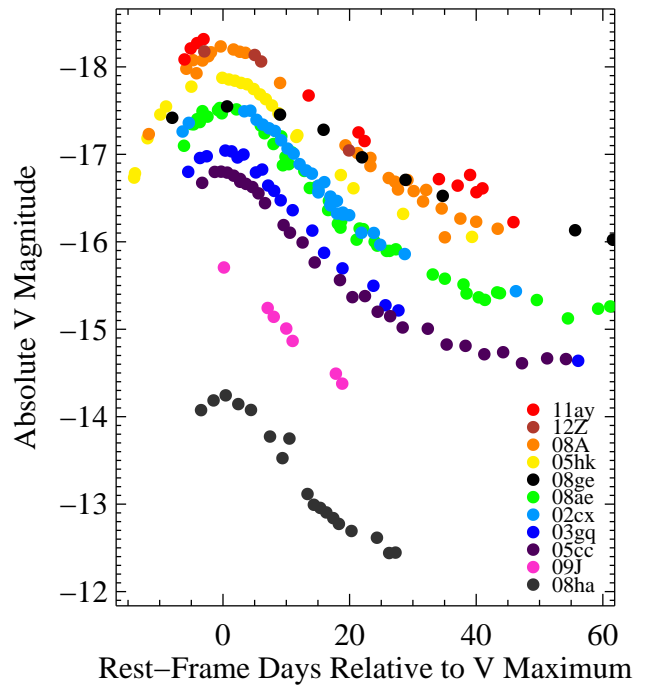}}
\caption{Absolute $V$-band light curves for a subset of SNe~Iax.  Each
  SN is plotted with a different color.}\label{f:abslc}
\end{center}
\end{figure}

Using the measured times of maximum, we can directly compare the
absolute-magnitude light curves of several SNe in the class.  We
present those light curves in Figure~\ref{f:abslc} and note that they
have not been corrected for host-galaxy extinction (see
Section~\ref{ss:cc}).  The figure shows the large range of peak
absolute magnitudes for the class.  It also displays the differing
photometric behavior.  Some SNe with very similar absolute magnitudes
can have very different decline rates (e.g., SNe~2002cx and
2008ge.\footnote{The SN~2008ge light curve presented by
\citet{Foley10:08ge} was observed unfiltered and converted to $V$,
and therefore some of the discrepancy may be related to the slightly
different bandpasses.})  Despite these outliers, there is a general
trend that the more luminous SNe at peak tend to have broader light
curves.  There is also some diversity in the rise times (although most
SNe have scarce pre-maximum data).  Contrary to the trend seen with
SNe~Ia, SN~2008A is brighter than SN~2005hk but has a slightly shorter
rise time.  Nonetheless, the faintest SNe~Iax, such as SN~2008ha and
SN~2004cs, have shorter rise times and faster decline rates than the
brightest SNe, such as SNe~2005hk and 2008A.

SNe~Iax span a range of decline rates that is similar to that of
normal SNe~Ia, although the SN~Iax range is somewhat larger.  The
rise-time range for SNe~Iax (from SN~2008ha at \about 10~days to
SN~2008ge, which might be $>$20~days) is also larger than for that of
SNe~Ia \citep{Ganeshalingam11}; based on current data, it appears that
the average SN~Iax has a shorter rise time than the average SN~Ia, but
few SNe have light curves sufficient for this measurement.  Despite
their rough similarity in light-curve shape, SNe~Iax have consistently
lower luminosity (even if that criterion is relaxed from our
classification scheme) than SNe~Ia.

For SNe~Iax, there are several clear trends in the derived photometric
parameters.  Peak brightness and decline rates are highly correlated
for a given object in all bands.  In other words, a SN that is bright
and declines slowly in $B$ is also bright and declines slowly in $R$.

Performing a Bayesian Monte-Carlo linear regression on the data
\citep{Kelly07}, we determine correlations between different
parameters in different bands.  The linear relationships and their
correlation coefficients are presented in Table~\ref{t:lc_fits}, where
the equations are all of the form
\begin{equation}
  p_{2} = \alpha p_{1} + \beta,
\end{equation}
where $p_{1}$ and $p_{2}$ are the two parameters, $\alpha$ is the
slope, and $\beta$ is the offset.

\begin{deluxetable*}{ccccccc}
\tabletypesize{\scriptsize}
\tablewidth{0pt}
\tablecaption{Light-Curve Relations for SNe~Iax\label{t:lc_fits}}
\tablehead{
\colhead{1st} &
\colhead{2nd} &
\colhead{Slope} &
\colhead{Offset} &
\colhead{Equation} &
\colhead{No.\ } &
\colhead{Correlation} \\
\colhead{Parameter} &
\colhead{Parameter} &
\colhead{($\alpha$)} &
\colhead{($\beta$)} &
\colhead{No.\ } &
\colhead{SNe} &
\colhead{Coefficient}}

$t_{\rm max} (B)$   & $t_{\rm max} (V)$   & 0.98 (0.03) & \phs5.17 (1.83) & \refstepcounter{equation}\label{e:tbtv}\ref{e:tbtv} & 7 & 0.9994 \\
$t_{\rm max} (R)$   & $t_{\rm max} (V)$   & 1.00 (0.02) &  $-1.47$ (1.52) & \refstepcounter{equation}\label{e:tbtr}\ref{e:tbtr} & 6 & 0.9997 \\
$t_{\rm max} (I)$   & $t_{\rm max} (V)$   & 1.00 (0.04) &  $-4.41$ (2.79) & \refstepcounter{equation}\label{e:tbti}\ref{e:tbti} & 6 & 0.9996 \\
$t_{\rm max} (r)$   & $t_{\rm max} (V)$   & \nodata     & \nodata         & \nodata                                             & 4 & 0.958 \\
$t_{\rm max} (i)$   & $t_{\rm max} (V)$   & \nodata     & \nodata         & \nodata                                             & 4 & 0.947 \\
\tableline
$\Delta m_{15} (B)$ & $\Delta m_{15} (V)$ & 0.40 (0.12) & \phs0.32 (0.20) & \refstepcounter{equation}\label{e:dbdv}\ref{e:dbdv} & 7 & 0.959 \\
$\Delta m_{15} (R)$ & $\Delta m_{15} (V)$ & 0.77 (0.26) & \phs0.46 (0.18) & \refstepcounter{equation}\label{e:drdv}\ref{e:drdv} & 6 & 0.966 \\
0$\Delta m_{15} (I)$ & $\Delta m_{15} (V)$ & 1.35 (0.66) & \phs0.27 (0.34) & \refstepcounter{equation}\label{e:didv}\ref{e:didv} & 6 & 0.929 \\
$\Delta m_{15} (r)$ & $\Delta m_{15} (V)$ & \nodata     & \nodata         & \nodata                                             & 4 & $-0.277$ \\
$\Delta m_{15} (i)$ & $\Delta m_{15} (V)$ & \nodata     & \nodata         & \nodata                                             & 4 & 0.471 \\
\tableline
$M_{B}$             & $M_{V}$             & 0.95 (0.09) & $-1.22$ (1.45)  & \refstepcounter{equation}\label{e:mbmv}\ref{e:mbmv} & 7 & 0.996 \\
$M_{R}$             & $M_{V}$             & 0.96 (0.10) & $-0.37$ (1.74)  & \refstepcounter{equation}\label{e:mrmv}\ref{e:mrmv} & 6 & 0.997 \\
$M_{I}$             & $M_{V}$             & 0.97 (0.12) & $-0.09$ (2.13)  & \refstepcounter{equation}\label{e:mimv}\ref{e:mimv} & 6 & 0.992 \\
$M_{r}$             & $M_{V}$             & \nodata     & \nodata         & \nodata & 4 & $-0.310$ \\
$M_{i}$             & $M_{V}$             & \nodata     & \nodata         & \nodata & 4 & $-0.311$ \\
\tableline
$\Delta m_{15} (V)$ & $M_{V}$             & 10.7 (2.4)  & $-27.4$ (2.3)   & \refstepcounter{equation}\label{e:dvmv}\ref{e:dvmv} & 9 & 0.980 \\
$\Delta m_{15} (R)$ & $M_{V}$             &  8.3 (2.0)  & $-22.3$ (1.3)   & \refstepcounter{equation}\label{e:drmv}\ref{e:drmv} & 6 & 0.982
\enddata

\end{deluxetable*}

Using the Equations in Table~\ref{t:lc_fits}, one can effectively
transform observations in one band into measurements in another.
However, we note that although the correlations are generally quite
strong, the uncertainties for the linear relations are relatively
large for several relations.

There are only four SNe with $V$ and $r$/$i$ light curves.  The small
number of measurements prevents robust determinations of relationships
between the parameters in $V$ and $r$/$i$.  However, the measurements
in $r$/$i$ are consistent with measurements in $R$/$I$ (modulo offsets
related to the filter response functions).  As such, we are able to
use the rough relations between $r$/$i$ and $V$ to estimate
light-curve parameters in $V$ for SN~2009ku (which was only observed
in PS1 bands).  Doing this, we find $M_{V} = -18.94 \pm 0.54$~mag, and
$\Delta m_{15} (V) = 0.58 \pm 0.17$~mag.  Both values are consistent
with those found by \citet{Narayan11}, who used only SN~2005hk to
provide scalings.  Using the same relations for SN~2012Z, we find
$M_{V} = -18.56 \pm 0.40$~mag, which is consistent with the direct
measurement in the $V$ band of $M_{V} \lesssim -18.18$~mag, and
$\Delta m_{15} (V) = 0.92 \pm 0.26$~mag.  We also convert the
parameters measured for the unfiltered light curve of SN~2004cs
assuming that the unfiltered light curve is equivalent to $R$.  This
conversion provides an estimate of $M_{V} = -16.2 \pm 0.3$~mag, and
$\Delta m_{15} (V) = 1.4 \pm 0.2$~mag for SN~2004cs.

There is a clear progression in time of maximum brightness from blue
to red, with the \vri\ bands peaking 5.1, 6.6, and 11.1~days after
$B$, respectively.  This progression is similar to that seen for
normal SNe~I and is indicative of the SN ejecta cooling with time.
There is no robust correlation between the time between maxima and
light-curve shape.

\begin{figure}
\begin{center}
\epsscale{1.1}
\rotatebox{90}{
\plotone{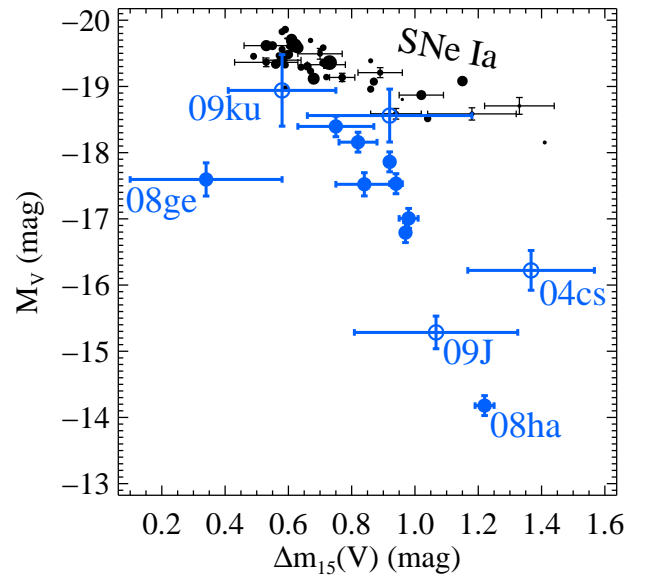}}
\caption{WLR ($\Delta m_{15} (V)$ vs.\ $M_{V}$) for SNe~Ia (black
  points) and SNe~Iax (blue points).  Filled points are direct
  measurements in $V$, while the open points (SNe~2004cs, 2009J, and
  2009ku) are an estimate by converting parameters measured from
  unfiltered (assumed to be $R$) or $r$ and $i$ to $V$.  The sizes of
  the SN~Ia points are inversely proportional to their uncertainty,
  with some points having representative errors bars to show the
  scaling.}\label{f:wlr}
\end{center}
\end{figure}

In Figure~\ref{f:wlr}, we show the relation between $\Delta m_{15}
(V)$ and $M_{V}$ (uncorrected for host-galaxy extinction) for the nine
SNe with measurements of these parameters and three SNe with estimates
of these parameters.  There is a general WLR, which is confirmed by a
linear relationship between the parameters (Equation~\ref{e:dvmv},
which did not include the estimates for SNe~2004cs, 2009J, or 2009ku).
The tight WLR for SNe~Ia is interpreted as the result of a homogeneous
class of objects with a single parameter (related to the $^{56}$Ni
mass) controlling multiple observables \citep[e.g.,][]{Mazzali07}.
Similar physics may be behind the SN~Iax WLR, but the large scatter
indicates that the class is not as homogeneous as SNe~Ia, with other
parameters (perhaps ejecta mass; see Section~\ref{ss:maxspec}) also
being important.

We note that there is a strong correlation between $\Delta m_{15} (R)$
and $M_{V}$ (uncorrected for host-galaxy extinction) for our small
sample.  \citet{Narayan11} found no strong correlation between these
two quantities using a slightly different sample, but this appears to
be because they included SN~2007qd in the relation.  Although
\citet{McClelland10} reported peak magnitudes and decline rates for
SN~2007qd (which were reproduced by \citealt{Narayan11}), their light
curves do not include any pre-maximum detections; consequently, their
derived quantities may be misestimated.  Using the
\citet{McClelland10} light curves, we present limits for SN~2007qd in
Table~\ref{t:lc}.

The basic parameters of time of maximum, peak absolute magnitude, and
light-curve shape for the SNe with $V$-band light curves are reported
in Table~\ref{t:prop}.  We also give estimates for SNe where we have
light curves in bands other than $V$.  Finally, we use the information
in the discovery reports (combined with the relations presented above)
to place approximate limits on $M_{V}$ at peak and $t_{\rm max} (V)$.

\subsection{Color Curves and Host-galaxy Reddening}\label{ss:cc}

\begin{figure}
\begin{center}
\epsscale{1.5}
\rotatebox{90}{
\plotone{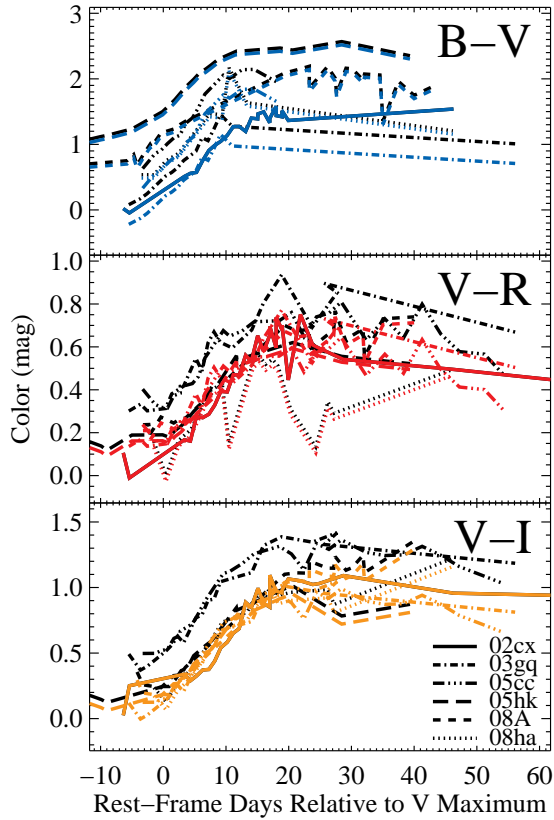}}
\caption{Color curves ($B-V$, $V-R$, and $V-I$ from top to bottom) of
  SNe~Iax.  Black lines correspond to the observed (corrected for
  Milky Way reddening) colors, while the colored curves correspond to
  an additional reddening correction corresponding to $E(B-V) = 0.05$,
  0.4, 0.4, 0, 0.2, and 0.1~mag for SNe~2002cx, 2003gq, 2005cc,
  2005hk, 2008A, and 2008ha, respectively.}\label{f:cc}
\end{center}
\end{figure}

In Figure~\ref{f:cc}, we present color curves for a subset of the
class.  The SNe all display the same general color evolution with
time.  SNe~Iax typically get redder from maximum brightness until
15--20~days after maximum, at which point they stay relatively
constant in color, but become slightly bluer with time.

Similar to what has been done for SNe~Ia \citep{Lira98}, we can create
a tight color locus in $V-I$ and a relatively tight locus in $V-R$
when we assume some host-galaxy extinction.  We find that reddening
corrections corresponding to $E(B-V) = 0.05$, 0.4, 0.4, 0, 0.2, and
0.1~mag for SNe~2002cx, 2003gq, 2005cc, 2005hk, 2008A, and 2008ha
(respectively) significantly reduce the scatter in the $V-R$ and $V-I$
colors at all times.  These corrections, however, do not improve the
$B-V$ scatter, which remains quite large.  It is therefore unclear if
this color shift is related to dust reddening, or if the scatter is
intrinsic to the SNe.  We therefore do not apply these corrections to
any of our estimates for the absolute magnitudes presented in
Table~\ref{t:lc}.  Furthermore, no dust reddening correction can
simultaneously make the $V-R$ and $V-I$ colors of SN~2008ha similar to
those of the other SNe; it is consistently bluer than all other SNe in
$V-R$ when matching the $V-I$ color.


\section{Spectroscopic Properties}\label{s:spec}

\subsection{Maximum-light Spectra}\label{ss:maxspec}

\begin{figure}
\begin{center}
\epsscale{1.4}
\rotatebox{90}{
\plotone{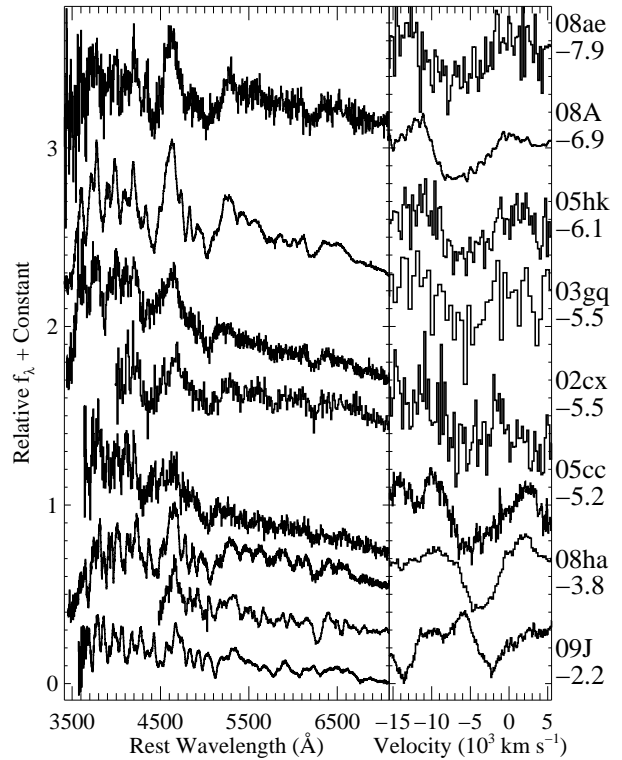}}
\caption{Near-maximum optical spectra of SNe~2008ae, 2008A, 2005hk,
  2003gq, 2002cx, 2005cc, 2008ha, and 2009J from top to bottom.  The
  respective phases relative to $V$ maximum are $-2.4$, $-2.7$,
  $-5.9$, $-2.4$, $-3.2$, $-0.1$, $3.2$, and 0.2~days.  The spectra
  have been ordered by their \ion{Si}{2} $\lambda 6355$ velocity.  The
  right-hand panel displays the \ion{Si}{2} $\lambda 6355$ feature on
  a velocity scale.  Narrow galactic features have been manually
  removed from some of the spectra.  Each spectrum is labeled with the
  SN name and its \ion{Si}{2} $\lambda 6355$ velocity (in
  $10^{-3}$~\kms).}\label{f:spec_cont}
\end{center}
\end{figure}

SNe~Iax have a large range of photospheric velocities.  For the
members with spectra near $t = -2$~days relative to $V$ maximum, we
show their spectra in Figure~\ref{f:spec_cont}.  The spectra are
ordered by \ion{Si}{2} $\lambda 6355$ velocity, and the \ion{Si}{2}
$\lambda 6355$ feature is shown in detail in the right panel of the
figure.  The spectra are at slightly different phases (from $t = -5.9$
to $-0.1$~days), so the comparison is appropriate for general patterns
rather than looking for specific differences.  Despite the different
ejecta velocities, the spectra look very similar --- the main
differences are that the lower-velocity SNe have otherwise blended
features resolved, with all spectra have features attributed to C/O
burning.  Furthermore, the extreme SN~2008ha does not appear to be
significantly different from the other members of the class, but
rather there seems to be a spectral sequence for the objects defined
primarily by ejecta velocity.

\begin{figure}
\begin{center}
\epsscale{0.82}
\rotatebox{90}{
\plotone{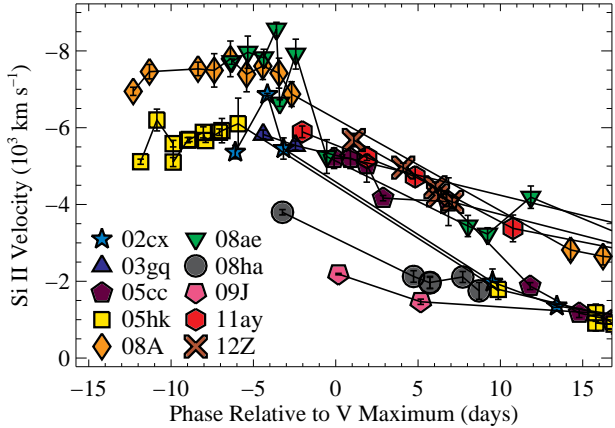}}
\caption{\ion{Si}{2} $\lambda 6355$ velocity as a function of time
  relative to $V$-band maximum brightness for SNe~Iax.}\label{f:vgrad}
\end{center}
\end{figure}

Using the method of \citet{Blondin06} to measure photospheric
velocities, we determine the temporal evolution of the \ion{Si}{2}
$\lambda 6355$ feature for several members of the class.
Figure~\ref{f:vgrad} shows the results.  Generally, the \ion{Si}{2}
velocity is relatively flat until a few days before $V$ maximum.  At
that time, the velocity decreases quickly, and around 10--15~days
after $V$ maximum the velocity starts to decline more gradually.  This
behavior is slightly different from that of normal SNe~Ia, where
high-velocity features at early times can cause steep velocity
declines at $t \lesssim -5$~days \citep[e.g.,][]{Foley12:09ig}, but is
similar to the velocity evolution of some potential
``Super-Chandrasekhar'' SNe~Ia \citep{Scalzo12}, for which a thin,
dense shell of material in the ejecta has been invoked to explain the
low velocity gradients.  For some SNe at times starting around
10--15~days after maximum brightness, the \ion{Si}{2} feature begins
to blend with other features, and the velocity is not a particularly
reliable measurement.

Fitting the velocity data for $-6 < t < 15$~days, we measured the
velocity gradients near maximum brightness.  Unlike normal SNe~Ia
\citep{Foley11:vgrad}, there is no evidence for a correlation between
the velocity gradient near maximum brightness and the velocity at
maximum brightness.  The individual velocity gradients are all similar
to the weighted mean of 250~km~s$^{-1}$~day$^{-1}$, which is at the
high end of velocity gradients for normal SNe~Ia
\citep{Foley11:vgrad}.

Using the data near maximum brightness, we simultaneously fit for the
velocity at $V$ maximum when fitting for the velocity gradient.  These
range from $-2210$ (for SN~2009J) to $-6350$~\kms\ (for SN~2008A).
Near maximum, there is a larger spread in velocities than at later
times.  And although SNe with relatively high velocities at early
times (like SNe~2008A and 2008ae) also have relatively high velocities
at later times, some SNe with intermediate velocities at early times
(like SNe~2002cx and 2005cc) at later times have velocities similar to
those of SN~2008ha.  Clearly the velocity/density structure of the
ejecta can be very different in these objects; nonetheless, we treat
the maximum-light velocity as an indication of the photospheric
velocity.

\citet{McClelland10} suggested that there is a progression from
extremely low-velocity, low-luminosity SNe like SN~2008ha through SNe
more similar to SN~2002cx to normal SNe~Ia with relatively high
velocity and high luminosity.  To do this, they measured the
photospheric velocity at 10~days after $B$ maximum for SNe~2002cx,
2005hk, 2007qd, and 2008ha.  (We note that we cannot reliably measure
the time of maximum for SN~2007qd.)  The velocity measurement was made
by cross correlating the spectra.  Although this method should produce
reasonable results, we prefer our direct measurements to provide a
reproducible result.

\citet{Narayan11} countered the claims of \citet{McClelland10} by
presenting observations of SN~2009ku, which had low velocity and a
high luminosity.  Although it is clear that SN~2009ku was a
low-velocity SN relative to SN~2002cx, its first spectrum was obtained
18~days after $B$ maximum, and we are unable to measure its
\ion{Si}{2} velocity at maximum brightness.  But based on spectral
similarities, SN~2009ku likely had a velocity similar to that of
SN~2008ha.  For SN~2009ku, we assign a velocity intermediate between
the first and third lowest-velocity SNe in our sample, SNe~2009J and
2005cc.

\begin{figure}
\begin{center}
\epsscale{0.9}
\rotatebox{90}{
\plotone{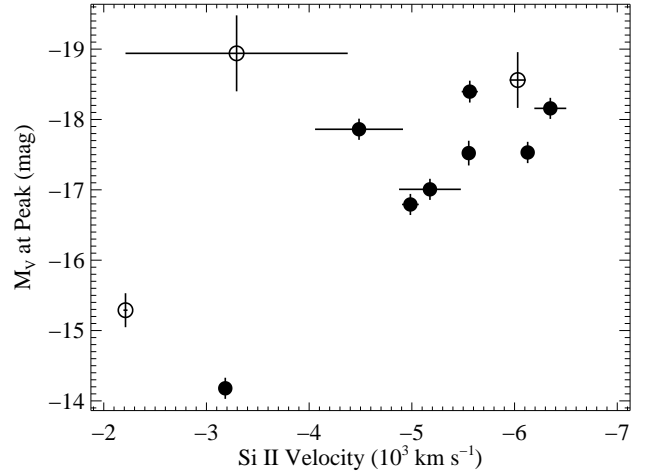}}
\caption{\ion{Si}{2} $\lambda 6355$ velocity vs.\ peak absolute $V$
  magnitude for SNe~Iax.  SNe~2009J, 2009ku, and 2012Z, for which
  $M_{V}$ is estimated using the relations derived in
  Section~\ref{ss:phot_rel}, are represented by the empty circles,
  indicating their approximate values.}\label{f:vel_mag}
\end{center}
\end{figure}

Figure~\ref{f:vel_mag} compares peak absolute magnitude and
maximum-light velocity for the SNe~Iax.  Ignoring SN~2009ku, there is
a slight trend where higher-velocity SNe are also more luminous.  This
trend may be stronger with additional data for SNe~2010ae and 2010el,
which both have low velocities and low peak magnitudes (Stritzinger
et~al., in preparation; Valenti et~al., in preparation).

\begin{figure}
\begin{center}
\epsscale{0.9}
\rotatebox{90}{
\plotone{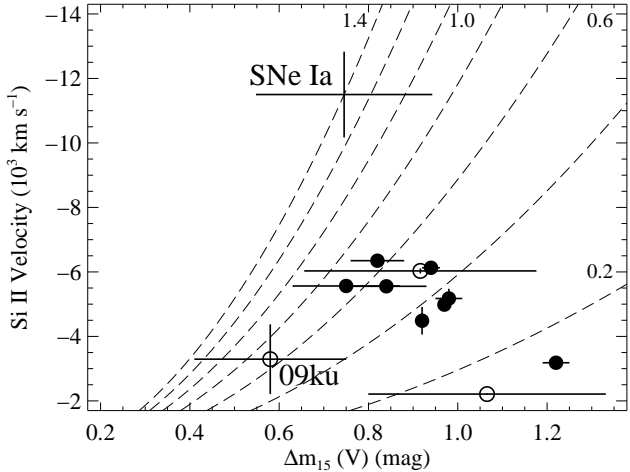}}
\caption{Light-curve shape vs.\ \ion{Si}{2} $\lambda 6355$ velocity
  for SNe~Iax.  SNe~2009J, 2009ku, and 2012Z, for which $\Delta m_{15}
  (V)$ is estimated using the relations derived in
  Section~\ref{ss:phot_rel}, are represented by the empty circles,
  indicating their approximate values.  The average and standard
  deviation for normal SNe~Ia are plotted as a single cross and
  labeled.  The dashed lines represent lines of equal ejecta mass,
  ranging from 1.4~$M_{\sun}$ in the upper left to 0.2~$M_{\sun}$ in
  the lower right in steps of 0.2~$M_{\sun}$.  The majority of SNe~Iax
  cluster at \about 0.5~$M_{\sun}$.}\label{f:vel_dm15}
\end{center}
\end{figure}

Similar to \citet{Narayan11}, we also compare the maximum-light
velocity to light-curve shape for SNe~Iax in Figure~\ref{f:vel_dm15}.
Again, ignoring SN~2009ku, there is a possible correlation, where
higher-velocity SNe also have slower declining light curves.  However,
besides SNe~2008ha, 2009J, and 2009ku, SNe~Iax occupy a relatively
small portion of the parameter space.

In Figure~\ref{f:vel_dm15}, we also plot the average and standard
deviation measurements for \ion{Si}{2} velocity \citep{Foley11:vgrad}
and $\Delta m_{15} (V)$ \citep{Hicken09:lc} for SNe~Ia.  As expected,
normal SNe~Ia have higher velocities than SNe~Iax, but also typically
have slightly slower decline rates.  As was done by \citet{Narayan11},
we estimate the ejecta mass using analytical expressions of
\citet{Arnett82}, the decline rate in $V$ as a proxy for the
bolometric rise time, and the ejecta velocity, and we scale the
relations such that a normal SN~Ia has an ejecta mass of
1.4~$M_{\sun}$.  From this relation, we see that the majority of
SNe~Iax have ejecta masses of \about $0.5 \pm 0.2 M_{\sun}$, and
SN~2008ha has an ejecta mass slightly below 0.2~$M_{\sun}$ (very
similar to the 0.15~$M_{\sun}$ of ejecta estimated by
\citealt{Foley09:08ha}).

From these simple relations and scalings, it appears that the majority
of SNe~Iax have significantly less than a Chandrasekhar mass of
ejecta.  The luminosity gap between the majority of SNe~Iax in our
sample and SNe~2008ha/2009J may also be reflected in the apparent
ejecta mass gap.  Perhaps more low and intermediate-luminosity SNe~Iax
will produce a continuum of ejecta masses.

\subsection{Presence and Frequency of Helium}

\begin{figure}
\begin{center}
\epsscale{1.3}
\rotatebox{90}{
\plotone{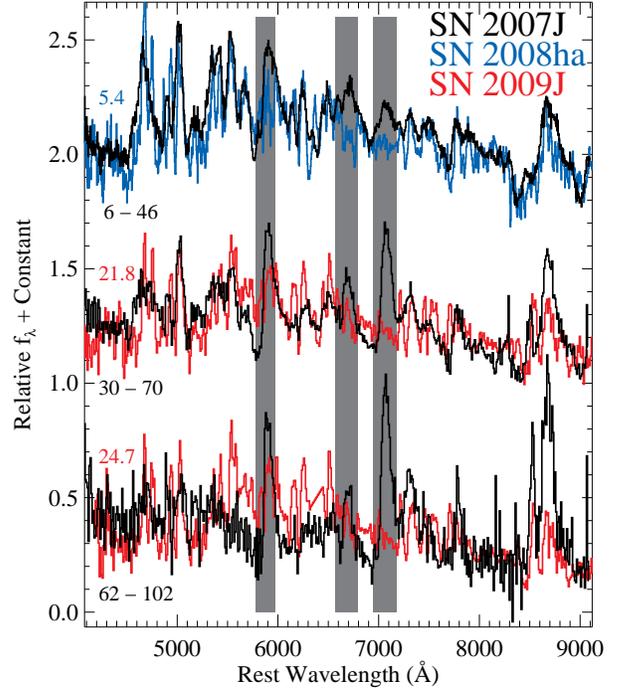}}
\caption{Spectral sequence of SN~2007J (black curves) compared to
  spectra of SNe~2008ha (blue) and 2009J (red).  The phases (and phase
  ranges) are noted next to each spectrum.  Wavelength regions
  corresponding to strong transitions of \ion{He}{1} and
  $\pm5000$~\kms\ around those wavelengths are shaded
  gray.}\label{f:spec_07j}
\end{center}
\end{figure}

As originally noted by \citet{Foley09:08ha}, SN~2007J is a SN~Iax with
strong \ion{He}{1} features.  These features developed over time, but
are present even in our first spectrum of SN~2007J.  Our spectral
series of SN~2007J is displayed in Figure~\ref{f:spec_07j}.  The
\ion{He}{1} lines are relatively weak in the first SN~2007J spectrum,
and it is not easy to distinguish this emission from normal SN~Iax
spectra \citep{Filippenko07:07J1}.  However, the \ion{He}{1} lines
clearly exist when comparing the spectrum to that of SN~2008ha.  In
our next spectrum, corresponding to a phase of 30--70~days after
maximum, the \ion{He}{1} features become quite strong.  In our final
spectrum, at a phase of 62--102~days, the strongest \ion{He}{1} line
is the most prominent feature in the optical range.  The spectrum of
SN~2004cs presented in Figure~\ref{f:04cs_spec} is at a phase of
42~days after $V$ maximum brightness, and also shows strong
\ion{He}{1} lines.

There is a phase where SN~2005E was spectroscopically similar to
SN~2007J.  \citet{Perets10:05e} presented three spectra of SN~2005E at
phases of 9, 29, and 62~days relative to maximum brightness.  The
first and last spectra are clearly distinct from SN~2007J.  Although a
careful examination of the 29~day spectrum shows that it is different
from those of SNe~Iax, it could potentially be confused with a SN~Iax,
or alternatively, perhaps a SN~Iax with helium can be confused with a
SN~2005E-like object.  The first and last SN~2007J spectra are
separated by 60~days, so we expect that one of the SN~2007J spectra
would resemble SN~2005E at another epoch if it were a SN~2005E-like
object.  Our SN~2004cs spectrum is at a phase of 42~days.  This is
intermediate between the second and third SN~2005E spectra.  It is
therefore possible that SN~2004cs is actually a SN~2005E-like object,
but it is spectroscopically more similar to other SNe~Iax than to
SN~2005E.

We have identified two SNe~Iax with \ion{He}{1} lines in their
spectra: SNe~2004cs and 2007J.  No other SN~Iax exhibits clear He
emission.  At a phase of \about 45~days after maximum brightness, the
spectra of both SNe have strong \ion{He}{1} lines.  Several SNe~Iax
have spectra at phases similar to the SN~2007J spectra that show He
features.  We estimate that 9--14 SNe~Iax have spectra which could
rule out He features as strong as those of SNe~2004cs and 2007J.  This
provides a very rough estimate that \about 15\% of SNe~Iax show He
features during the phases that SNe~2004cs and 2007J do.

As SN~2007J demonstrates, the frequency of He features is highly
dependent on the phases of spectra obtained.  It is also possible that
many SNe~Iax with \ion{He}{1} lines are misclassified as SNe~IIb and
SNe~Ib like SNe~2004cs and 2007J were \citep{Rajala05,
  Filippenko07:07J2}.  If that is the case, then the fraction would
increase.  Additionally, the strengths of \ion{He}{1} features are
highly dependent on density/mixing/temperature/ionization effects in
addition to abundance.  For instance, an outstanding question is how
much helium exists in the ejecta of SNe~Ic
\citep[e.g.,][]{Hachinger12:he}.  Additionally, models of helium-shell
(on top of a C/O WD) explosions tend to be asymmetric
\citep[e.g.,][]{Townsley12}, and one might expect that the helium
distribution is asymmetric in a SN~Iax explosion, resulting in
viewing-angle-dependent spectral features.

Additional data, particularly spectral sequences covering a large
phase range, as well as detailed modeling, will be necessary to
determine the fraction of SNe~Iax with \ion{He}{1} features and
ultimately the mass and distribution of helium in the ejecta.

\subsection{Presence and Frequency of Carbon}

The amount and distribution of carbon in SN~Ia ejecta is a key
discriminant between various explosion models \citep{Gamezo03}.  The
presence of carbon also restricts certain progenitor systems.  For
instance, there should be little carbon in the ejecta of an
electron-capture SN of an O-Mg-Ne WD.  Depending on the temperature,
it is possible to see \ion{C}{1}, \ion{C}{2}, and \ion{C}{3} lines in
an optical SN spectrum \citep{Hatano99:ion}, although \ion{C}{1} has
mostly weak optical lines blueward of \about 8000~\AA.

Carbon lines have been detected in several normal SNe~Ia, and \about
30\% of high signal-to-noise ratio (S/N) SN~Ia spectra obtained before
maximum brightness show an indication of carbon absorption
\citep{Parrent11, Thomas11:carbon, Folatelli12, Silverman12:carbon}.

\begin{figure}
\begin{center}
\epsscale{1.35}
\rotatebox{90}{
\plotone{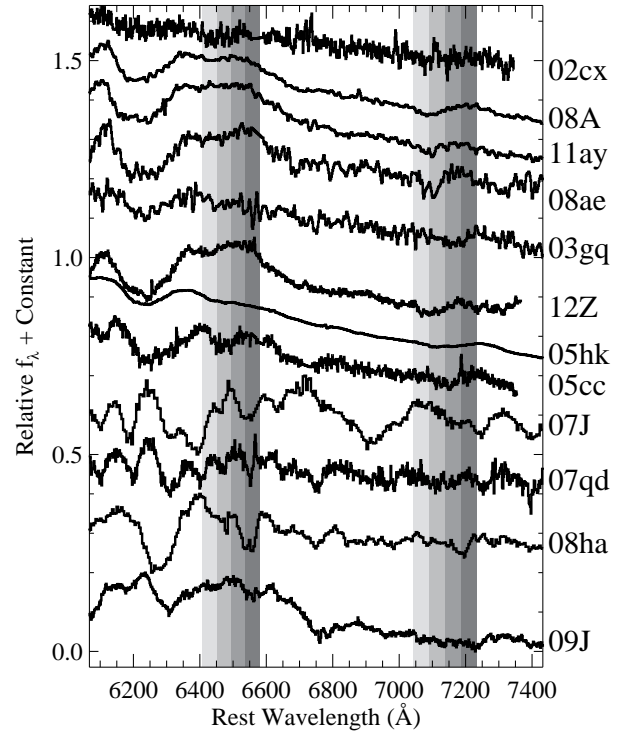}}
\caption{Optical spectra of SNe~Iax showing the region covering the
  \ion{C}{2} $\lambda 6580$ and $\lambda 7234$ lines.  These spectra
  were chosen to best show the \ion{C}{2} features.  Each spectrum is
  labeled and ordered by approximate \ion{Si}{2} velocity.  The grey
  regions correspond to velocities for the \ion{C}{2} lines spanning a
  width of 2000~\kms\ each, and starting at 0, $-2000$, $-4000$, and
  $-6000$~\kms, respectively.}\label{f:carbon}
\end{center}
\end{figure}

\ion{C}{2} lines have been reported in SN~2008ha \citep{Foley10:08ha},
and possibly identified in SNe~2002cx \citep{Parrent11}, 2005hk
\citep{Chornock06}, and 2007qd \citep{McClelland10}.  In
Figure~\ref{f:carbon}, we present these SNe along with several other
SNe~Iax that show some indication of \ion{C}{2}.  Specifically,
SNe~2005cc, 2005hk, 2007qd, 2008A, 2008ae, 2008ha, 2009J, 2011ay, and
2012Z have clear detections of \ion{C}{2} $\lambda 6580$ and $\lambda
7234$.  SNe~2002cx and 2003gq also have features suggestive of
\ion{C}{2}.  {\it Every} SN~Iax with a spectrum before or around
maximum light has some indication of carbon absorption.

The phase range for the spectra is from $-9.9$ to 1.1~days relative to
$V$ maximum.  Since some SNe~Iax have very hot photospheres at early
times, one might expect there to be relatively strong \ion{C}{3} in
those spectra.  There is some indication that the early spectra of
SNe~2005hk and 2012Z contain \ion{C}{3} $\lambda 4647$ and $\lambda
5696$.  Other spectra may also contain \ion{C}{3} features, but
spectral modeling may be required to clearly identify the features.

The high fraction (82--100\%) of SNe~Iax with carbon visible in their
pre-maximum spectra is extremely interesting.  There is no indication
that our spectra have a significantly higher S/N than those used in
previous studies to determine the fraction of SNe~Ia with carbon in
their spectra.  With their lower velocities reducing the blending of
\ion{C}{2} $\lambda 6580$ into the \ion{Si}{2} $\lambda 6355$ feature,
it should be somewhat easier to detect \ion{C}{2} in SNe~Iax; however,
only a small number (\about 10\%) of SNe~Ia have velocities high
enough to be affected by blending \citep{Silverman12:carbon}.

Clearly there is a significant amount of carbon present in the ejecta
of all SNe~Iax.  Using a modified model from \citet{Tanaka11},
\citet{Folatelli12} determined the mass of carbon in SNe~Ia by
measuring the pseudo-equivalent width (pEW) of \ion{C}{2} $\lambda
6580$.  The upper range of the measured pEWs is \about 4~\AA\ at about
5~days before $B$ maximum and \about 1~\AA\ near maximum.  Several
SNe~Iax have stronger \ion{C}{2} $\lambda 6580$; for instance,
SNe~2005cc and 2008ha have pEWs of 24 and 5~\AA\ at phases of $-3.2$
and $-0.1$~days relative to $V$ maximum (\about $-1.2$ and 5.6~days
relative to $B$ maximum), respectively.  SN~2008ae, which has a
relatively weak line, has a pEW of 3~\AA\ at $-5.3$ ($-0.2$)~days
relative to $V$ ($B$) maximum.  These values are all significantly
higher than the observations of SNe~Ia at similar phases.  Although we
do not produce additional models, it is clear that SNe~Iax require
significantly more carbon (by mass fraction) than do SNe~Ia.

\subsection{Late-time spectra}\label{ss:late}

SNe~Iax have very different spectral sequences than SNe~Ia
\citep[e.g.,][]{Jha06:02cx, Sahu08, Foley10:08ge}.  Specifically, we
have not yet seen a truly nebular spectrum (one which is dominated by
broad forbidden lines and lacks P-Cygni lines indicative of a
photosphere) for any member of the class, despite observing these SNe
$>$300~days after maximum brightness.  All other known classes of SNe
have nebular spectra at these phases.

\begin{figure*}
\begin{center}
\epsscale{0.5}
\rotatebox{90}{
\plotone{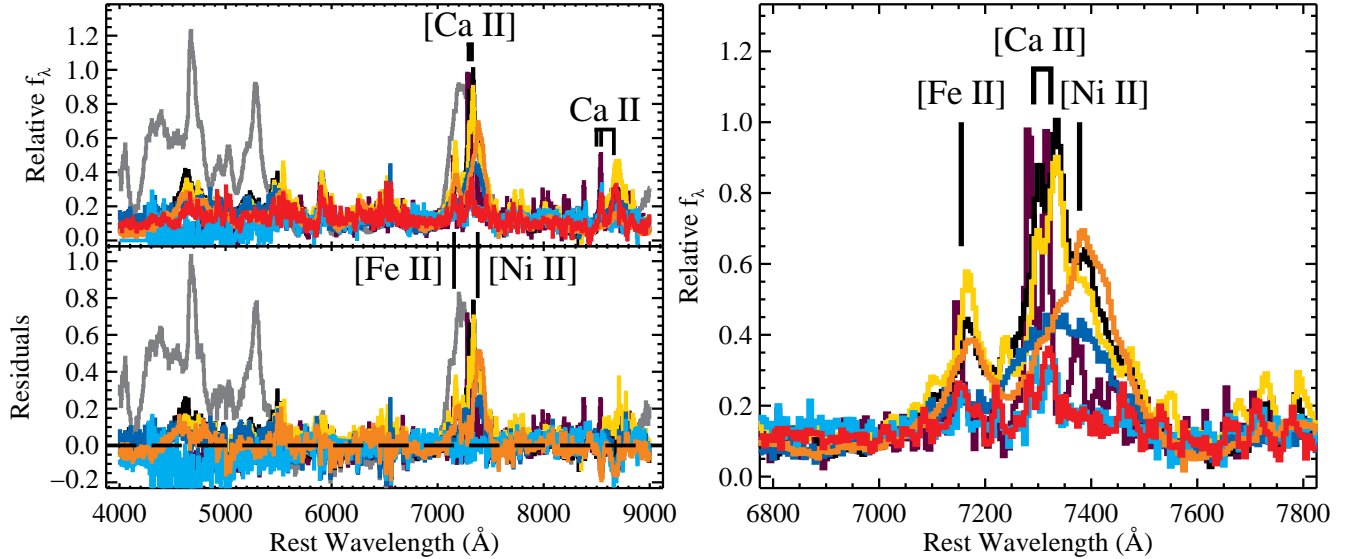}}
\caption{Late-time spectra of SNe~Iax.  The red, yellow, purple,
  black, gold, dark blue, and light-blue spectra correspond to
  SNe~2002cx, 2005P, 2005hk, 2008A, 2008ge, 2011ay, and 2011ce at
  phases of 227, an unknown phase, 224, 220, 225, 176, and 371~days,
  respectively.  The grey spectrum is that of the SN~1991bg-like
  SN~1999by.  The bottom-left panel displays the residual spectra
  relative to SN~2002cx.  SNe~2005P, 2008A, 2008ge, and 2011ay have
  significant [\ion{Ni}{2}] emission relative to SN~2002cx.  The right
  panel shows the region near 7300~\AA, and does not show the spectrum
  of SN~1999by.}\label{f:late_spec}
\end{center}
\end{figure*}

In Figure~\ref{f:late_spec}, we present late-time spectra of the five
SNe~Iax with a well-determined time of maximum and spectra at $t >
150$~days.  We also include SN~2011ce, which has a late-time spectrum
at a phase of about 383~days (see Section~\ref{ss:11ce}), and
SN~2005P, which does not have a well-defined time of maximum, but
whose spectrum is very similar to other late-time spectra of SNe~Iax
\citep{Jha06:02cx, Foley10:08ha}.  All spectra are similar, with the
differences mainly being the observed velocities and the strengths of
[\ion{Ca}{2}], [\ion{Fe}{2}], and [\ion{Ni}{2}] features.  None of
these spectra are similar to those of SNe~Ia, including low-luminosity
SNe like SN~1991bg, or to SNe~Ic.  It is worth noting that SNe~2005hk,
2008A, 2011ay, and 2011ce were originally classified as members of
this class from their first near-maximum spectra, and despite having
relatively high ejecta velocities and peak absolute magnitudes similar
to those of low-luminosity SNe~Ia, their late-time spectra bear no
resemblance to any SN~Ia.  \citet{Valenti09} suggested that a
late-time spectrum of SN~2005hk was similar to a late-time spectrum of
the low-luminosity SN~II~1997D (except lacking any hydrogen features).
However, this similarity is likely because these objects cool through
similar lines.

Despite generally looking similar, the SN~Iax late-time spectra are
still relatively distinct.  The right-hand panel of
Figure~\ref{f:late_spec} shows the spectra in the region of the
[\ion{Ca}{2}]-[\ion{Fe}{2}]-[\ion{Ni}{2}] complex (around 7300~\AA).
SNe~2002cx, 2005hk, and 2011ce have low velocities for all features
and very little emission in forbidden lines from Fe-group elements.
SNe~2005P and 2008A have very similar spectra, with both having strong
emission in forbidden lines from Fe-group elements and from
[\ion{Ca}{2}].  The spectrum of SN~2005P has slightly stronger
[\ion{Fe}{2}] compared to [\ion{Ni}{2}] than SN~2008A.  SN~2008ge has
barely detectable [\ion{Ca}{2}] emission and very strong [\ion{Ni}{2}]
emission.  SN~2011ay is somewhat similar to SN~2008ge (lacking strong
[\ion{Ca}{2}] emission), but its phase is a bit earlier than that of
the other SNe.  Interestingly, the spectrum of SN~2008A looks
extremely similar to that of SN~2008ge, except for additional
[\ion{Ca}{2}] emission.  Moreover, the spectrum of SN~2005hk is very
similar to those of SNe~2002cx and 2011ce, but it has exceptionally
strong emission lines.  Of the six SNe~Iax with particularly late-time
spectra (all SNe above except SN~2011ay), there are two equal-sized
groups in velocity, with SNe~2002cx, 2005hk, and 2011ce all having
similarly low velocities and SNe~2005P, 2008A, and 2008ge all having
similarly high velocities.

In all cases where we can measure both [\ion{Fe}{2}] and
[\ion{Ni}{2}], the [\ion{Ni}{2}] emission is stronger than the
[\ion{Fe}{2}] emission.  The opposite occurs in normal SNe~Ia.  By
this phase, nearly all $^{56}$Ni has radioactively decayed.  To have
such a large [\ion{Ni}{2}] to [\ion{Fe}{2}] ratio, SNe~Iax must
produce significantly more stable Ni relative to radioactive Ni than
SNe~Ia, or significantly less stable Fe relative to radioactive Ni
(and Co).  Low-energy explosions of low-density/low-temperature
material are predicted to produce such a trend
\citep[e.g.,][]{Travaglio04}.

The [\ion{Fe}{2}] $\lambda 7155$ line is the strongest [\ion{Fe}{2}]
feature in SN~Iax spectra at late times.  Normal SNe~Ia tend to have
stronger bluer Fe features such as [\ion{Fe}{2}] $\lambda 5262$ and
[\ion{Fe}{3}] $\lambda 4701$, that are not (or perhaps weakly)
detected in SNe~Iax.  The ejecta in SNe~Iax must be relatively cold to
produce these spectra (McCully et~al., in preparation).

For SNe~2002cx, 2008A, and 2008ge, we are able to measure the width of
the [\ion{Ca}{2}] emission lines.  The lines have full widths at
half-maximum intensity (FWHMs) of 600, 1000, and 500~\kms,
respectively.  For SNe~2005hk and 2011ce, the FWHM is at or below the
resolution of the spectra, 300~\kms.  The measurement of SN~2008ge is
particularly uncertain since the lines are weak.  For SNe~2002cx,
2008A, 2008ge, and 2011ce, we measure FWHMs for [\ion{Fe}{2}] $\lambda
7155$ of 1200, 1200, 2900, and 500~\kms, respectively, with the line
again being $\lesssim$300~\kms\ for SN~2005hk.  The [\ion{Fe}{2}]
lines are consistently broader than the [\ion{Ca}{2}] lines,
indicating that the Ca is {\it interior} to the Fe.  Furthermore, the
[\ion{Ca}{2}] and [\ion{Fe}{2}] FWHMs are not well correlated from
object to object.

In addition to a diverse set of line widths, the nebular lines have
diverse velocity shifts.  If the ejecta are completely transparent
(because we see P-Cygni features, the ejecta are not completely
transparent, but we assume that they are mostly transparent), the line
shifts should indicate any bulk offsets in the ejecta mapped by the
lines relative to the center of mass of the ejecta.  For SNe~2005hk,
2008A, and 2008ge, we are able to measure line shifts of [\ion{Fe}{2}]
$\lambda 7155$, [\ion{Ca}{2}] $\lambda\lambda 7291$, 7324, and
[\ion{Ni}{2}] $\lambda 7378$.  For SNe~2002cx and 2011ce, we could
measure line shifts of the first two features.  For SNe~2002cx,
2005hk, 2008A, 2008ge, and 2011ce we measure shifts of $+60$, $+250$,
$-560$, $-810$, and $+100$~\kms\ for [\ion{Fe}{2}], and $-220$,
$-380$, $+450$, $+410$, and $+20$~\kms\ for [\ion{Ca}{2}],
respectively.

In all but one case, the [\ion{Fe}{2}] and [\ion{Ca}{2}] features have
shifts in opposite directions, while in the three cases where
[\ion{Ni}{2}] was measured, the shift was in the same direction and
had approximately the same magnitude as [\ion{Fe}{2}].  The one
outlier, SN~2011ce, has velocity shifts very close to zero.
Interestingly, the two cases with blueshifted [\ion{Ca}{2}] are
SNe~2002cx and 2005hk, which have low-velocity late-time spectra,
while SNe~2008A and 2008ge have redshifted [\ion{Ca}{2}] and have
high-velocity late-time spectra.  SN~2011ce, which has a very small
shift, also has low-velocities.  The difference between the
[\ion{Fe}{2}] and [\ion{Ca}{2}] velocities also roughly correlates
with FWHM; SNe with narrower lines also have a smaller difference
between [\ion{Fe}{2}] and [\ion{Ca}{2}] velocities.

\citet{Maeda10:neb} suggested that [\ion{Ni}{2}] $\lambda 7178$ comes
primarily from the innermost ejecta of SNe~Ia, tracing out
electron-capture processes prevalent during the deflagration phase.
The [\ion{Fe}{2}] $\lambda 7155$ line is supposed to track the
high-density region directly outside the electron-capture zone and
also is a tracer of the deflagration phase.  Strong [\ion{Ca}{2}]
lines are not seen in nebular spectra of SNe~Ia, but Ca should not be
the result of supersonic burning.  If Fe/Ni track the deflagration
phase of C/O burning, perhaps Ca tracks lower-energy or lower-density
burning on the opposite side of the WD from the deflagration flame.

Although our current data are limited, it is clear that the velocity
structure of SNe~Iax is probably quite complex.  Detailed spectral
synthesis of various explosion models will be necessary to explain all
behavior seen in the late-time spectra.  Additionally, there is no
clear correlation between late-time spectral features (velocities,
line ratios, line strengths) and maximum-light properties (peak
luminosity, light-curve shape, photospheric velocity).  Models will
have to simultaneously describe both the diversity of observables and
their uncorrelated nature.


\section{Rates}\label{s:rates}

SNe~Iax are the most common ``peculiar'' class of SNe.  Thus far, they
have been mostly ignored in SN rate calculations and quantities
derived from those rates.  \citet{Li11:rate2} examined the prevalence
of SNe~Iax in the LOSS sample, finding that SNe~Iax have a relative
rate of $1.6\err{2.0}{1.2}$ and $5.7\err{5.5}{3.8}$ SNe~Iax for every
100 SNe~Ia in a magnitude-limited and volume-limited survey,
respectively.  However, this study was hampered by small numbers and
the SNe used in the KAIT sample were all more luminous than $M =
-16.7$~mag at peak.  The KAIT study did not account for the
low-luminosity SNe~Iax, and thus, underestimated the true rate of
SNe~Iax.  Meanwhile, \citet{Foley09:08ha} estimated that
SN~2008ha-like events could have a relative rate of \about 10 SNe~Iax
for every 100 SNe~Ia.  Simply combining the \citet{Li11:rate2}
estimate of the higher-luminosity SNe~Iax and the \citet{Foley09:08ha}
estimate of SN~2008ha-like events results in an estimate of \about 17
SNe~Iax for every 100 SNe~Ia.

Thus far, only measurements of the fraction of SNe~Ia being SNe~Iax
have been attempted.  With our dataset, we do not have the proper
information to derive a true SN rate.  However, we can estimate the
relative rate of SNe~Iax to normal SNe~Ia.  We caution that our
approach is limited by our heterogeneous dataset and assumptions built
into the analysis.  A significant uncertainty is the classification:
several SNe~Iax may exist without being properly classified.  Since
the discovery of SN~2002cx, 8 of the 21 recent members of the SN~Iax
class were originally misclassified, and one member was not classified
for more than a year after discovery.  Even within the BSNIP data
release, which provides spectra for 11 members and classifications
were performed recently, there was a single member (SN~2006hn) which
was simply classified as a ``SN~Ia'' \citep{Silverman12:bsnip}.
Furthermore, BSNIP did not include SN~2007J because of its He lines.
Nonetheless, \citet{Silverman12:bsnip} identified a previously
unclassified SN as a member (SN~2002bp).  Clearly, classification is
difficult and some true members (even within the last 10 years) are
not included in our current sample.  As a result, our relative rate
estimates are likely underestimates of the true relative rate.

With the above caveats, we wish to estimate the relative rate, or
ratio, of SNe~Iax to normal SNe~Ia in a given volume.  This is
expressed as
\begin{equation}
  r_{\rm Iax} = C_{\rm Iax} \frac{N_{\rm Iax} (\mu \le \mu_{\rm max})}{N_{\rm Ia} (\mu \le \mu_{\rm max})},
\end{equation}
where $\mu_{\rm max}$ is the maximum distance modulus for which one
can detect SNe~Iax in a survey, $N(\mu \le \mu_{\rm max})$ is the
number of SNe with a distance modulus less than $\mu_{\rm max}$, and
$C_{\rm Iax}$ is a correction factor to account for various
observational effects.  For a given survey, we have
\begin{equation}
  \mu_{\rm max} = m_{\rm lim} - M,
\end{equation}
where $m_{\rm lim}$ is the limiting magnitude of the survey and $M$ is
that absolute magnitude of a SN~Iax.  For this study, we will focus on
$R_{\rm Iax} = 100 r_{\rm Iax}$, the number of SNe~Iax expected for
every 100 SNe~Ia in a given volume.

For our calculation, we will focus on SNe discovered with nearby
surveys.  We therefore exclude from our analysis SNe~2007ie, 2007qd,
and 2009ku, which were discovered (exclusively) by the SDSS-II SN
Survey and PS1.  For nearby surveys, the typical limit is $m_{\rm lim}
\approx 18.5$~mag.  Nearby surveys have varying cadences, but are
typically 5--15~days, and are usually conducted in a band similar to
$R$.  We therefore define
\begin{equation}
  M = M_{R \rm{, peak}} + A_{R \rm{, MW}} + \Delta m_{15} (R).
\end{equation}
We adjust the peak absolute magnitude by $\Delta m_{15}$ to guarantee
that the SN will be detected in at least two images.  We also correct
for Milky Way extinction since dust obscuration will affect a survey's
ability to discover SNe.

Although we do not know the SN~Iax luminosity function, we have
measured the peak absolute magnitude and decline rate for several
members.  From these measurements, we determined a set of $\mu_{\rm
max}$, which are the maximum distance moduli at which we would
detect each of those SNe.

The least luminous SN in the class is SN~2008ha, which is at $\mu =
31.65$~mag.  The nominal scheme outlined above yields $\mu_{\lim} =
31.60$~mag for SN~2008ha.  Having $\mu_{\lim} < \mu$ is likely the
result of faster cadences for very nearby galaxies; thus we choose
$\mu_{\rm lim} = 31.65$~mag for SN~2008ha.  If there are no SNe
significantly fainter than SN~2008ha, we expect the sample within this
volume to be reasonably complete, resulting in a relatively unbiased
first look at the relative rate.

In the 10 years following the discovery of SN~2002cx, there have been
22 normal SNe~Ia and 4 SNe~Iax (SNe~2008ge, 2008ha, 2010ae, and
2010el) discovered within $\mu_{\lim} = 31.65$~mag (from the Asiago SN
catalog; \citealt{Barbon89}).  Unless there are significantly more
SNe~Iax with luminosities less than SN~2008ha, this sample should be
relatively complete for monitored galaxies (with the exception of
misclassification, cadence, seasons, and weather).  Using Poisson
statistics within this volume, the uncorrected ($C_{\rm Iax} = 1$)
relative rate is $R_{\rm Iax} = 18\err{12}{9}$.  SNe~2010ae and 2010el
have peak luminosities similar to that of SN~2008ha (Stritzinger
et~al., in preparation; Valenti et~al., in preparation), while
SN~2008ge is much more luminous than SN~2008ha \citep{Foley10:08ge}.
If SNe~2008ha, 2010ae, or 2010el had exploded when they were in close
proximity to the Sun, we would not have detected the SNe, while any
SN~Ia within that distance could have been discovered more than six
months after explosion.  Because of that and other factors
(host-galaxy dust reddening, weather, etc.), we perform the same
analysis, but with $C_{\rm Iax} = 2$ for SNe~2008ha, 2010ae, and
2010el, and $C_{\rm Iax} = 1$ for SN~2008ge.  This then gives a
corrected relative rate of $R_{\rm Iax} = 30\err{21}{15}$.

We can also place a relatively robust lower limit on the relative rate
using all 19 members found by nearby surveys since the discovery of
SN~2002cx.  Of the members of the class with good $R$-band light
curves, SN~2008A is the most luminous.  Within the volume defined by
its $\mu_{\rm lim}$, there have been 573 spectroscopically normal
SNe~Ia as derived from the Asiago catalog with distances cross-matched
to NED.  Within this volume, all SNe~Ia (with the exception of those
with extreme dust extinction) should be detectable for several months.
We exclude the peculiar SNe~2006bt \citep{Foley10:06bt} and 2009dc
\citep[e.g.,][]{Yamanaka09, Tanaka10, Silverman11:09dc}, as well as
SNe with redshifts only derived from SN features (since these SNe
could be potential SNe~Iax).  Without any corrections for luminosity
or effects related to the surveys, we find $R_{\rm Iax} =
3.3\err{0.8}{0.7}$.  This method is similar to what was done by
\citet{Li11:rate2}, where only the most luminous SNe~Iax were included
in the rate calculation.  Unsurprisingly, our derived fraction is
similar to the fraction from \citet{Li11:rate2}.

We now perform a more thorough treatment with the 19 members found by
nearby surveys since the discovery of SN~2002cx, including SN~2002cx.
For this, we assume a luminosity function that has equal probability
of a SN having the luminosity of any of the 6 SNe with measured $M_{R
\rm{, peak}}$ and another, SN~2009J, which has a first measurement
that appears to be at peak.  Our sample, which is composed of SNe
discovered by magnitude-limited surveys, almost certainly
underrepresents the fraction of low-luminosity SNe~Iax, which should
result in an underestimated relative rate.

The flat luminosity function produces 7 nested volumes.  If a member
lies within the first volume (defined by SN~2008ha), it could have a
luminosity similar to any of the SNe defining the luminosity function.
We therefore assign it a fractional membership to each volume (in this
case, 1/7th of a SN per volume).  For each SN without a well-defined
peak magnitude, we attempt to determine limits on its absolute
magnitude from photometry reported in the literature and its distance
modulus.  We use this information to place further constraints on the
fraction of SNe in each volume.  For example, SN~2008ge, which was
discovered within the volume defined by SN~2008ha, is much more
luminous ($M_{V \rm{, peak}} = -17.60$~mag) than SN~2008ha.  It
therefore does not contribute to the rate of ``SN~2008ha-like'' SNe.
SNe~2010ae and 2010el have $M_{V \rm{, peak}} \approx -15.0$ and
$-14.7$~mag, respectively (Stritzinger et~al., in prep.; Valenti
et~al., in prep.).  These magnitudes were estimated from $B$ and $r$
light curves and converted using the relations outlined in
Section~\ref{s:phot}.  However, these relations may not adequately
describe SNe similar to SN~2008ha.  For instance, from the $B$-band
light curve, SN~2010ae has $M_{B \rm{, peak}} \approx -13.9$~mag, only
0.2~mag brighter than SN~2008ha, but after converting the estimated
$M_{V \rm{, peak}}$ for SN~2010ae is 0.8~mag brighter than SN~2008ha.
Because of the potential errors in the peak-magnitude estimates along
with uncertainties in reddening and distance moduli, we include
SNe~2010ae and 2010el as having equal chance of being in the first two
volumes.

\begin{figure}
\begin{center}
\epsscale{0.9}
\rotatebox{90}{
\plotone{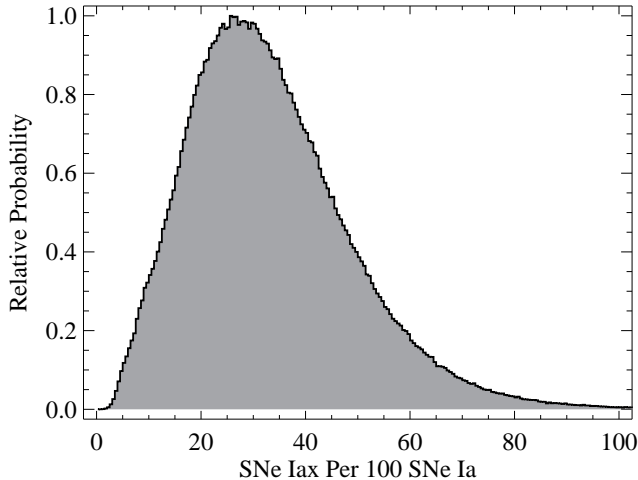}}
\caption{Probability density function for the SN~Iax rate relative to
  the SN~Ia rate.}\label{f:rate_pdf}
\end{center}
\end{figure}

For the reasons listed above as well as possible misclassifications
for fainter SNe, we assume $C_{\rm Iax} = 2$.  However, if $C_{\rm
Iax} = 2$ for the three closest volumes and $C_{\rm Iax} = 1$ for
the rest or if $C_{\rm Iax} = 1.5$ for all volumes, the relative rate
decreases by 13\% and 8\%, respectively.  The corrected relative rate
(with $C_{\rm Iax} = 2$) is $R_{\rm Iax} = 31\err{17}{13}$; however,
as seen in Figure~\ref{f:rate_pdf}, the probability distribution
function is slightly skewed with there being 3\% and 14\% chances that
the relative rate is $R_{\rm Iax} < 10$ and $R_{\rm Iax} > 50$,
respectively.

Our final, best relative rate of $R_{\rm Iax} = 31\err{17}{13}$ is
very similar to what was found using only the most nearby SNe: $R_{\rm
Iax} = 30\err{21}{15}$.  This agreement provides an additional level
of confidence for our values.  Both values are marginally consistent
with the combined \citet{Foley09:08ha} and \citet{Li11:rate2} rates of
$R_{\rm Iax} \approx 17$.

We can now estimate the amount of Fe generated by SN~Iax explosions
relative to that of SNe~Ia.  Our estimate relies on the assumption
that the luminosity of SNe~Iax are powered primarily by $^{56}$Ni and
its radioactive products; if they are powered by other radioactive
elements, the scalings assumed here would not apply.  The other
assumption is that the ratio of $^{56}$Ni to stable $^{54}$Fe is the
same (on average) for SNe Ia and SNe Iax.  For our estimate, we use
the $^{56}$Ni masses estimated by \citet{Foley09:08ha} for SN~2008ha
($(3.0 \pm 0.9) \times 10^{-3}$~$M_{\sun}$) and \citet{Phillips07} for
SN~2005hk ($0.2 \pm 0.05$~$M_{\sun}$).  Since SN~2005hk has
well-sampled UV through NIR light curves, its $^{56}$Ni mass is
relatively robust.  We then estimate the $^{56}$Ni mass for all other
SNe in our luminosity-function sample by scaling the $^{56}$Ni mass of
SN~2005hk by the ratio of their peak $R$-band luminosity.  These
estimates will be somewhat incorrect (by a factor of at most 2) if the
rise times of those SNe differ significantly from that of SN~2005hk or
if they have significantly different colors.  For our sample, the
$^{56}$Ni mass ranges from 0.003 to 0.27~$M_{\sun}$.

To determine the amount of $^{56}$Ni produced by SNe~Ia, we use the
\citet{Stritzinger06:prog} sample of SNe~Ia.  This sample, which may
be biased slightly, consists primarily of nearby SNe, and therefore
should be similar to a volume-limited sample.  From this sample, we
exclude the peculiar SN~2000cx \citep{Li01:00cx}.  \citet{Li11:rate2}
found that 74.0\%, 10.0\%, and 16.1\% of SNe~Ia in a volume-limited
sample are ``normal'', similar to SN~1991T, and similar to SN~1991bg,
respectively.  The \citet{Stritzinger06:prog} sample has similar
percentages of 67\%, 20\%, and 13\%, respectively, but slightly
over-represents and under-represents SNe similar to SNe~1991T and
1991bg, respectively.

The SNe~Ia in the \citet{Stritzinger06:prog} sample produce on
average, $0.24 \pm 0.02$~$M_{\sun}$ of $^{56}$Ni.  The median
$^{56}$Ni in the sample is $0.66 \pm 0.18$~$M_{\sun}$.  When comparing
the SN~Iax sample to SNe~Ia, we must assume that the ratio of
$^{56}$Ni generated in the explosion to the total amount of Fe
produced (both stable Fe in the explosion and that which decays from
$^{56}$Ni) is the same for both classes.  However, there is evidence
from the late-time spectra that this is not the case
(Section~\ref{ss:late}).

Using the rate calculation performed above and the average $^{56}$Ni
value for SNe~Ia, but weighting each luminosity bin and the SN~Ia
population by the appropriate $^{56}$Ni mass, we determine that
SNe~Iax produce $0.052\err{0.016}{0.014}$~$M_{\sun}$ of Fe for every
$M_{\sun}$ of Fe produced by SNe~Ia.  Using the median $^{56}$Ni value
for SNe~Ia, we find that SNe~Iax produce
$0.019\err{0.010}{0.006}$~$M_{\sun}$ of Fe for every $M_{\sun}$ of Fe
produced by SNe~Ia.  The values based on the different assumptions
about the SN~Ia $^{56}$Ni production differ by $<$2~$\sigma$.

Because of the low $^{56}$Ni yield of SN~2008ha-like objects, this
estimate is relatively unaffected by their inclusion in the rate
calculation.  Of course, the estimate of relative Fe production is
fairly uncertain.  Besides the caveats related to the rates above and
the normalization of $^{56}$Ni mass, there is also the possibility
that SNe~Iax produce significantly less stable Fe for the amount of
$^{56}$Ni produced (see Section~\ref{ss:late}).  More importantly, we
only estimate the Fe production for $z = 0$.  The SN~Ia rate rises
with redshift to at least $z \approx 1$ \citep{Graur11}, but since
SNe~Iax are associated with star-forming galaxies, one would expect
the ratio of SNe~Iax to SNe~Ia (although perhaps not the Fe
generation) to increase with redshift.  However, we estimate that at
$z = 0$, SNe~Ia produce about 0.036~$M_{\sun}$ of Fe for every
$M_{\sun}$ of Fe produced by SNe~Ia.


\section{Implications for the Progenitor System}\label{s:prog}

In the preceding sections, we have identified a number of
observational characteristics for the SN~Iax class.  Here we summarize
these observations and explore their implications for a progenitor
system.  Specifically, we have noted the following.

\begin{enumerate}

\item SNe~Iax have spectral and photometric properties that are
  similar to those of SNe~Ia, but are distinct in several ways.

\item The class has a continuum in both luminosity and ejecta
  velocity.  SN~2008ha, although on the extreme end of the spectrum
  for both quantities, is an extension of this continuum and not an
  outlier.

\item There is a large range of peak luminosity ($-14.2 \ge M_{V, {\rm
  peak}} \gtrsim -18.4$~mag) and velocity at maximum brightness
  ($2000 \lesssim |v| \lesssim 8000$~\kms).  These ranges indicate a
  large range in explosion energies, ejecta masses, and $^{56}$Ni
  masses.

\item SN~Iax light-curve shape and luminosity are correlated (a
  width-luminosity relation, or WLR), similar to that of SNe~Ia, but
  with larger scatter.

\item The color curves of SNe~Iax are similar in shape, but a simple
  reddening correction cannot reduce the scatter in all colors at
  once.  This indicates a more complicated spectral energy
  distribution at intermediate phases for SNe~Iax than for SNe~Ia.
  The lack of host-galaxy reddening corrections probably contributes
  to some of the additional scatter in the WLR.

\item There is a slight correlation between peak luminosity and ejecta
  velocity.

\item Using basic assumptions, we estimate a wide range of ejecta
  masses for SNe~Iax (from \about0.2 to possibly
  \about1.4~$M_{\sun}$), but most SNe~Iax have \about0.5~$M_{\sun}$ of
  ejecta.

\item All SNe~Iax with maximum-light spectra, including SN~2008ha,
  show clear signs of C/O burning in their maximum-light spectra.

\item There is significant mixing of the ejecta with both IMEs and
  Fe-group elements in all layers.

\item There are two SNe~Iax that show helium lines in their spectra.
  Because of radiative-transfer effects, viewing-angle dependencies,
  and other effects, perhaps all SNe~Iax have significant amounts of
  helium in their ejecta.  Since helium should not be synthesized in
  the explosion, there must be helium somewhere in the progenitor
  systems of these two events.

\item Every SN~Iax with a premaximum spectrum has some indication of
  carbon in the spectrum, with 82\% having clear absorption.  SNe~Ia,
  on the other hand, show carbon in only ~30\% of SNe Ia with
  premaximum spectra.  By mass fraction, SNe~Iax should have
  significantly more carbon than SNe~Ia.

\item Late-time spectra of SNe~Iax are very different from those of
  SNe~Ia, with no truly nebular spectrum of a SN~Iax yet being
  observed.

\item In detail, the late-time spectra of SNe~Iax differ significantly
  from object to object.  Specifically, the strengths of various
  forbidden lines and the velocities of both the permitted and
  forbidden lines vary significantly between different SNe.

\item Unlike SNe~2002cx and 2005hk \citep{Jha06:02cx, Sahu08}, not all
  SNe~Iax have exclusively low-velocity features in their late-time
  spectra.  Some have [\ion{Fe}{2}] FWHMs of \about 3000~\kms.

\item All SNe~Iax for which we have late-time spectra have calcium
  interior to iron.  This is the opposite of what is seen in SNe~Ia.

\item In the late-time spectra, the [\ion{Ni}{2}] $\lambda 7378$
  feature is typically stronger than the [\ion{Fe}{2}] $\lambda 7155$
  feature.  This indicates that SNe~Iax either produce much more
  stable Ni or much less stable Fe than SNe~Ia.  This is a prediction
  of a low-energy, low-density explosion \citep[e.g.,][]{Travaglio04}.

\item The large-velocity line shifts of forbidden nebular lines
  suggest that the explosion was asymmetric.

\item The late-time spectra can be divided into distinct groups having
  lower and higher velocities.  The lower-velocity group tends to have
  blueshifted Ca and redshifted Fe, as determined by velocity shifts
  in forbidden lines, along our line of site.  The opposite is true
  for the higher-velocity group.  There is no evidence for a
  correlation between the line shifts (or equivalently velocities at
  late times) and maximum-light observables.

\item Because of their large range in luminosity, SNe~Iax have been
  undercounted in previous rate estimates.  Through a basic analysis,
  we find that in a given volume, for every 100 SNe~Ia there should be
  $31\err{17}{13}$ SNe~Iax, although even this number is likely a low
  estimate.

\item Although SNe~Iax are relatively numerous, a large fraction of
  the population consists of low-luminosity events which likely
  produce little Fe.  We estimate that SNe~Iax produce \about
  0.036~$M_{\sun}$ of Fe for every $M_{\sun}$ of Fe produced by
  SNe~Ia.

\end{enumerate}

In addition to these findings, others have noted the following.

\begin{enumerate}
  \setcounter{enumi}{20}

\item A lack of a second maximum in the NIR \citep{Li03:02cx}.

\item A host-galaxy morphology distribution highly skewed to late-type
  galaxies \citep{Foley09:08ha, Valenti09}.

\item SN~2005hk had very low polarization near maximum brightness,
  indicating nearly spherical outer ejecta \citep{Chornock06,
  Maund10:05hk}.

\item A lack of X-ray detections with 5 SNe~Ia having 3$\sigma$
  luminosity limits below $9 \times 10^{39}$~erg~s$^{-1}$ and 3
  (SNe~2008ge, 2008ha, and 2010ae) below $2.7 \times
  10^{39}$~erg~s$^{-1}$ \citep{Russell12}.  These limits are below
  most, but not all, detections and limits for stripped-envelope
  core-collapse SNe at early times \citep{Perna08}.

\item The host galaxy of SN~2008ge is an S0 galaxy with no signs of
  star formation to deep limits.  Additionally, pre-explosion {\it
  HST} images indicate that there are no massive stars at or near
  the SN site to deep limits.

\end{enumerate}

It is remarkable that a long list of common properties applies to a
large sample given the small number of observational criteria used to
define the class.  This is a clear indication that the SNe classified
as Type Iax are physically similar, representing a physical class and
not simply an observational class.

The above list of properties for SNe~Iax provides robust tests for
progenitor/explosion models.  Below we examine the various properties
to constrain the set of possible scenarios.  During this analysis, we
assume that all members identified as SNe~Iax share a common
progenitor system and explosion mechanism, although the details of
that explosion may vary.  However, it is possible that we have
incorrectly grouped physically distinct SNe into this single class.
There may also be multiple progenitor paths to create a SN~Iax.
Because of the relatively small data, we do not attempt to further
subdivide the class.

First, there are many reasons to believe that the progenitor star was
a WD.  The progenitor of SN~2008ge was most likely a WD with no
indication of massive stars near the SN site and strict limits on star
formation in its host galaxy.  More broadly, the estimated ejecta
masses are all equal to or below a Chandrasekhar mass.  If the
progenitors were massive stars, then one would expect that at least
occasionally this limit would be exceeded.  Additionally, SNe~Iax over
the full luminosity range shows clear signs of C/O burning, which is
not expected in a core-collapse SN.

However, complete thermonuclear burning of a WD cannot reproduce the
velocities observed and ejecta masses inferred.  Roughly, a
0.5~$M_{\sun}$ WD will be completely unbound and have ejecta
velocities $>$2000 (6000)~\kms\ for pure helium to nickel or pure
carbon to nickel burning efficiencies of $>$4\% and 7\% (15\% and
24\%), respectively.  That is, if 10\% of the carbon in a
0.5~$M_{\sun}$ WD is burned to $^{56}$Ni, then the star will be
unbound and the ejecta velocity will be larger than 6000~\kms.  This
implies incredibly inefficient burning, either by traditional burning
of a small percentage of the WD mass or by a more exotic, less
efficient nuclear burning through the entire star.

Furthermore, SNe~2004cs and 2007J have helium in their spectra, and
therefore there must be helium in the progenitor system.  There is no
hydrogen in any spectra, suggesting that hydrogen is most likely
absent in the progenitor system.  This indicates that either a He WD
or a nondegenerate He star is in the system.  The ejecta mass from a
He WD explosion should not exceed \about 0.4~$M_{\sun}$, making it an
unlikely progenitor star.  It is possible that a He WD could be the
binary companion of the primary star, but then one would expect there
to be many SNe~Iax in early-type host galaxies.  Having any He WD in
the progenitor system is unlikely, and instead, a He-burning-star
companion to a C/O WD is the most likely progenitor system.

\citet{Tutukov96} predicted that C/O-He-burning-star progenitor
systems could have a rate similar to the Galactic SN~Ia rate, but
would not explain the SNe~Ia in older stellar populations.  Performing
a detailed population-synthesis analysis, \citet{Ruiter11} found that
13\% of sub-Chandrasekhar mass thermonuclear SNe could come from a C/O
WD-He star channel.  Assuming that this channel produces SNe~Iax and
other channels produce SNe~Ia, the population-synthesis rate is
roughly consistent with our measured rate.  This channel has a
relatively short delay time of \about 500~Myr, which should restrict
the host-galaxy population to mostly spiral galaxies, but could
include some SNe in S0 galaxies (like SN~2008ge).

A single system has been identified as a C/O WD with a He-burning-star
companion for which mass transfer will begin before the end of the
He-burning phase \citep{Geier12, Vennes12}.  This system should be
able to transfer mass for \about 50~Myr at a rate of \about $10^{-9}
M_{\sun}$~year$^{-1}$ for a potential total mass of \about
0.05~$M_{\sun}$ of material.

At certain accretion rates, He mass transfer to a C/O WD can occur
without periodic He flashes, resulting in a massive He shell.  In this
scenario \citep[e.g.,][]{Iben91, Tutukov96}, after the WD acquires a
significant He layer, the He layer can detonate, possibly causing a
thermonuclear explosion within the C/O layer \citep{Livne90,
Woosley94, Livne95}; this is sometimes called an ``edge-lit''
explosion.

For C/O WDs with large He layers, most theoretical studies have
focused on ``double-detonation'' models where the He layer detonates
\citep[e.g.,][]{Nomoto82, Sim12, Townsley12}, which in turn causes a
detonation with the C/O layer \citep[e.g.,][]{Nomoto82, Fink10,
Sim12}.  Although there is still some discussion about the details
of these explosions, the consensus is that it is difficult to
reproduce the full diversity of SNe~Iax with this mechanism.

In addition to ``double-detonation'' models, it is possible to
detonate the He layer without causing the rest of the star to explode.
This ``.Ia'' model \citep{Bildsten07, Shen10} shares many
characteristics with SNe~Iax.  \citet{Shen10} and \citet{Waldman11}
examine the details of this model for high-mass and low-mass C/O
cores, respectively.  These models roughly match the characteristics
and the diversity of SNe~Iax, but it is unclear if they reproduce the
explosions in detail.  In particular, it is difficult to produce IMEs
in these explosions, particularly for those on top of a high-mass
core.  These explosions also have low luminosity, so it appears
difficult to reproduce SN~2002cx and more luminous SNe using this
model.  \citet{Foley09:08ha} examine how these models match
observations in detail.

The single C/O WD/He-burning-star system should produce a helium layer
that is \about 0.05~$M_{\sun}$ to perhaps 0.3~$M_{\sun}$
\citep{Iben91}.  Although this mass may not be enough to account for
all of the ejecta in SNe~Iax, it should be massive enough to detonate,
perhaps triggering a double-detonation on a sub-Chandrasekhar mass SN
\citep{Fink10}.  \citet{Sim12} examined double-detonation models for
relatively low-mass sub-Chandrasekhar C/O WDs with \about
0.2~$M_{\sun}$ He shells.  The resulting observables were somewhat
consistent with some SNe~Iax ($\Delta m_{15} \approx 2$~mag, $t_{\rm
  rise} \approx 10$~days, low luminosity, low velocity).

\citet{Woosley11} explored possible sub-Chandrasekhar-mass explosion
models, some of which had a He-layer detonation or deflagration.  The
models, which only burned the He layer, and are somewhat similar to
the .Ia model of \citet{Bildsten07}, roughly match the observables of
SNe~Iax.  However, these models should be exclusively low luminosity,
low ejecta mass, and contain a significant amount of unburned He.
These models have difficulty reproducing all of the characteristics of
all SNe~Iax and the full ranges of those characteristics, particularly
the luminosity range.  Because of non-thermal effects that are not fully modeled in the radiative transfer, it is
unclear if He lines should be strong in the spectra of these events.

A promising model for SNe~Iax is a failed deflagration of a C/O WD.
In this model, a burning flame within the WD rises to the surface,
expelling \about 0.2--0.4~$M_{\sun}$ of material, some of which is the
ash from the burning.  In this model, the full WD is not consumed by
the nuclear burning, and the entire star is not disrupted.  This model
was first proposed by \citet{Foley09:08ha} as a possible explanation
for SNe~Iax.  Model luminosity, ejecta mass, and velocity are all
consistent with SNe~Iax \citep{Jordan12, Kromer12}.  It is unclear if
this scenario requires a thick He shell or if the He shell potentially
changes the explosion properties, but it is possible that a He-shell
explosion could trigger a failed deflagration within the C/O core.
Stable helium accretion occurs at a higher rate than stable hydrogen
accretion, and thus the WD density and temperature structure are
different for the two scenarios.  These differences may be important
in understanding the nature of SN~Iax explosions.  Alternatively,
since a He layer appears to be important in creating SNe~Iax, one
might assume that an initial He-layer explosion is necessary to
produce SNe~Iax.  Detailed modeling of nucleosynthetic products and
the resulting light curves and spectra are still required, but one
expects products similar to those of a normal SN~Ia with perhaps
additional unburned material.  This model continues to be the most
promising yet proposed.

Assuming that all SNe~Iax come from the same progenitor system and
have the same explosion mechanism, the most promising model for
SNe~Iax is the deflagration of a C/O WD, either triggered by an
explosion in an outer He layer or under conditions only met when
stable He accretion occurs.  The C/O WD would have a He-star binary
companion.  A scenario very similar to this was first proposed by
\citet{Iben91}.  At least some of the time, and perhaps in all cases,
the deflagration fails to unbind the WD, leaving a stellar remnant
that may be observationally distinct from most WDs.  Specifically,
given the large velocity shifts for nebular lines in the late-time
spectra of SNe~Iax, one might expect a large kick for the remnant.
Similarly, the energy injection and chemical enrichment may produce a
puffed-up WD with peculiar abundances \citep{Jordan12, Kromer12}.
This remnant may have particular observational signatures.

\citet{Perets10:05e} and \citet{Sullivan11:09dav} both suggested
similar explosion mechanisms, particularly He detonations on C/O WDs,
for SN~2005E and PTF~09dav, respectively.  These SNe share many
characteristics with SNe~Iax, including luminosity, velocity, and
ejecta mass.  However, both SN~2005E and PTF~09dav appear to come from
old stellar populations (although PTF~09dav had hydrogen in its
late-time spectrum; \citealt{Kasliwal12}).  In particular, the
host-galaxy morphology distribution of SNe similar to SN~2005E is
highly skewed to early-type galaxies \citep{Perets10:05e}, the
opposite of SNe~Iax.  SN~2005E and similar objects all show relatively
strong He at early times, again indicating He in the progenitor
system.  Perhaps SNe~Iax are created in the above scenario and
SN~2005E was produced in a very similar scenario, except with a He WD
companion instead of a nondegenerate He star.  This difference would
account for the difference in stellar populations, and the accretion
of degenerate and nondegenerate He could change the stable accretion
rates and/or nucleosynthesis enough to create noticeably different
events.


\section{Conclusions}\label{s:conc}

We have described in detail the observational properties of the Type
Iax class of SNe.  A SN is a member of this class if it has four
observational properties, three of which can be determined from a
single maximum-light spectrum: (1) no evidence of hydrogen in any
spectrum, (2) relatively low ejecta velocity near maximum brightness
($|v| \lesssim 8000$~\kms), (3) an absolute magnitude that is low
relative to a SN~Ia with the same light-curve shape, and (4) spectra
that are similar to SN~2002cx at similar epochs.  Using these
criteria, we have identified 25 members of this class.  We presented
optical photometry and spectroscopy for various members of this class.

SNe~Iax represent a distinct stellar endpoint from SNe~Ia and other
common SNe.  The class has a large range of luminosities, ejecta
velocities, and inferred ejecta masses.  Although there are clear
signs of C/O burning in their ejecta, some SNe~Iax also have
\ion{He}{1} lines in their spectra.  As a possible progenitor system,
a C/O WD that accretes from a nondegenerate He star and undergoes a
(sometimes partial) deflagration matches the SN observables.  The
lifetimes of such systems are consistent with the SN~Iax host-galaxy
morphology distribution.  Similarly, the rates of explosions from
these systems roughly match the measured rates of $31\err{17}{13}$
SNe~Iax for every 100 SNe~Ia.  This is perhaps a simplified view of
the real physical picture, and future observations will test this
model.

Additional studies of the details of the properties of SNe~Iax should
improve our knowledge of both this class of SN as well as providing
insight into normal SNe~Ia.  For instance, if C/O WD-He-star systems
exclusively produce SNe~Iax, then we can rule out these systems as
normal SN~Ia progenitors.

SNe~Iax are relatively common (about a third as common as SNe~Ia at $z
= 0$), and deep transient surveys of the local universe should
discover a significant number.  LSST is expected to discover \about
$10^{6}$ SNe over its lifetime.  Even taking the conservative,
magnitude-limited rate for SNe~Iax, LSST should discover more than
$10^{4}$ SNe~Iax, similar to the number of SNe~Ia discovered to this
day.  Additionally, because of their frequency, we will continue to
discover many in the very local universe ($D \lesssim 20$~Mpc),
allowing detailed studies of individual objects and their stellar
environments.  The combination of thorough studies of local events and
an upcoming large sample will undoubtedly improve our understanding of
this recently discovered class of stellar explosion.

\begin{acknowledgments} 

  {\it Facilities:} \facility{du~Pont(Tek5)},
  \facility{ESO:3.6m(EFOSC)}, \facility{FLWO:1.5m(FAST)},
  \facility{Keck:I(LRIS)}, \facility{KAIT}, \facility{Shane(Kast)},
  \facility{Magellan:Baade(IMACS)}, \facility{Magellan:Clay(LDSS3)},
  \facility{NTT(EMMI)}, \facility{PROMPT}, \facility{Swope(SITe3)}

  \bigskip R.J.F.\ is supported by a Clay Fellowship.  Supernova
  research at Harvard is supported by NSF grant AST--1211196.
  A.V.F.'s supernova group at U.C. Berkeley is supported by Gary \&
  Cynthia Bengier, the Richard \& Rhoda Goldman Fund, the Christopher
  R. Redlich Fund, the TABASGO Foundation, and NSF grants AST-0908886
  and AST-1211916. KAIT and its ongoing operation were made possible
  by donations from Sun Microsystems, Inc., the Hewlett-Packard
  Company, AutoScope Corporation, Lick Observatory, the NSF, the
  University of California, the Sylvia \& Jim Katzman Foundation, and
  the TABASGO Foundation.  CSP material is based upon work supported
  by the NSF under grants AST-0306969, AST-0607438, and AST-1008343.
  M.S.\ acknowledges generous support provided by the Danish Agency
  for Science and Technology and Innovation through a Sapere Aude
  Level 2 grant.  J.A., M.H., and G.P.\ acknowledge support provided
  by the Millennium Center for Supernova Science through grant
  P10-064-F (funded by ``Programa Bicentenario de Ciencia y
  Tecnolog\'{i}a de CONICYT'' and ``Programa Iniciativa Cient\'{i}fica
  Milenio de MIDEPLAN''), with input from `Fondo de Innovaci\'{o}n
  para la Competitividad, del Ministerio de Econom\'{i}a, Fomento y
  Turismo de Chile'.  J.A.\ and G.P.\ also acknowledge support by
  CONICYT through FONDECYT grants 3110142 and 11090421, respectively.
  Support for this research at Rutgers University was provided in part
  by NSF CAREER award AST-0847157 to S.W.J.
  
  We thank all of the amateur astronomers who have searched for
  supernovae.  The discovery of many SNe~Iax is the result of their
  tireless effort.  We especially thank C.\ Moore, discoverer of
  SN~2008ha.  We gratefully acknowledge L.\ Bildsten, A.\ Gal-Yam, M.\
  Hicken, D.\ Leonard, R.\ Margutti, T.\ Matheson, K.\ Moore, J.\
  Nordin, L.\ \"{O}stman, H.\ Perets, T.\ Piro, E.\ Ramirez-Ruiz, D.\
  Sahu, K.\ Shen, V.\ Stanishev, and S.\ Valenti for providing data,
  insights, and/or help for this study.  R.\ Assef, A.\ Barth, V.\
  Bennert, P.\ Berlind, P.\ Blanchard, G.\ Canalizo, B.\ Cenko, K.\
  Clubb, D.\ Cohen, A.\ Diamond-Stanic, G.\ Folatelli, S.\ Hoenig, J.\
  Irwin, M.\ Lazarova, N.\ Lee, A.\ Miller, A.\ Morgan, P.\ Nugent,
  D.\ Poznanski, J.\ Rex, A.\ Sonnenfeld, T.\ Steele, B.\ Tucker, and
  J.\ Walsh helped obtain some data presented in this paper; we thank
  them for their time. We are indebted to the staffs at the La Silla,
  Las Campanas, Lick, Keck, and Fred L.\ Whipple Observatories for
  their dedicated services.

  Some of the data presented herein were obtained at the W.~M. Keck
  Observatory, which is operated as a scientific partnership among the
  California Institute of Technology, the University of California,
  and the National Aeronautics and Space Administration (NASA); the
  observatory was made possible by the generous financial support of
  the W.~M. Keck Foundation.  This research has made use of the
  NASA/IPAC Extragalactic Database (NED), which is operated by the Jet
  Propulsion Laboratory, California Institute of Technology, under
  contract with NASA.  The analysis pipeline used to reduce the DEIMOS
  data was developed at UC Berkeley with support from NSF grant
  AST-0071048.  NTT data was obtained under ESO Programme was
  080.A-0516.

  Part of the analysis occurred when R.J.F.\ visited the Aspen Center
  for Physics during the Summer 2012 workshops, ``Non-Gaussianity as a
  Window to the Primordial Universe'' and ``The Evolution of Massive
  Stars and Progenitors of GRBs'', as well as during his time at the
  Center as an independent researcher.  This material is based upon
  work supported in part by the National Science Foundation under
  Grant No.\ 1066293 and the hospitality of the Aspen Center for
  Physics.  Some analysis was performed at the Woody Creek Community
  Center in Woody Creek, CO.

\end{acknowledgments}

\bibliographystyle{../fapj}
\bibliography{../astro_refs}


\end{document}